\documentclass[11pt,a4paper,jhep]{article}
\usepackage{jheppub}
\usepackage{mathtools}
\usepackage{amsmath,amsthm,amssymb,scrextend}
\usepackage{bbold}
\usepackage{slashed}
\usepackage{dsfont}
\usepackage[colorlinks=true]{hyperref}
\hypersetup{linkcolor=blue}
\usepackage{comment}
\usepackage{caption}
\usepackage{braket}
\usepackage{multirow}
\usepackage{graphicx}
\usepackage{subcaption}
\usepackage[capitalise,noabbrev,nameinlink]{cleveref}
\usepackage{diagbox}

\newcommand{\myparagraph}[1]{\paragraph{#1}\mbox{}\\}
\newcommand{\kp}{\boldsymbol{k}_\perp}

\newcommand{\Tr}{{\rm Tr}\,}

\newcommand\eqtr{\stackrel{\mathclap{\normalfont\mbox{\Tr}}}{~\equiv ~}}

\renewcommand{\arraystretch}{1.2}

\title{  Mean field theory for strongly coupled systems: Holographic approach}
\author[a]{Supalert Sukrakarn,}
\author[a]{Taewon Yuk}
\author[a]{and Sang-Jin Sin}
\affiliation[a]{Department of Physics, Hanyang University, Seoul 04763, South Korea}
\emailAdd{supalertee@gmail.com, tae1yuk@gmail.com, sangjin.sin@gmail.com}
\abstract{ In this paper, we develop the holographic mean field theory  for strongly interacting fermion systems.  We investigate  various types of the symmetry-breakings and their effect on the spectral function. We found analytic expressions of fermion Green's functions in the probe-limit for all types of  tensor order parameter fields.    We classified  the spectral shapes and singularity types  from the analytic Green's function. We   calculated the fermions spectral function in the full backreacted background and then compared it with the analytic results to show the reliability of analytic results in the probe limit.}

\begin{document}
  \newpage
	\maketitle
\section{Introduction}

Study on strongly interacting system has been the frontier of physics for more than  half century. It has been the  continuous source of motivation  for a new theoretical framework.  Along this line,  the gravity dual picture  \cite{Maldacena:1997re,Witten:1998qj,Gubser:1998bc,Alvarez:1998wr,Balasubramanian:1999jd,deBoer:1999tgo} has been suggested 
as a new   paradigm  to understand strange metals \cite{Hartnoll:2018xxg,zaanen2015holographic} and quantum critical  points (QCP) \cite{Sachdev:2011mz,Wilson:1971bg,Wilson:1974mb} of strongly coupled systems.  A few experimental data were also compared with the theory showing   agreements,  especially in quark gluon plasma \cite{Kovtun:2004de}, in clean graphene \cite{doi:10.1126/science.aad0343,Lucas:2015sya,Seo:2016vks,Seo:2017oyh,Seo:2017yux} and surface of topological insulators \cite{PhysRevLett.108.036805,PhysRevB.86.205127}.  

 On the other hand,  QCP itself    is a phenomena  at zero temperature taking a measure zero sector of the phase diagram. Therefore for the theory to be compared with  experiment, it is essential to consider   the symmetry  broken phases as well as the unbroken phase near the QCP.  
Such symmetry breaking is an essential step for new collective phenomena, because  a phenomenon associated with manybody collective phenonena is  possible only when we have an order that leads to a  gap which  protect the new ground state from the   excitations causing disorder.  Otherwise, what we will have is just fermi liquid. 
While  gap calculation is  usually a consequence of a mean field theory which is seeking for  value of the condensation corresponding to the certain fermion bilinear, its validity is limited to   the weakly interacting  cases. Therefore a dream for  theorists is to develop a mean field theory  which is valid for the strongly interacting system  maintaining  the  spirit of Landau-Ginzburg and Wilson.  

The  strongly coupled  systems share many similarities with the gravity system   in the sense that both  have unreasonable speed of equilibration: the  Plankian dissipation near the QCP   is very similar to the exponentially fast scrambling power \cite{Shenker:2014cwa,Lashkari:2011yi} of black hole horizon. Therefore it is not  very surprising to expect a mean field theory for the strongly interacting system out of the gravity dual description. While it is hard to find the exact gravity dual of a given system, it is reasonable to assume the presence 
of approximate dual   theory for a strongly interacting system.  It can serve as a representative theory for the purpose of studying various types of the  gaps  and  condensations  as well as the singularity types  of the Green functions for the strongly interacting systems. 

Notice that in the holographic superconductor theory  \cite{Hartnoll:2008kx,Gubser:2008px},   the gap calculation was exemplified by considering scalar-vector-gravity theory, which is an important ingredient of the holographic mean field theory (MFT). 
However, the heart of the MFT is  to study the gap creation and  the fermion spectrum  together, which has not been done   systematically yet for different types of condensations. 
 Furthermore, because the holographic theory as a continuum field  theory does not encode  condensed matter system's detail it is useful to study all possible  type of interactions  together and classify their spectral behavior  to match   with  physical system's spectral pattern.   In ref. \cite{AdS4}  we put forwarded  a step to such  direction  by considering all possible couplings of the bulk fermion bilinears with the various tensorial fields representing the  different condensations  in holographic set-up, which is analogue  of the femion bilinear coupled  to the Hubbard-Stratonovich field in the usual mean field theory. 
 
In our previous work \cite{AdS4}, the fermion spectral function (SF) \cite{Lee:2008xf,Cubrovic:2009ye,Nonfermi,Iqbal:2009fd} of each  broken symmetry phase was  calculated  and classified. 
Subsequently, we realized that \cite{Strangemetal} the  gapless mode in the scalar coupling  is the AdS analogue of the surface mode of topological insulator whose topological stability might be related to amazing stability of the Fermi Liquid. 
The holographic  mean field theory has   universal structure so that it has the power to accommodate various different phenomena in the same fashion. Just as the  superconductivity is described by the condensation of itinerant electrons $\Delta_S \sim cc$, one may also expect a new quantum ground state of manybody coming from the condensation of the itinerant-localized electrons with spin. 
Such many Kondo phenomena would be  also   described by considering the scalar condensation $\Delta_K\sim f^\dagger c$ if the mean field theory works for strongly interacting case like Kondo system. 
Recently, we realized  such idea   in the holographic set-up \cite{Im:2023ffg}  by  finding  that  the Kondo condensation $\Delta_K$ produces a  gap, which was also  observed experimentally \cite{Im:2023ffg}.  

However, the ref.   \cite{AdS4} has several limitations: i) it is completely numerical so that it is hard to  characterize the singularity types of the spectral  functions. ii) the back reaction was not taken care of.  iii) it was confined to the AdS$_4$ so that it is not possible to discuss the Weyl semi-metal spectrum from the beginning. 
In this paper,   we will find the analytic expressions of the  
Green's function for all tensor types of symmetry breaking in the probe limit.  
 Three  aspects of the spectral function will be emphasized: i) gap/gapless, ii) presence/absence of flat bands of various types, iii)  presence/absence of split Dirac cones. Especially interesting coupling turned out to be  the scalar type  coupling which can give both  gapped and gapless spectrum depending on the sign of the coupling, which can never be possible in flat space theory. 

From the analytic expressions, we noticed that most of the Green functions have branch cut singularities but some of them have poles and we  can now understand why various different types of flat band exist.  
 The  definition of  the  non-fermi liquid is  the vanishing of the quasi-particle weight  assuming that the Green function has a pole type singularity. 
 In our holographic mean field theory, it turns out that the generic form of the fermion propagator  does not have pole   singularity  but has the branch cut singularity. That is, the fluid appearing in our theory is  generically  non-fermi liquid. Interestingly, in some cases, such behavior can happen even after order parameter is turned on. 
  However, for a few types of the order parameter, the pole type singularity appears. Typically,   it happens  when 
 the flat band appears. 
 
 At this moment, our analytic expressions   are only  for the case where  order parameter fields  are alternatively quantized where only the leading term is nonvanishing    in the pure AdS background.  In standard quantization where only the subleading term of the  matter field is  nonvanishing, we still  have  not find exact Green functions even in the probe limit.  
To back up this limitation, we performed and presented  the numerical analysis to find fully back reacted solutions of antisymmetric 2-tensor field, which is the most important and complicated case. We  calculated the spectral function of the fermions based on such backreacted solution, and   compared with the analytic result of the probe limit. 
In case the singularity is of pole type, we observed a detailed agreement  between the approximate analytic result and the fully back reacted numerical result, which suggests the stability of Green function with the pole type singularity. In contrast, for the propagators with branch-cut type singularity, the detail of the spectrum is relatively vulnerable to the deformation by back-reaction, although qualitative similarity remains.  
  
The rest of this paper goes as following. 
In section 2, we give a short description  of  formalism of the holographic mean field theory and fermion Green function calculation. 
In section 3, we  calculate Green's function analytically and draw spectral function for all types of the Lorentz symmetry breaking. 
In section 4, we draw spectral functions and discuss features of them.
In section 5, we do extensive numerical calculation to find the fully back-reacted order parameter and metric and use them to calculate the  fermion spectral function again to 
support the probe limit analytic results. In section 6, we discuss and conclude.

%
\section{Holographic mean field theory  with symmetry breaking}\label{section1}
Our  holographic mean field theory has four components: 1st, bulk fermions $\psi$ which are dual to the boundary fermions  with strong interactions. 2nd,  order parameter fields $\Phi^I$ describing the  condensations of fermion bilinear which is the analogue of  the Hubbard-Statonovich field of the  usual mean field theory.   3rd,  the   gravity describing the interactions between  electrons, and finally the gauge field which should  be included  to describe the density or chemical potential of the fermions. In this work, we do not include the vector for the analytic work but we can do it when we work numerically. 
For simplicity, we assume that they can be described by local fields in the bulk not in the boundary, because we should allow the  strongly interacting system in the boundary  be described by  nonlocal field theory.  
 %
 The total action is given by   \cite{Iqbal:2009fd}
\begin{align}
    S_{total} &= S_\psi +S_{bdy}+S_{g,\Phi} +S_{int},\\
    S_\psi &= \int d^{5}x \sum_{j=1}^2 \sqrt{-g}~ \bar\psi^{(j)}\Big(\frac1{2} ({\overrightarrow{\slashed{D}}-\overleftarrow{\slashed{D}}})-m^{(j)}\Big)\psi^{(j)},\label{eq:setup1} \\
    S_{g,\Phi} &= \int d^{5}x \sqrt{-g}\Big(R-2\Lambda + |D_M\Phi_I|^2-m^2_{\Phi}|\Phi|^2\Big),\\
  S_{bdy} &= \frac{i}{2} \int_{bdy} d^4x \sqrt{-h} \Big(\bar{\psi}^{(1)}\psi^{(1)}\pm \bar{\psi}^{(2)}\psi^{(2)}\Big),\label{eq:setup2}\\
    S_{int} &= \int d^{5}x \sqrt{-g} \Big(\bar\psi^{(1)}\Phi\cdot\Gamma \psi^{(2)}+h.c \Big).
\end{align}
where $\slashed{D} =  \Gamma^M(\partial_M + \frac{1}{4}\omega_{M\alpha\beta}\Gamma^{\alpha\beta}$), $\omega_{M\alpha\beta}$ is the spin connection, $\Phi\cdot \Gamma = \Gamma^{\underline{\mu_1}\underline{\mu_2}\cdots \underline{\mu_I}}\Phi_{\underline{\mu_1}\underline{\mu_2}\cdots \underline{\mu_I}}$. $\Phi^I$ is the order parameter field which couples with bilinear spinor in the bulk, leading to the symmetry breaking under the presence of the source or  its condensation. Additionally, we will turn on just one component of field $\Phi$  to calculate the spectral function. The gamma matrix convention and the geometry are chosen and given as follows,
\begin{align}
\Gamma^{\underline{t}} &= \sigma_1 \otimes i\sigma_2, \quad \Gamma^{\underline{x}} = \sigma_1 \otimes \sigma_1,\quad \Gamma^{\underline{y}} = \sigma_1 \otimes \sigma_3,\quad
\Gamma^{\underline{z}} = \sigma_2 \otimes \sigma_0,\quad
\Gamma^{\underline{u}} = \sigma_3 \otimes \sigma_0\\
\centering
ds^2 &= \frac{1}{u^2}(dt^2+\sum_{i=1}^{3}d\vec{x_i}^2+du^2),\quad f(u) = 1, \quad h = gg^{uu}, \quad u_h = \infty, \label{Gammamu}
\end{align}
where the underlined indices represent tangent space ones. Under this convention, the boundary locates at $u = 0$.

Notice that  in $\text{AdS}_5$,  including   two-flavors of fermions  is   mandatory  because holography   projects out half of the fermion degrees of freedom while  we need a full 4 component spinor in the 4 dimensional boundary. On the other hand, in $\text{AdS}_4$, considering one flavor is still allowed since the boundary is of 2+1 dimension where  spinors are of  two components. 
 
 We will analytically determine and analyze the fermions' Green's function in the presence of an order parameter field. So, we consider the absence of gauge field  for simplicity. Furthermore, we will begin our analysis in the probe limit  and later we will eventually calculate the spectral function in the full back-reacted background. We will compare it with the probe limit analytic results to check the reliability of the latter.

\subsection{Variational analysis and boundary actions}
In this section, we will perform the variational analysis in   detail to show the boundary fermions in   different quantization choices.  The standard-standard (SS) and standard-alternative (SA) quantization   can be   distinguished by the sign of the boundary action (\ref{eq:setup2}). We first simplify the action by introducing $\zeta^{(j)}$,
\begin{align}
\psi^{(j)} = (-gg^{uu})^{-1/4}\zeta^{(j)}e^{-i\omega t + ik_x x +ik_yy +ik_z z}.
\end{align}
Then,   the variation of bulk fermions action (\ref{eq:setup1}), after the equation of motion is imposed can be written as a boundary term given below. 
\begin{align}
 \delta S_{bulk} &= \frac{i}{2} \sum_{i=1}^{2} \int d^4x \Big[ \bar\zeta_-^{(i)}\delta\zeta^{(i)}_+ - \bar\zeta^{(i)}_+\delta\zeta^{(i)}_- - \delta\bar\zeta_-^{(i)}\zeta^{(i)}_+ + \delta\bar\zeta^{(i)}_+\zeta_-^{(i)}\Big].
\end{align}
If we add the variation of boundary action with sign $\pm$ (\ref{eq:setup2}) depending on   SS and SA quantization  respectively, the variation of total action is given by
\begin{align}
\delta S_{tot}^{(SS)} &= \frac{i}{2} \int_{bdy} d^4 x \Big( \bar\zeta^{(1)}_-\delta\zeta^{(1)}_+ + \delta\bar\zeta^{(1)}_+\zeta^{(1)}_- + \bar\zeta^{(2)}_-\delta\zeta^{(2)}_+ + \delta\bar\zeta^{(2)}_+\zeta^{(2)}_-\Big) \label{tot1},\\
\delta S_{tot}^{(SA)}  &= \frac{i}{2} \int_{bdy} d^4 x \Big( \bar\zeta^{(1)}_-\delta\zeta^{(1)}_+ + \delta\bar\zeta^{(1)}_+\zeta^{(1)}_- - \bar\zeta^{(2)}_+\delta\zeta^{(2)}_- - \delta\zeta^{(2)}_-\zeta^{(2)}_+ \Big)\label{tot2}.
\end{align} 
From this expression, we see what are chosen as independent 
degree of freedom to makes the variation of total action zero. We call such independent fermions  as the source fermions.  We define a 4 component spinor $\xi$'s by 
\begin{align}
 \xi^{(SS)}_{S} \coloneqq (\zeta^{(1)}_+, \zeta^{(2)}_+),  \quad \hbox{
and } \quad \xi^{(SS)}_C \coloneqq  (\zeta^{(1)}_- , \zeta^{(2)}_-),
\end{align}
as the boundary spinors for SS-quantization.  The indices $S, C$ in  $\xi_S$ and $\xi_C$ are adopted since they correspond to the source and condensation terms. The  $\xi_C$'s supposed to be determined by $\xi_S$'s. 
Similarly, for the SA-quantization we  define  
\begin{align}
 \xi^{(SA)}_{S} \coloneqq (\zeta^{(1)}_+,\zeta^{(2)}_-),  \quad \hbox{
and } \quad \xi^{(SA)}_C \coloneqq (\zeta^{(1)}_- , \zeta^{(2)}_+).
\end{align}
One should remember that all $\zeta_\pm$ are two component spinors while $\xi$'s are 4 component ones. 
The extension $\xi(u)$ of these fermions  to the bulk of the AdS can also be considered  by the original $\zeta(u)$'s whose boundary values were used above as $\zeta$, so that 
\begin{equation}
\xi(u)= u^{m_s}\xi_{S} + u^{m_c}\xi_{C} + \cdots . 
\end{equation}

 We can  now   rewrite  the on-shell effective action   as  
\begin{align}
S^{(SS)}_{tot} &= \frac{1}{2} \int_{bdy}d^{4}x \Big(\xi^{(SS)\dagger}_{S}(-\sigma_0\otimes \sigma_2)\xi^{(SS)}_{C}+h.c\Big)\label{totbdySS},\\
S^{(SA)}_{tot} &= \frac{1}{2} \int_{bdy}d^{4}x \Big(\xi^{(SA)\dagger}_{S}(-\sigma_3\otimes \sigma_2)\xi^{(SA)}_{C}+h.c\Big)\label{totbdySA}.
\end{align}

\subsection{Green's function}
We can determine the boundary Green's functions from the  the effective actions (\ref{totbdySS})-(\ref{totbdySA}) and the definition of source and condensation fermions.  
Since we have 4 components of $\xi(u)$, there must be 4 independent solutions ${\bf \Psi}_i(u)$ which can span the  space of spinor solutions. Our $\xi(u)$ with prescribed boundary value should be a linear combination of these, so that 
\begin{equation}
\xi(u)= \sum_i c^i {\bf \Psi}_i(u) . 
\end{equation}
By taking the   $a$-th component of this equation, we have $\xi^a(u)=\Psi^a_i(u)c^i $, which can be written as a matrix equation $ \xi(u)=   {\bf \Psi}(u){\bf c}$. Here   $\Psi^a_i(u)$  is the $a$-th component of $i$-th solution and we considered $ \xi(u)$ and ${\bf c}$ as  column matrices. By expanding the matrix 
 $ {\bf \Psi}(u)$ near the boundary, 
 \begin{equation}
 {\bf \Psi}(u)=  u^{m_s}\mathbb{S} + u^{m_c} \mathbb{C} + \cdots 
\end{equation}

 We now can write the source and condensation fermions depending on the quantization choice as follows, 
\begin{align}
\xi^{(\mathbb{Q})}_S = \mathbb{S}^{(\mathbb{Q})} \mathbf{c} , \quad 
\xi^{(\mathbb{Q})}_C = \mathbb{C}^{(\mathbb{Q})} \mathbf{c} . \label{xiSC}
\end{align}
where $\mathbb{Q}$ stands for quantization choice.
From eqs. (\ref{totbdySS})-(\ref{totbdySA}) and using (\ref{xiSC}),
\begin{align}
S_{total}^{(\mathbb{Q})}\Big|_{bdy}&= \frac{1}{2} \int_{bdy} d^4x \Big(\xi_S^{\dagger}\Gamma_{bdy}\xi_C + h.c\Big)^{(\mathbb{Q})},\\
&=  \frac{1}{2} \int_{bdy}d^4x ( \xi_S^{\dagger} \Gamma_{bdy}\mathbb{C} \mathbb{S} ^{-1} \xi_S +h.c)^{(\mathbb{Q})},\\
&= \frac{1}{2} \int_{bdy}d^4x (\xi_S^{\dagger} \mathbb{G} \xi_S+h.c)^{(\mathbb{Q})}.
\end{align}
the boundary Green's function can be defined as follow
\begin{align}
\mathbb{G}^{(SS)} &=  -(\sigma_0 \otimes \sigma_2)\mathbb{C} \mathbb{S} ^{-1},    \\\mathbb{G}^{(SA)}&= -(\sigma_3 \otimes \sigma_2)\mathbb{C} \mathbb{S} ^{-1}. \label{GF}
\end{align} 
 Notice  that the definition of Green's function remains valid even  for the zero bulk    fermion mass. Consequently, as far as we can extract the leading-order terms of the bulk fermions near the boundary, the Green's  function calculation remains solvable.
This   is helpful because, in the zero fermion  mass, we will be able to obtain    Green's function analytically for all Lorentz symmetry  breaking  interaction with proper choice of scaling dimension of the order parameter field. 
 {The term suitable here refers to choosing scaling dimensions for the source term, in which any $u$-dependence in the interacting term will be eliminated after fully expressing the vierbein and spin connection. This ensures the results remain u-independent at the interaction terms and  allows solvable Dirac equations.}

\section{Analytic  Green's function of fermions in symmetry broken phases} \label{section2}
We now consider  zero bulk mass fermions with a holographic order parameters having only  the leading term by setting $\langle \mathcal{O}_{\Phi} \rangle = 0$.
This is so called alternative quantization of the $\Phi$. This setup allows us to   derive Green's function for all types of Lorentz symmetry-breaking analytically. We already performed  numerical calculations  to study the case of non-zero fermion bulk mass and for the case with condensation in previous works \cite{Yuk:2022lof,Lieb,ABC}.
 From the expressions of $\Gamma$ in \eqref{Gammamu} and 
 \begin{eqnarray}
\Gamma^{tx} &=& \sigma_0 \otimes \sigma_3,\; 
 \Gamma^{ty} = \sigma_0 \otimes -\sigma_1 ,\;
 \Gamma^{tz} = \sigma_3 \otimes -\sigma_2, \; \\
 \Gamma^{xy} &=& \sigma_0 \otimes -i\sigma_2,\; 
\Gamma^{xz} = \sigma_3 \otimes i\sigma_1,\; 
 \Gamma^{yz} = \sigma_3 \otimes i\sigma_3,\;\\
\Gamma^{ut} &=& i\sigma_2 \otimes i\sigma_2, \;
\Gamma^{ux} = i\sigma_2 \otimes \sigma_1.\;
\Gamma^{uy} = i\sigma_2 \otimes \sigma_3, \;
\Gamma^{uz} = \sigma_1 \otimes -i\sigma_0
\end{eqnarray}
 our gamma matrices can be expressed in the following decomposed form:
\begin{align}
    \Gamma^{\underline{\mu}} = \begin{pmatrix}
        0  & \gamma^{\underline{\mu}}\\
        \gamma^{\underline{\mu}*} & 0
    \end{pmatrix},~~ \Gamma^{\underline{\mu\nu}} = \begin{pmatrix}
         \gamma^{\underline{\mu\nu}} & 0 \\
        0& \gamma^{\underline{\mu\nu}*} 
    \end{pmatrix},
    \Gamma^{\underline{\mu u}} = \begin{pmatrix}
          0 & \gamma^{\underline{\mu}} \\
        -\gamma^{\underline{\mu}*} & 0 
    \end{pmatrix},
\end{align}
where $\gamma^{\underline{\mu}} =(i\sigma_2,\sigma_1,\sigma_3,-i\mathbb{1})$ with $\mu=t,x,y,z$ and $\gamma^{\underline{\mu}\underline{\nu}}=\gamma^{\underline{\mu}}\gamma^{\underline{\nu}}$.  The complex conjugation appears due to $\Gamma^{z}$'s being pure imaginary. In $\text{AdS}_4$, this decomposition is still   valid with $\mu=t,x,y$, because apart from $\Gamma^5$, all gamma matrices are real. So that the complex conjugate disappears in $\text{AdS}_4$. Such decomposition will be   utilized for analytic solutions. 

\myparagraph{Classification of interaction types }
Since the boundary represents the physical world, we will classify the interaction type  from the   boundary point of view; 
\begin{itemize}
	\item	2 types of scalar: $\mathds{1}$, $\Gamma^{u}$(radial scalar).
	\item	2 types of vector: $\Gamma^{\mu}$(polar vector), $\Gamma^{\mu u}$(radial vector).
	\item	1 type of tensor: $\Gamma^{\mu\nu}$(polar antisymmetric 2-tensor).
\end{itemize}
Although $B^u$ and $B^{\mu u}$ are component of vector and tensor, respectively, 
they are scalar and vector  from  the  boundary  point of view,  


\subsection{Scalar: $\mathcal{L}_{int} = i\Phi(\bar{\psi}^{(2)}\psi^{(1)}+h.c)$}
For scalar interaction, the solvable solution can be obtained by choosing 
\begin{align*}
\Phi(u) = M_0 u+\underbrace{M}_{=0} u^3 \qquad ; \qquad m^2_{\Phi} = -3 
\end{align*}
\paragraph{ SS case:}
The bulk equations of motion are given by 
\begin{align}
\big[\partial_u- M_0(\sigma_1 \otimes \sigma_0)\big]\xi_{S}^{(SS)} + i (\sigma_0 \otimes \gamma^{\underline{\mu}}k_{\mu}) \xi_{C}^{(SS)}  &= 0,\\
\big[\partial_u+ M_0 (\sigma_1 \otimes \sigma_0)\big]\xi_{C}^{(SS)} - i (\sigma_0 \otimes \gamma^{\underline\mu *}k_{\mu}) \xi_{S}^{(SS)}  &= 0,
\end{align}
due to the simple commutation relation between $(\sigma_0 \otimes \gamma^{\underline\mu *}k_{\mu})$, and its conjugate, one can get the simple fully diagonalized decoupled equations \cite{Strangemetal}, which reads
\begin{align}
(\partial_u^2-M_0^2-\boldsymbol{k}^2+\omega^2)\xi^{(SS)}_{S,C} = 0, \label{eq:scalar}
\end{align} 
The solutions are well-known and decay  exponentially since the growing  terms are removed  by imposing in-falling boundary condition (BC).   As a result, the asymptotic solutions  near the AdS boundary located at $u=0$  are given by 
\begin{align}
\xi_{S}^{(SS)} = [\mathbb{S}_0(k)+\mathbb{S}_1(k) u +\cdots]\mathbf{c}~~,~~\xi_{C}^{(SS)}| \simeq [\mathbb{C}_0(k)+\mathbb{C}_1(k) u+\cdots]\mathbf{c},\label{eq:sol}
\end{align}
where $\mathbb{S}_{n},\mathbb{C}_{n}$ are $u$-independent but momentum dependent $4\times4$ matrices. But apart from the leading term $\mathbb{S}_0(k), \mathbb{C}_0(k)$, they will not contribute to the boundary Green's function. Therefore we will write them simply as 
$\mathbb{S}(k), \mathbb{C}(k)$ by deleting the index 0. 
Since $\xi_{S}$ and $\xi_{C}$ are solved independently, one can plug-in one of the solution on the Dirac equations to find the relation between them \cite{Nonfermi,Iqbal:2009fd}.  The condensation term is determined by the source term by solving the Dirac equation and it is given as follows:
\begin{align}
\mathbb{C}(k) = i \frac{\sigma_0 \otimes \gamma^{\underline{\mu} *}k_\mu}{\boldsymbol{k}^2-\omega^2}[\mathcal{T}(k)-M_0(\sigma_1 \otimes \sigma_0)] \mathbb{S}(k),
\end{align}
where we define the matrix $\mathcal{T} = \mathbb{S}_1\mathbb{S}_0^{-1}$. For the scalar interaction, it is given by
\begin{align}
	\mathcal{T}(k) = -\sqrt{\boldsymbol{k}^2-\omega^2+M_0^2}~\mathds{1}_{4\times4}, \label{Tscalar}
\end{align}
From the definition of boundary Green's function (\ref{GF}), one gets 4 by 4 retarded Green's function as follows,
\begin{align}
\mathbb{G}(k )_R &= -\frac{\Gamma^{\underline{t}}}{\boldsymbol{k}^2-\omega^2}(\sigma_1 \otimes \gamma^{\underline{\mu} *}k_\mu)[\mathcal{T}(k)-M_0(\sigma_1 \otimes \sigma_0)],\\
&= \frac{1}{\boldsymbol{k}^2-\omega^2} \Big[\sqrt{\boldsymbol{k}^2-\omega^2+M_0^2} ~\sigma_0 + M_0 ~\sigma_1\Big] \otimes \mathbb{K}.\label{G:SSscalar}
\end{align}
where  $\boldsymbol{k}^2 = k_x^2+k_y^2+k_z^2$$, \mathbb{K} = (\gamma^{\underline{t}}\gamma^{\underline{\mu}}k_\mu)^T =
\begin{pmatrix}
	k_x+\omega & -k_y+ik_z\\
	-k_y-ik_z & -k_x+\omega
\end{pmatrix}
$. It is important to note that $\Tr\mathbb{K} = 2\omega$. This matrix, $\mathbb{K}$, will play a consistent role in our subsequent calculations. We will discuss the trace result of the Green's functions in the next section.

\paragraph{SA case:} The bulk equations of motion are given by
\begin{align}
\partial_u\xi^{(SA)}_S -(\sigma_2 \otimes \sigma_0)[\Gamma^{\underline{\mu}*}k_\mu+iM_0]\xi_C^{(SA)} &= 0\label{eq:SSscalar},\\
\partial_u\xi^{(SA)}_C +(\sigma_2 \otimes \sigma_0)[\Gamma^{\underline{\mu}}k_\mu-iM_0]\xi_S^{(SA)} &= 0,
\end{align}
similar to the SS case, one can decouple above equations which again yields (\ref{eq:scalar})-(\ref{eq:sol}). Plugging the asymptotic solution into (\ref{eq:SSscalar}), 
 we get
\begin{align}
\mathbb{C}(k) &= [\Gamma^{\underline{\mu}*}k_\mu+iM_0]^{-1}(\sigma_2 \otimes \sigma_0)\mathcal{T}(k)\mathbb{S}(k),
\end{align}
following the definition of Green's function, and the $\mathcal{T}(k)$ given in (\ref{Tscalar}), one can get the general form of it  as follows.
\begin{align}
\mathbb{G}(k_\mu)_R &=  -(\sigma_3 \otimes \sigma_2) [\Gamma^{\underline{\mu}*}k_\mu+iM_0]^{-1}(\sigma_2 \otimes \sigma_0)\mathcal{T}(k)\\
&= \frac{1}{\sqrt{\boldsymbol{k^2}-\omega^2+M_0^2}} \Big[\sigma_0 \otimes \mathbb{K} +M_0  ~\sigma_1 \otimes \sigma_1  \Big]. \label{G:SAscalar}
\end{align}
One can see that the Green's function contains off-diagonal terms, which are absent in intra-flavor interaction case. Moreover, calculating AA (alternative-alternative) or AS(alternative-standard) quantization cases yields the results with the  propagator replaced by the complex conjugation.

\subsection{Radial scalar: $\mathcal{L}_{int}= B_{u}(\bar{\psi}^{(2)}\Gamma^{u}\psi^{(1)}+h.c)$}
The solvable solution can be obtained by choosing
\begin{align*}
B_u(u) = b+ \underbrace{b_{u}^{(2)}}_{=0} u^2\qquad 	&; \qquad m^2_{B_u} = 0,
\end{align*}
where $b$ is the  constant measuring the symmetry breaking strength.
\paragraph{SS case:}The bulk equations of motion are given by
\begin{align}
[\partial_u - i (\sigma_1 \otimes \sigma_0)b]\xi^{(SS)}_{S} +  i (\sigma_0 \otimes \gamma^{\underline{\mu}}k_{\mu}) \xi^{(SS)}_C &= 0, \\
[\partial_u - i (\sigma_1 \otimes \sigma_0)b]\xi^{(SS)}_{C} - i (\sigma_0 \otimes \gamma^{\underline\mu *}k_{\mu}) \xi^{(SS)}_S &= 0. 
\end{align}
Decoupled   differential equations (DEs) are given by 
\begin{align}
[\partial_u^2 - 2ib (\sigma_1 \otimes \sigma_0)\partial_u + (\boldsymbol{k}^2-\omega^2+b^2)]\xi_{S,C}^{SS} = 0,
\end{align}
Then, we can diagonalize the system by using the similarity transformation $\mathcal{P}$ defined by the eigenvectors matrix of $\sigma_1 \otimes \sigma_0 $ which yields 
\begin{align}
[\partial_u^2 - 2ib\partial_u + (\boldsymbol{k}^2-\omega^2+b^2)] \mathcal{P}^{-1}\xi_{S,C}^{SS} = 0,
\end{align}
This yields the solution  of the exponential form   even after mapping the solution back by the inverse similarity transformation. So, overall,  nothing new for $B_u$ case. We can still write the Green's function as follows:
\begin{align}
\mathbb{C} (k) = i \frac{(\sigma_0 \otimes \gamma^{\underline{\mu}*}k_\mu)}{\boldsymbol{k}^2-\omega^2} [\mathcal{T}(k)-i (\sigma_1 \otimes \sigma_0) b]\mathbb{S} (k) ,
\end{align}
The surprise of this interaction types is structure of $\mathcal{T}(k)$, which is given by
\begin{align}
\mathcal{T}(k) = -\sqrt{\boldsymbol{k}^2 - \omega^2} +  i (\sigma_1 \otimes \sigma_0)b, \label{bucancelation}
\end{align}
So the Green's function  reduces  into the non-interacting case, 
\begin{align}
\mathbb{G}(k) =  i(\sigma_0 \otimes \sigma_2)\frac{(\sigma_0 \otimes \gamma^{\underline{\mu}*}k_\mu)}{\sqrt{\boldsymbol{k}^2-\omega^2}}
=\frac{\sigma_0 \otimes \mathbb{K}}{\sqrt{\boldsymbol{k}^2-\omega^2}}.\label{G:SSbu}
\end{align}
\paragraph{SA case:} the bulk dirac equation is given by
\begin{align*}
\partial_u \xi^{SA}_S - [(\sigma_2 \otimes \sigma_0)\Gamma^{\underline{\mu}*}k_\mu + i (\sigma_1 \otimes \sigma_0) b]\xi^{SA}_C &= 0,\\
\partial_u \xi^{SA}_C + [(\sigma_2 \otimes \sigma_0)\Gamma^{\underline{\mu}}k_\mu - i (\sigma_1 \otimes \sigma_0) b]\xi^{SA}_S &= 0,
\end{align*}
Since the product of $[(\sigma_2 \otimes \sigma_0)\Gamma^{\underline{\mu}*}k_\mu + i (\sigma_1 \otimes \sigma_0) b]$ and its complex conjugation is a non-diagonal matrix, a similarity transformation is needed for the diagonalization, which yields 
\begin{align}
[\partial_u^2 + (b-i\sqrt{\boldsymbol{k}^2-\omega^2})^2]P^{-1}\xi_{\pm}^{SA} = 0.
\end{align}
The fermion Green's functions  for SS and SA quantization with    $B_u$  interactions are found to be the same. This discovery raises the question of why SS and SA lead to identical Green's functions despite the  differences in their equations of motion. The answer lies in two critical factors that influence the structure of the Green's function. Firstly, it depends on the combination of gamma matrices present in the Dirac equation, which varies with the choice of quantization. Secondly, the solutions are affected by   the proper in-falling boundary conditions (BC). As a result, despite the apparent difference in the initial appearance of the Dirac equations, the solutions surviving the BC  end up with the same Green's function.

\subsection{Polar vectors: $\mathcal{L}_{int}= B_\mu(\bar{\psi}^{(2)}\Gamma^{\mu}\psi^{(1)}+h.c)$}
The solvable solution can be obtained by choosing 
\begin{align}
B_\mu(u) = B_{\mu}^{(0)} + \underbrace{B_{\mu}^{(2)}}_{=0} u^2 ; & ,\qquad m^2_{B_\mu} = 0. \label{ansatzforBmu}
\end{align}
where $B_{\mu}^{(0)}$ is a vectors with a single nonzero component $b_{(t,i)}$ which is nothing other than a constant order parameter.
\paragraph{SS case:} The bulk equations of motion are given by
\begin{align}
\partial_u\xi^{(SS)}_S +i[\sigma_0 \otimes \gamma^{\underline{\mu}} k_\mu-\sigma_1 \otimes \gamma^{\underline{\mu}} B_{\mu}^{(0)}]\xi_C^{(SS)} &= 0 \label{eq:SSvector},\\
\partial_u\xi^{(SS)}_C -i[\sigma_0 \otimes \gamma^{\underline{\mu* }} k_\mu-\sigma_1 \otimes \gamma^{\underline{\mu *}} B_{\mu}^{(0)}]\xi_S^{(SS)} &= 0,
\end{align}
Unlike the scalar case, polar vector type interactions cannot be decoupled simply in  this basis. However, we can transform the equations  to a suitable basis by a u-independent similarity transformation,  $\mathcal{P}$, where
\begin{align*}
\mathcal{P}^{-1} \mathcal{B}^*\mathcal{B}\mathcal{P} = (\mathcal{K}_{\mu_-}^2 \sigma_0 \oplus \mathcal{K}_{\mu_+}^2\sigma_0),
\end{align*} 
where $\mathcal{B} \coloneqq [\tilde\Gamma^{\underline\mu}k_\mu-\sigma_1 \otimes \gamma^\mu B_{\mu}^{(0)}]$, and $\mathcal{K}_{\mu_\pm}^2 \coloneqq \kp^2 +(b_\mu \pm k_\mu)^2$. Under this transformation, we can get the decoupled equations, 
\begin{align}
[\partial_u^2 -(\mathcal{K}^2_{\mu_-} \sigma_0\oplus \mathcal{K}^2_{\mu_+}\sigma_0)]\tilde{\xi}^{(SS)}_{S,C} &= 0, \label{eq:bx}
\end{align}
where $\tilde\xi_{S,C} = P^{-1} \xi_{S,C}$. We obtain simple solutions even after transforming the solutions back to the original basis, providing exponential decay. Consequently, we can express the asymptotic solution   similar to scalar case (\ref{eq:sol}). By substituting the solution into the  (\ref{eq:SSvector}), we obtain
\begin{align}
\mathbb{C}(k) = i[\sigma_0 \otimes \gamma^{\underline{\mu}} k_\mu - \sigma_1 \otimes \gamma^{\underline{\mu}}  B_{\mu}^{(0)}]^{-1}\mathcal{T}(k) \mathbb{S}(k),
\end{align}
Then, the Green's function is given by 
\begin{align}
\mathbb{G}(k) = -i(\sigma_0 \otimes \sigma_2)[\sigma_0 \otimes \gamma^{\underline\mu} k_\mu - \sigma_1 \otimes \gamma^{\underline\mu}  B_{\mu}^{(0)}]^{-1}\mathcal{T}(k).\label{G:SSbx}
\end{align}
The  Green's functions can be determined straightforwardly by plugging in  $\mathcal{T}(k)$ into (\ref{G:SSbx}).  Now let us calculate the explicit expression of the Green's function.\\

$\boldmath{B_{t}/\textbf{SS}}$ : By solving the Dirac equations, one gets 
\begin{align}
	\mathcal{T}(k)_{B_t^{(0)}}^{(SS)} &=   -\frac{1}{2}\Big((\mathcal{K}_{t_-} + \mathcal{K}_{t_+})\mathds{1}_{4\times 4} +(\mathcal{K}_{t_+} - \mathcal{K}_{t_-})\sigma_1 \otimes \sigma_0\Big),
\end{align}
where $\mathcal{K}_{t_\pm} = \sqrt{\boldsymbol{k}^2 - (b_t \pm \omega)^2}$ and $b_t = B_{t}^{(0)} $ . By plugging the above result into (\ref{G:SSbx}), we can get the Green's function:
\begin{align}
	\mathbb{G}(k)^{(SS)}_{B_t^{(0)}} &= \frac{1}{2 \mathcal{K}_{t_+}\mathcal{K}_{t_-}}
	\begin{pmatrix}
		g_{11} & g_{12}\\
		g_{21} & g_{22}
	\end{pmatrix},\\
	\text{with}	\quad g_{11} = g_{22} &= \Big((\mathcal{K}_{t_-} - \mathcal{K}_{t_+}) b_t \sigma_0 +(\mathcal{K}_{t_-} + \mathcal{K}_{t_+}) \mathbb{K}\Big),\notag\\
	g_{12} = g_{21} &= \Big((\mathcal{K}_{t_-} + \mathcal{K}_{t_+}) b_t \sigma_0 +(\mathcal{K}_{t_-} - \mathcal{K}_{t_+}) \mathbb{K}\Big).
\end{align}

$\boldmath{B_{x}/\textbf{SS}}$ : By follwing the same calculation of time-like case, one gets
 \begin{align}
 	\mathcal{T}(k)_{B_x^{(0)}}^{(SS)} &=   -\frac{1}{2}\Big((\mathcal{K}_{x_-} + \mathcal{K}_{x_+})\mathds{1}_{4\times 4} +(\mathcal{K}_{x_-} - \mathcal{K}_{x_+}) \sigma_1 \otimes \sigma_0\Big),
 \end{align}
 where $\mathcal{K}_{x_\pm} = \sqrt{(b_x \pm k_x)^2 +\kp^2 -\omega^2}$, $\kp^2 = k_y^2+k_z^2$, and $b_x = B_{x}^{(0)} $. By plugging the $\mathcal{T}(k)$ in the above result into (\ref{G:SSbx}), one gets
\begin{align}
	\mathbb{G}(k)^{(SS)}_{B_x^{(0)}} &= \frac{1}{2 \mathcal{K}_{+x}\mathcal{K}_{-x}}
	\begin{pmatrix}
		g_{11} & g_{12}\\
		g_{21} & g_{22}
	\end{pmatrix},\\
	\text{with}	\quad g_{11} = g_{22} &= \Big((\mathcal{K}_{-x} - \mathcal{K}_{+x}) b_x \sigma_3 +(\mathcal{K}_{-x} + \mathcal{K}_{+x}) \mathbb{K}\Big),\notag\\
	g_{12} = g_{21} &= -\Big((\mathcal{K}_{-x} + \mathcal{K}_{+x}) b_x \sigma_3 +(\mathcal{K}_{-x} - \mathcal{K}_{+x}) \mathbb{K}\Big).
\end{align}

One can see that the Green's function of polar vectors together with SS quantization contains branch-cut singularity by the presence of $\mathcal{K}_{\mu \pm}^{-1}$ as the denominator terms. The singularity type does not change after tracing the Green's function matrix, which we will discuss in the next section. 

After this case, we will no longer show the full expression of $\mathcal{T}(k)$, because it will become more complicated while lacking meaningful content. However, one can obtain the Green's function by following the same logic and calculations. 

\paragraph{SA case:} The bulk equations of motion are given by
\begin{align}
(\partial_u - i\Gamma^{\underline{u\mu}}B_{\mu}^{(0)})\xi^{(SA)}_S -(\sigma_2 \otimes \sigma_0)\Gamma^{\underline{\mu}*}k_\mu\xi^{(SA)}_{C} &= 0\label{eq:SAvector},  \\
(\partial_u + i\Gamma^{\underline{u\mu} *}B_{\mu}^{(0)})\xi^{(SA)}_C + (\sigma_2 \otimes \sigma_0)\Gamma^{\underline{\mu}}k_\mu\xi^{(SA)}_{S} &= 0,
\end{align}
In this case, the differential equations cannot be fully decoupled by   similarity transformation. However, the DE  is nothing but a linear Ordinary DE  system. Similar to other cases, the solutions satisfying the infalling BCs   exponential decay (\ref{eq:sol}). We, therefore, substitute the solution back to the (\ref{eq:SAvector}) and get  the condensation   in terms of the source: 
\begin{align}
\mathbb{C}(k) = \frac{\Gamma^{\underline{\mu}*} k_\mu}{\boldsymbol{k}^2-\omega^2}(\sigma_2 \otimes \sigma_0 )[ \mathcal{T}(k)-i\Gamma^{\underline{u \mu}}B_{\mu}^{(0)} ]\mathbb{S}(k),
\end{align}
 Therefore, the algebraic Green's function for this case is given by 
\begin{align}
\mathbb{G}(k)_R&= -\frac{\Gamma^{\underline{t}}}{\boldsymbol{k}^2-\omega^2}\Gamma^{\underline{\mu}} k_\mu[ \mathcal{T}(k)-i\Gamma^{\underline{u \mu}}B_{\mu}^{(0)} ].\label{G:SAbx}
\end{align}\\

$\boldmath{B_{t}/\textbf{SA}}$: After getting $\mathcal{T}(k)$ by solving the Dirac equation, and plugging in (\ref{G:SAbx}), one can get
\begin{align}
	\qquad\quad& \mathbb{G}(k)_{B^{(0)}_t}^{(SA)} ~=~ \frac{1}{2b_t \boldsymbol{k^2}} \begin{pmatrix}
		g_{11} & g_{12}\\
		g_{21} & g_{22}
	\end{pmatrix},\\
	\text{with} \quad
	g_{11} = g_{22}^* &=  \Big(\omega(\mathcal{K}_{t_-}-\mathcal{K}_{t_+}) + b_t(\mathcal{K}_{t_-}+\mathcal{K}_{t_+})\Big)\mathbb{K}\notag\\
	&\quad\quad -\Big(b_t \omega(\mathcal{K}_{t_-}+\mathcal{K}_{t_+})-\mathcal{E}(\mathcal{K}_{t_-}-\mathcal{K}_{t_+})\Big)\sigma_0,\notag\\
	g_{12} = g_{21}^* &= i( k_x \sigma_1 + \sigma_0 \gamma^{\underline{\mu}}k_{\mu})(b^2_t-\mathcal{K}_{t_-}\mathcal{K}_{t_+}+\mathcal{E}).
\end{align}
where $\mathcal{K}_{t_\pm} = \sqrt{\boldsymbol{k}^2 - (b_t \pm \omega)^2}$, $\mathcal{E} = \boldsymbol{k}^2 - \omega^2$, and $b_t = B_{t}^{(0)} $. One can easily check that the pole type singularity $\boldsymbol{k}^{-2}$ will be canceled out after we take the trace of this Green's function. Since the cancellation of the pole makes the trace of the Green's function becomes non-singularity type Green's function. Therefore, the presence of singularities in the 4 by 4 expression of the Green's function does not guarantee the presence of singularity in spectral function.\\

 $\boldmath{B_{x}/\textbf{SA}}$: By the same calculation in the previous cases, the Green's function reads
\begin{align}
	\qquad\quad& \mathbb{G}(k)_{B^{(0)}_x}^{(SA)} ~=~ \frac{1}{2b_x(\kp^2-\omega^2)} \begin{pmatrix}
		g_{11} & g_{12}\\
		g_{21} & g_{22}
	\end{pmatrix},\\
	\text{with} \quad
	g_{11} = g_{22}^* &=  \Big(k_x(\mathcal{K}_{x_-}-\mathcal{K}_{x_+}) + b_x(\mathcal{K}_{x_-}+\mathcal{K}_{x_+})\Big)\mathbb{K}\notag\\
	&\quad\quad -\Big(b_xk_x(\mathcal{K}_{x_-}+\mathcal{K}_{x_+})+\mathcal{E}(\mathcal{K}_{x_-}-\mathcal{K}_{x_+})\Big)\sigma_3,\notag\\
	g_{12} = g_{21}^* &= (k_x \sigma_2 -i\sigma_3 \gamma^{\underline{\mu}}k_{\mu})(b^2_x+\mathcal{K}_{x_-}\mathcal{K}_{x_+}-\mathcal{E}).
\end{align}
where $\mathcal{K}_{x_\pm} = \sqrt{(b_x \pm k_x)^2 +\kp^2 -\omega^2}$, $\mathcal{E} = \boldsymbol{k}^2 - \omega^2$, and $b_x = B_{x}^{(0)} $. In this case, the singularity $(\kp^2 - \omega^2)^{-1}$ is not changed or canceled by the trace, so that the trace of the Green's function has pole type singularity. We will back to discuss the trace of these Green's functions in the next sections.

\subsection{Radial vectors: $\mathcal{L}_{int}= B_{\mu u}(\bar{\psi}^{(2)}\Gamma^{\mu u}\psi^{(1)}+h.c)$}
The solvable solution can be obtained by choosing 
\begin{align*}
B_{\mu u}(u) = \frac{B_{\mu u}^{(-1)}}{u} + \underbrace{B_{\mu u}^{(1)}}_0 u  ; \qquad m^2_{B_{\mu u}} = 1,
\end{align*}
where $B_{\mu u}^{(-1)}$ is a tensor in AdS bulk with a single nonzero component. However, from the boundary point of view, its physical role  can be classified  as a vector.
\paragraph{SS case:} The bulk Dirac equations are given by 
\begin{align}
\partial_u \xi^{(SS)}_S +i[ \sigma_0 \otimes \gamma^{\underline\mu}k_\mu +\sigma_1 \otimes \gamma^{\underline\mu} B_{\mu u}^{(-1)}]\xi^{(SS)}_C &= 0,\\
\partial_u \xi^{(SS)}_C -i[ \sigma_0 \otimes \gamma^{\underline\mu*}k_\mu -\sigma_1 \otimes \gamma^{\underline\mu*} B_{\mu u}^{(-1)}]\xi^{(SS)}_S &= 0,
\end{align}
The main procedure of this type   is the same with other interactions. We get the condensation in terms of the source as follows,
\begin{align}
\mathbb{C} (k) = i [ \sigma_0 \otimes \gamma^{\underline\mu}k_\mu +\sigma_1 \otimes \gamma^{\underline\mu} B_{\mu u}^{(-1)}]^{-1} \mathcal{T}(k)\mathbb{S} (k),
\end{align} 
This yields   following Green's function,
\begin{align}
\mathbb{G}(k_\mu) = -i(\sigma_0 \otimes \sigma_2)[\sigma_0 \otimes \gamma^\mu k_\mu + \sigma_1 \otimes \gamma^\mu B_{\mu u}^{(-1)}]^{-1}\mathcal{T}(k).\label{G:SSbux}
\end{align}

$\boldmath{B_{tu}/\textbf{SS}}$: By the same calculation in the previous cases, the Green's function is given by
\begin{align}
	\qquad\quad& \mathbb{G}(k)_{B^{(-1)}_{tu}}^{(SS)} ~=~ \frac{1}{2 \boldsymbol{k}\mathcal{K}_{tu+}^2\mathcal{K}_{tu-}^2} \begin{pmatrix}
		g_{11} & g_{12}\\
		g_{21} & g_{22}
	\end{pmatrix},\\
	\text{with} \quad
	g_{11} = g_{22} &=  \Big(b(\boldsymbol{k}^2+\omega^2-b^2)(\mathcal{K}_{-}-\mathcal{K}_{tu+})+|\boldsymbol{k}|(\mathcal{E}-b^2)(\mathcal{K}_{tu-}+\mathcal{K}_{tu+})\Big)\mathbb{K}\notag\\
&\quad\quad +b\omega\Big(2b|\boldsymbol{k}|(\mathcal{K}_{tu-}+\mathcal{K}_{tu+}) + (\mathcal{E}+b^2)(\mathcal{K}_{tu-}-\mathcal{K}_{tu+})\Big)\sigma_0 ,\notag\\
	g_{12} = g_{21} &= -\omega\Big(2b|\boldsymbol{k}| (\mathcal{K}_{tu-}+\mathcal{K}_{tu+})+(\mathcal{E}+b^2)(\mathcal{K}_{tu-}-\mathcal{K}_{tu+})\Big)\mathbb{K}\notag\\
&\quad\quad + \Big((b^2(\boldsymbol{k}^2+\omega^2)-\mathcal{E}^2)(\mathcal{K}_{tu-}-\mathcal{K}_{tu+})-b |\boldsymbol{k}| (\mathcal{E}-b^2)(\mathcal{K}_{tu-}+\mathcal{K}_{tu+})\Big)\sigma_0.
\end{align}
where $\mathcal{K}_{tu\pm} = \sqrt{(b\pm|\boldsymbol{k}|)^2-\omega^2}$ , $\mathcal{E} = \boldsymbol{k}^2 - \omega^2$, and $b = B_{tu}^{(-1)}$. The trace of the Green's function turn into a simple form which will be discussed in the following section.

\paragraph{SA case: } The bulk Dirac equations are given by
\begin{align}
(\partial_u - i\Gamma^{\underline{\mu}}B_{\mu u}^{(-1)})\xi^{(SA)}_S -(\sigma_2 \otimes \sigma_0)\Gamma^{\underline{\mu}*}k_\mu\xi^{SA}_{C} &= 0,  \\
(\partial_u - i\Gamma^{\underline{\mu} *}B_{\mu u}^{(-1)})\xi^{(SA)}_C + (\sigma_2 \otimes \sigma_0)\Gamma^{\underline{\mu}}k_\mu\xi^{SA}_{S} &= 0.
\end{align}
The   condensation   are given as follows,
\begin{align}
\mathbb{C} (k) = \frac{\Gamma^{\underline{\mu}*} k_\mu}{\boldsymbol{k}^2-\omega^2}(\sigma_2 \otimes \sigma_0 )[ \mathcal{T}(k)-i\Gamma^{\underline{\mu}}B_{\mu u}^{(-1)} ]\mathbb{S}(k),
\end{align}
which yields following retarded Green's function matrix,
\begin{align}
\mathbb{G}(k)_R &= -\frac{\Gamma^{\underline{t}}}{\boldsymbol{k}^2-\omega^2}\Gamma^{\underline{\mu}} k_\mu[ \mathcal{T}(k)-i\Gamma^{\underline{\mu}}B_{\mu u}^{(-1)}].\label{G:SAbux}
\end{align}

$\boldmath{B_{tu}/\textbf{SA}}$: By the same calculation we have done, the Green's function is given by
\begin{align}
	\qquad\quad& \mathbb{G}(k)_{B^{(-1)}_{tu}}^{(SS)} ~=~ -\frac{1}{2b \omega \boldsymbol{k}} \begin{pmatrix}
		g_{11} & g_{12}\\
		g_{21} & g_{22}
	\end{pmatrix},\\
	\text{with} \quad
	g_{11} = g_{22}^* &= \omega(\mathcal{K}_{tu-}-\mathcal{K}_{tu+})\mathbb{K}+\Big(\mathcal{E}(\mathcal{K}_{tu-}-\mathcal{K}_{tu+})+b \boldsymbol{k}(\mathcal{K}_{tu-}+\mathcal{K}_{tu+})\Big)\sigma_0 ,\notag\\
	g_{12} = g_{21} &= \boldsymbol{k} (b^2 + \mathcal{K}_{tu-}\mathcal{K}_{tu+} -\mathcal{E}) \sigma_2
\end{align}

where $\mathcal{K}_{tu\pm} = \sqrt{(b\pm|\boldsymbol{k}|)^2-\omega^2}$ , $\mathcal{E} = \boldsymbol{k}^2 - \omega^2$, and $b = B_{tu}^{(-1)}$. We will discuss the trace of the Green's function in the coming section.\\

$\boldmath{B_{xu}/\textbf{SA}/ \textbf{SS}}$: The decomposition and simplification of the Green's functions for this case pose significant challenges. Consequently, we will focus only on the trace of the Green's function, which included in the following section.

\subsection{Antisymmetric 2-tensors: $\mathcal{L}_{int}= B_{\mu\nu}(\bar{\psi}^{(2)}\Gamma^{\mu\nu}\psi^{(1)}+h.c)$}
The solvable solution can be obtained by choosing 
\begin{align*}
B_{\mu\nu}(u) = \frac{B_{\mu\nu}^{(-1)}}{u} + \underbrace{B_{\mu\nu}^{(1)}}_0 u
;\qquad m^2_{B_{\mu\nu}} = 1,
\end{align*}
where $B^{(-1)}_{\mu\nu}$ is   symmetry breaking strength constant. 
\paragraph{SS case:} The bulk Dirac equations are given by
\begin{align}
\big[\partial_u- (i\sigma_1 \otimes \gamma^{\underline{\mu\nu}}B_{\mu\nu}^{(-1)})\big]\xi_S^{(SS)}  + i (\sigma_0 \otimes \gamma^{\underline{\mu}}k_{\mu}) \xi_C^{(SS)}  &= 0,\\
\big[\partial_u+ (i\sigma_1 \otimes \gamma^{\underline{\mu\nu} *}B_{\mu\nu}^{(-1)})\big]\xi_C^{(SS)} - i (\sigma_0 \otimes \gamma^{\underline\mu *}k_{\mu}) \xi_S^{(SS)}  &= 0,
\end{align}
Following a similar approach to $B_\mu$'s  SA-case, we can decouple above equation and get a system of decoupled linear equations. So, we then plugin the solution into the above equation  to get  
\begin{align}
\mathbb{C}(k) = i \frac{\sigma_0 \otimes \gamma^{\underline{\mu} *}k_\mu}{\boldsymbol{k}^2-\omega^2}[ \mathcal{T}(k)-(i\sigma_1\otimes \gamma^{\underline{\mu\nu}})B_{\mu\nu}^{(-1)} ]\mathbb{S}(k),
\end{align}
leads to the Green's function which is given by
\begin{align}
\mathbb{G}(k)_R  &=-\frac{\Gamma^{\underline{t}}}{\boldsymbol{k}^2-\omega^2}(\sigma_1 \otimes \gamma^{\mu *}k_\mu)[ \mathcal{T}(k)-(i\sigma_1\otimes \gamma^{\underline{\mu\nu}})B_{\mu\nu}^{(-1)} ].\label{G:SSbxy}
\end{align}

$\boldmath{B_{xy}/\textbf{SS}}$: By solving the Dirac equations, and plugging-in into (\ref{G:SSbxy}), we get
\begin{align}
	\qquad\quad& \mathbb{G}(k)_{B^{(-1)}_{xy}}^{(SS)} ~=~ \frac{1}{2b_{xy}(k_z^2-\omega^2)} \begin{pmatrix}
		g_{11} & g_{12}\\
		g_{21} & g_{22}
	\end{pmatrix},\\
	\text{with} \quad
	g_{11} = g_{22} &=  \Big(|\boldsymbol{k}|(\mathcal{K}_{xy-}-\mathcal{K}_{xy+})+b(\mathcal{K}_{xy-}+\mathcal{K}_{xy+})\Big)(k_z\sigma_2-\omega\sigma_0)\notag\\
	&\quad\quad +\frac{k_z^2-\omega^2}{|\kp|}(\mathcal{K}_{xy-}-\mathcal{K}_{xy+})(k_x \sigma_3 -k_y\sigma_1),\notag\\
	g_{12} = g_{21} &= (k_x\sigma_3-k_y\sigma_1)(b_{xy}^2+\mathcal{K}_{xy-}\mathcal{K}_{xy+}-\mathcal{E}).
\end{align}
where $\kp^2 = k_x^2+k_y^2$, $\mathcal{K}_{xy_\pm} = \sqrt{(b_{xy} \pm |\kp|)^2 +k_z^2 -\omega^2}$, $b_{xy} = B_{xy}^{(-1)}$, and $\mathcal{E} = \boldsymbol{k}^2 - \omega^2$. In this case, the singularity $(k_z^2 - \omega^2)^{-1}$ is not changed or canceled by the trace, so that the trace of the Green's function has pole type singularity. We will back to discuss the trace of these Green's functions in the next sections.\\

\paragraph{SA case:} The bulk Dirac equations are given by
\begin{align}
\partial_u \xi^{(SA)}_{S} - (\sigma_2 \otimes \sigma_0) [\Gamma^{\underline{\mu}*}k_\mu-\Gamma^{\underline{\mu\nu}}B_{\mu\nu}^{(-1)}]\xi^{(SA)}_C &= 0,\\
\partial_u \xi^{(SA)}_{C} + (\sigma_2 \otimes \sigma_0) [\Gamma^{\underline{\mu}}k_\mu-\Gamma^{\underline{\mu\nu}*}B_{\mu\nu}^{(-1)}]\xi^{(SA)}_S &= 0. 
\end{align}
The above coupled equations can be decoupled to get  exact solution of exponential form.
\begin{align}
\mathbb{C} (k) = [\Gamma^{\underline{\mu}*}k_\mu-\Gamma^{\underline{\mu\nu}}B_{\mu\nu}^{(-1)}]^{-1}(\sigma_2 \otimes \sigma_0)\mathcal{T}(k)\mathbb{S} (k). 
\end{align}
The Green's function for this case is given by
\begin{align}
\mathbb{G}(k)_R &= -(\sigma_3\otimes\sigma_2)[\Gamma^{\underline{\mu}*}k_\mu-\Gamma^{\underline{\mu\nu}}B_{\mu\nu}^{(-1)}]^{-1}(\sigma_2 \otimes \sigma_0)\mathcal{T}(k).\label{G:SAbxy}
\end{align}

$\boldmath{B_{xy}/\textbf{SA}}$: By the same calculation, the Green's function is given by
\begin{align}
	\qquad\quad& \mathbb{G}(k)_{B^{(-1)}_{xy}}^{(SA)} ~=~ \frac{1}{2\mathcal{K}_{xy+}\mathcal{K}_{xy-}} \begin{pmatrix}
		g_{11} & g_{12}\\
		g_{21} & g_{22}
	\end{pmatrix},\\
	\text{with} \quad
	g_{11} = g_{22}^* &=  (\mathcal{K}_{xy-}+\mathcal{K}_{xy+})(\mathbb{K}^T-\frac{b_{xy}k_y}{|\kp|}\sigma_1)+\frac{b_{xy}k_x}{|\kp|}(\mathcal{K}_{xy-}-\mathcal{K}_{xy+})\sigma_3, \\
	g_{12} = g_{21}^* &= -\frac{1}{2|\kp|}(\mathcal{K}_{xy-}-\mathcal{K}_{xy+})\tilde{\mathbb{K}} - b_{xy}(\mathcal{K}_{xy-}-\mathcal{K}_{xy+})\sigma_0.
\end{align}
where $\kp^2 = k_x^2+k_y^2$, $\mathcal{K}_{xy_\pm} = \sqrt{(b_{xy} \pm |\kp|)^2 +k_z^2 -\omega^2}$,  $b_{xy} = B_{xy}^{(-1)}$,\\ $\tilde{\mathbb{K}} = \mathbb{K}U\mathbb{K}$, $\displaystyle U= \frac{2}{\boldsymbol{k^2}-\omega^2}
\begin{pmatrix}
	k_x(k_x-\omega)+k_y(k_y+i k_z) & ik_xk_z + k_y \omega\\
	ik_xk_z + k_y \omega & k_x(k_x+\omega) + k_y(k_y-i k_z).
\end{pmatrix}
$.\\

$\boldmath{B_{tz}/\textbf{SS}/\textbf{SA}}$: The decomposition and simplification of the Green's functions for this case pose significant challenges. Consequently, we will focus only on the trace of the Green's function, which included in the following section.

\newpage
\section{Features of spectral functions}
The spectral functions (SF) can be determined by the imaginary part of the traced Green's functions:
\begin{equation}
A(\omega,k) = \text{Im}[\Tr(\mathbb{G})].
\end{equation}   
Since the analytic results can be obtained  when the order parameter fields  have only leading  terms   we will analyze only such cases. 
\subsection{Scalar}
\myparagraph{SS}
The essential part of the Green's functions is given by the trace  
  (\ref{G:SSscalar}),
\begin{align}
\Tr\mathbb{G}_{M_0}^{(SS)} &= 4\omega\frac{\sqrt{\boldsymbol{k}^2-\omega^2 +M_0^2}}{\boldsymbol{k}^2-\omega^2 -i\epsilon}.\label{TrGscalarSS}
\end{align} 
where $\boldsymbol{k}^2 = \sum_{i=1}^{d-1}k_i^2$, and $M_0$ is the  scalar source. The simple pole is located at the surface of the $d$ dimensional cone where $d$ is dimension of the $\text{AdS}$ boundary. 
Notice that the symmetry breaking strength $M_0$ does not affect the pole structure but only contributes   to the gap size. In $\text{AdS}_4$, the pair of the gapped spectrum with  $M_0$, $M_{50}$ was reported \cite{AdS4}. In $\text{AdS}_5$, we do not have the chiral dynamics of the boundary although we should have corresponding spectrum from the boundary point of view. The difficulty lies in the fact that  the chirality cannot be defined in odd dimensions. 
We postpone this problem to the future work. 

\begin{figure}[t!]
	\centering
	\begin{subfigure}{0.22\textwidth}
		\centering
		\includegraphics[width=3.0cm]{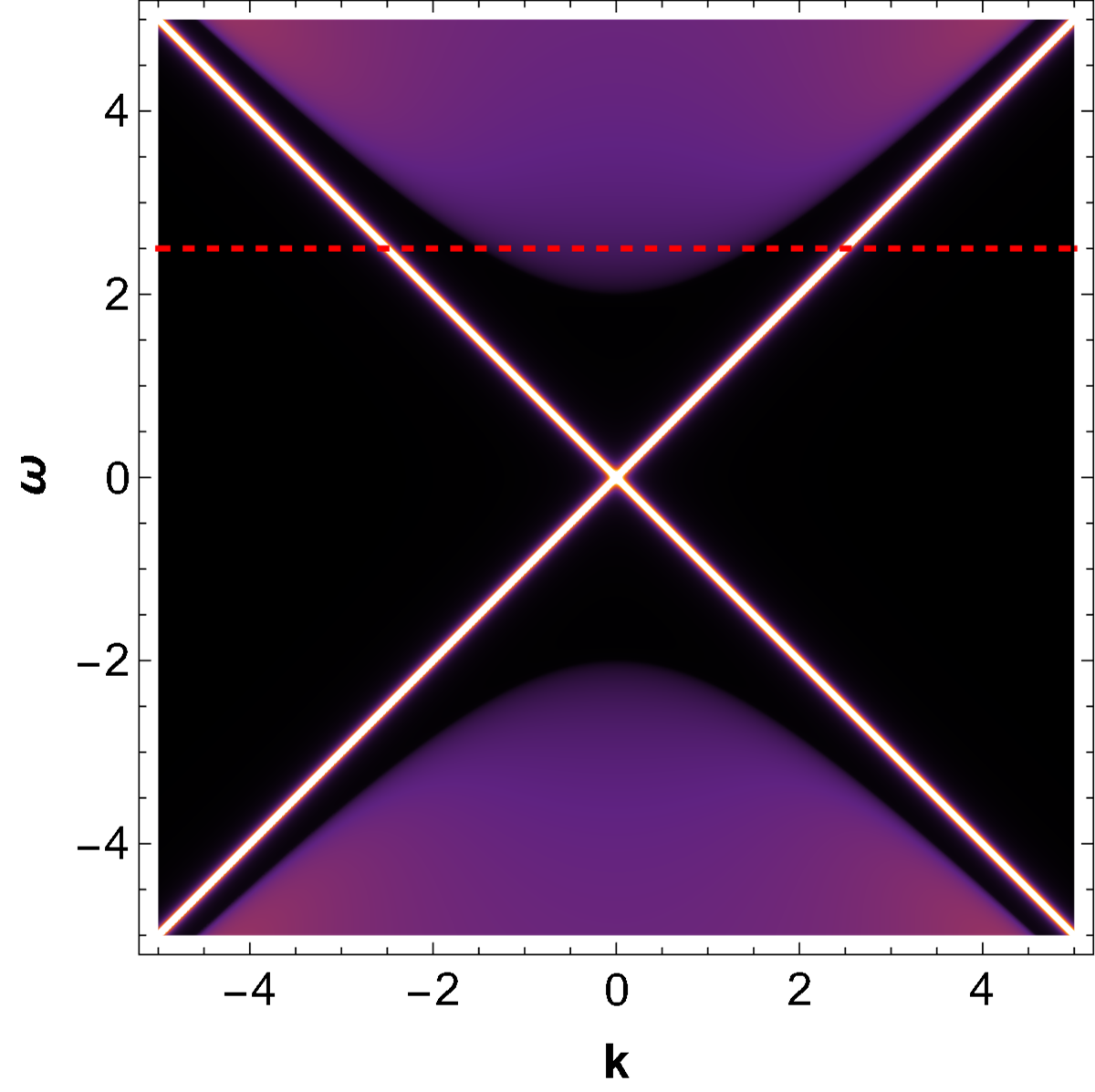}
		\caption{$M_0^{(SS)},\omega\text{-}\boldsymbol{k}$}
	\end{subfigure}
	\begin{subfigure}{0.22\textwidth}
		\centering
		\includegraphics[width=3.0cm]{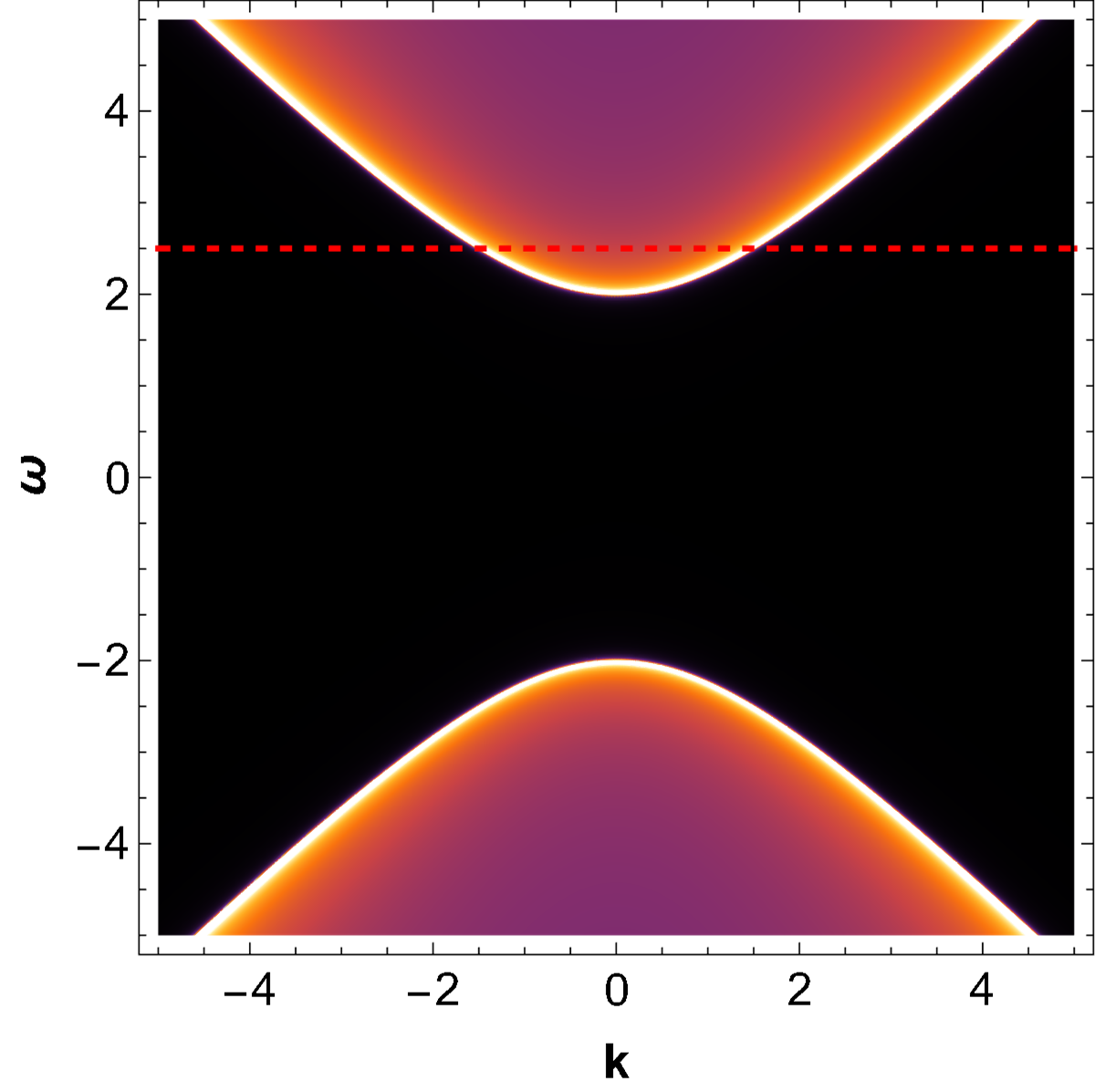}
		\caption{$M_0^{(SA)},\omega\text{-}\boldsymbol{k}$}
	\end{subfigure}
	\begin{subfigure}{0.22\textwidth}
	\centering
	\vspace{-0.39cm}
	\includegraphics[width=3.0cm]{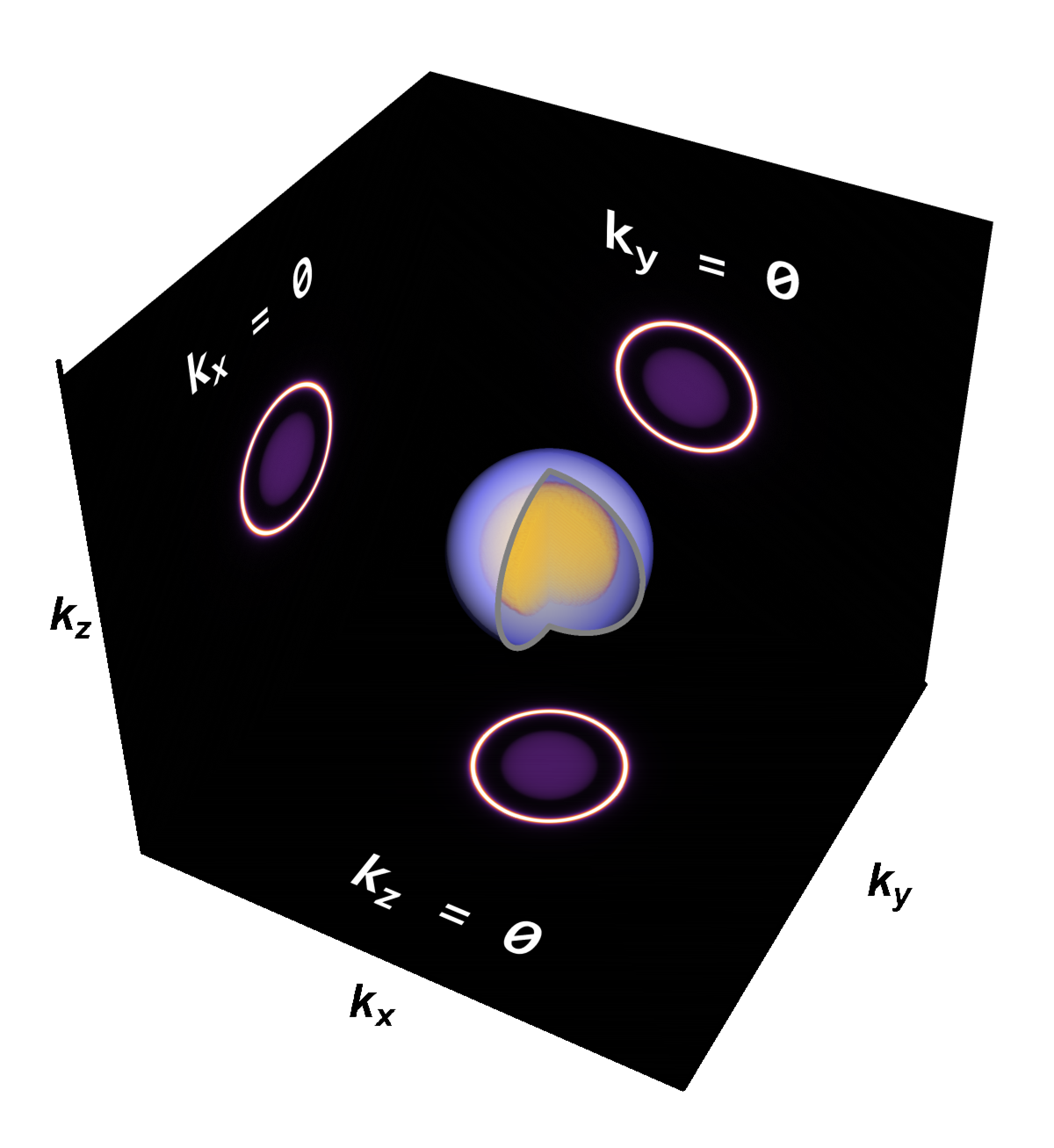}
	\caption{$\omega = 2.5$}
	\end{subfigure}
	\begin{subfigure}{0.22\textwidth}
		\centering
		\vspace{-0.39cm}
		\includegraphics[width=3.0cm]{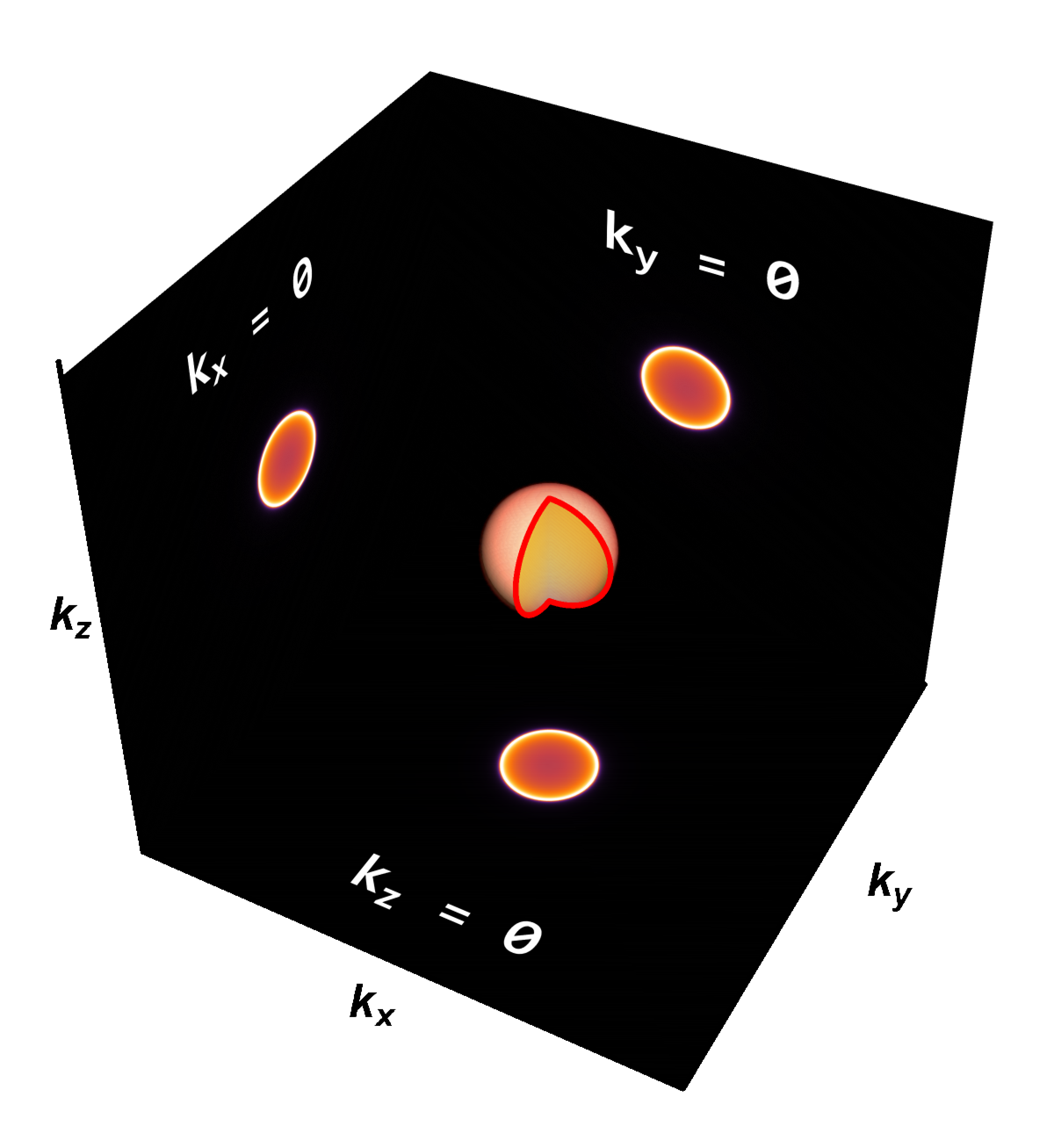}
		\caption{$\omega = 2.5$}
	\end{subfigure}
	\caption{\small Spectral Functions (SFs) of scalar source for both quantization choices. (a,b) SFs in $\omega\text{-}\boldsymbol{k}$ plane. (c,d) SFs in $\omega = 2.5$ slice, the dashed red line in figure (a), is the three dimensional object  in (c,d). The figure at   each plane is its projection to each plane. The blue (c) and red (d) surfaces represent  the pole and the branch-cut type singularity, respectively.
}
	\label{fig:scalar}
\end{figure}

\myparagraph{SA}
 The analytic expression is given by
\begin{align}
\Tr \mathbb{G}_{M_0}^{(SA)} &= \frac{4\omega}{\sqrt{\boldsymbol{k}^2-\omega^2+M_0^2}},\label{TrGscalarSA}
\end{align}
The main feature of this interaction is  the gap generation,  as it was noticed in \cite{AdS4,Yuk:2022lof, Lieb}.  Therefore, the scalar source in this case can be interpreted as the mass of boundary fermions. In $\text{AdS}_5$, only scalar SA quantization can generate the gap, while in $\text{AdS}_4$ case,  both $M_0,M_{05}^2$ can do that. See figure \ref{fig:scalar}(b,d).

\myparagraph{Diagonal interaction in fermion flavors}
 For the scalar,  we consider the case where the fermion-scalar interaction is diagonal type, namely, 
$$  S_{int} = \int d^{5}x \sqrt{-g} \Big(\bar\psi^{(i)}\Phi\cdot\Gamma \psi^{(i)}+h.c \Big).$$
In this case, we have independent sum of two flavors and the result is following.  
\begin{align}
	\Tr \mathbb{G}_{M_0}&= \frac{2\omega}{- M_0+\sqrt{\boldsymbol{k}^2-\omega^2 -i\epsilon+M_0^2}},\label{TrGscalarSA}
\end{align}
Notice that  the sign of $M_0$ is important:  for ${\rm sign}(M_0) >0$ we have gapless spectrum while for negative case we have gapped one. 
Therefore, in the intra-flavor case with $\mathcal{L}_{int} = i\Phi(\bar{\psi}^{(1)}\psi^{(1)})$, the massless-gapped phase transition depends on the changing sign of $M_0$ \cite{Strangemetal}.  However, in our inter-flavor with  $\mathcal{L}_{int} = i\Phi(\bar{\psi}^{(2)}\psi^{(1)}+h.c)$, there is no   phase transition  under the    sign change  of $M_0$. See figure \ref{fig:scalar}(a,c).  It turns out that for all  interaction types other than the scalar-fermion, there is no such phase transition between the gap-gapless phases in the spectral function.

\subsubsection{Radial scalar $B_{u}$}
\myparagraph{SS and SA}
For this interaction,   there is no effect from the order parameter $b$ due to the cancelation that happened during calculation of the Green's function, see (\ref{bucancelation}). In fact, this has been a puzzle from the view of the numerical calculation. 
As a result, the trace of the Green's function, regardless the quantization choice, is given by  
\begin{align}
\Tr\mathbb{G}_{B_{u}^{(0)}}^{(SS,SA)} = \frac{4\omega}{\sqrt{\boldsymbol{k}^2-\omega^2}} , \label{TrGBu}
\end{align}
which is the same as that of critical point where $B_u=0$. 
 
\subsection{Vectors}
\subsubsection{Time-like polar vector, $B_{t}$}
\begin{figure}[t!]
	\centering
	\begin{subfigure}{0.22\textwidth}
		\centering
		\includegraphics[width=3.0cm]{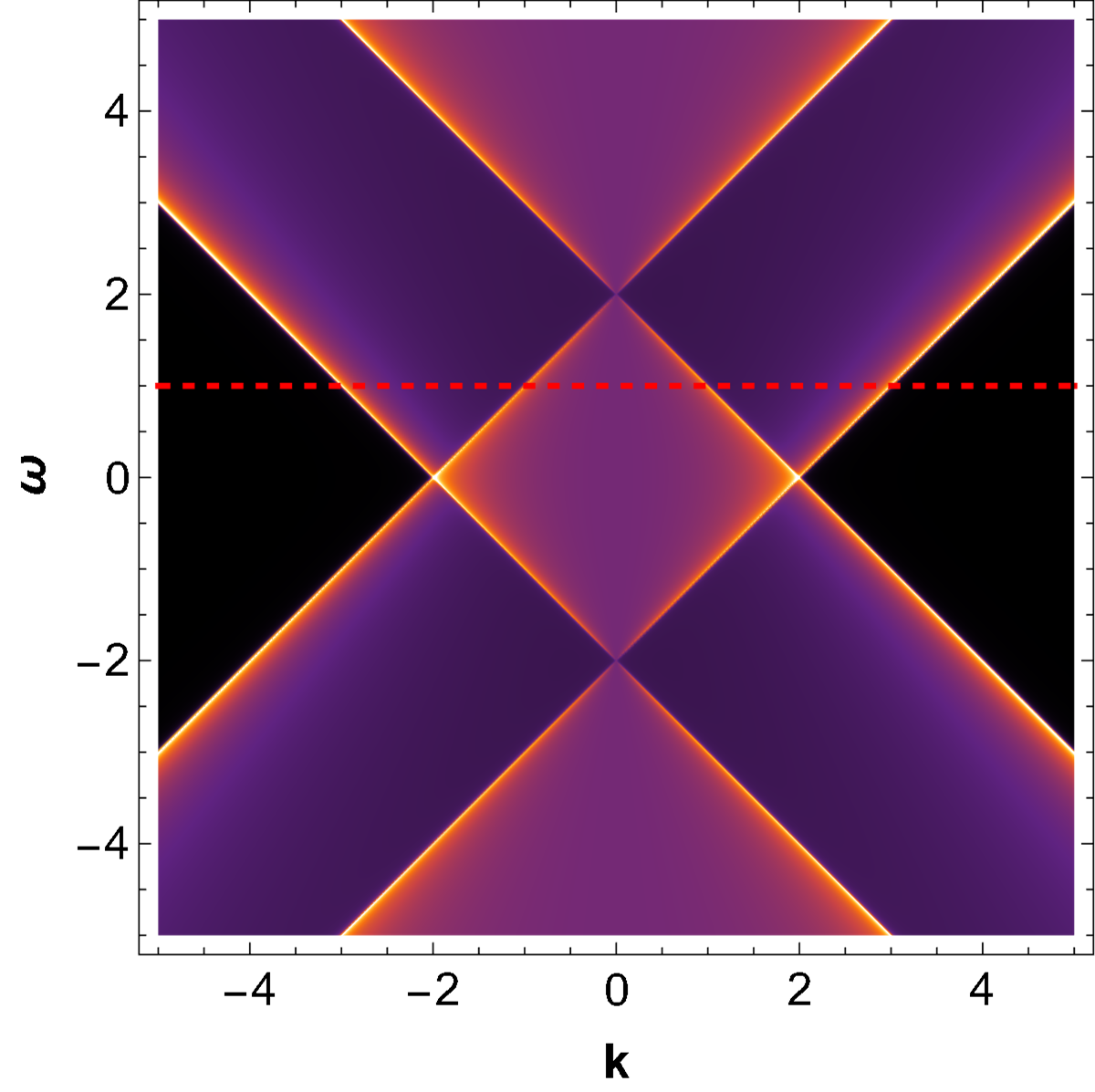}
		\caption{$B^{(0)(SS)}_{t},\omega\text{-}\boldsymbol{k}$}
	\end{subfigure}
	\begin{subfigure}{0.22\textwidth}
		\centering
		\vspace{-0.39cm}
		\includegraphics[width=3.0cm]{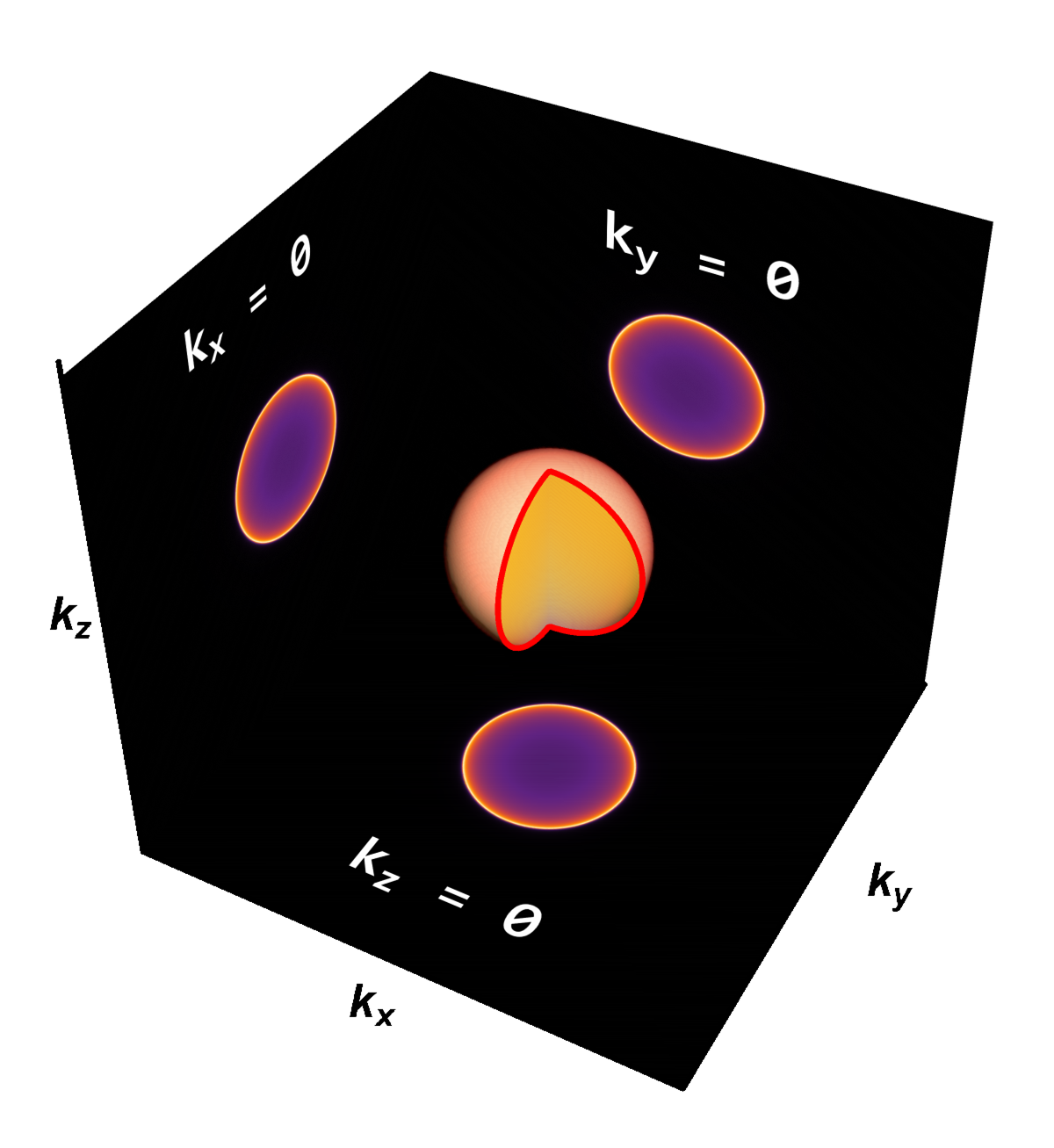}
		\caption{$\omega = 0$}
	\end{subfigure}
	\begin{subfigure}{0.22\textwidth}
		\centering
		\vspace{-0.39cm}
		\includegraphics[width=3.1cm]{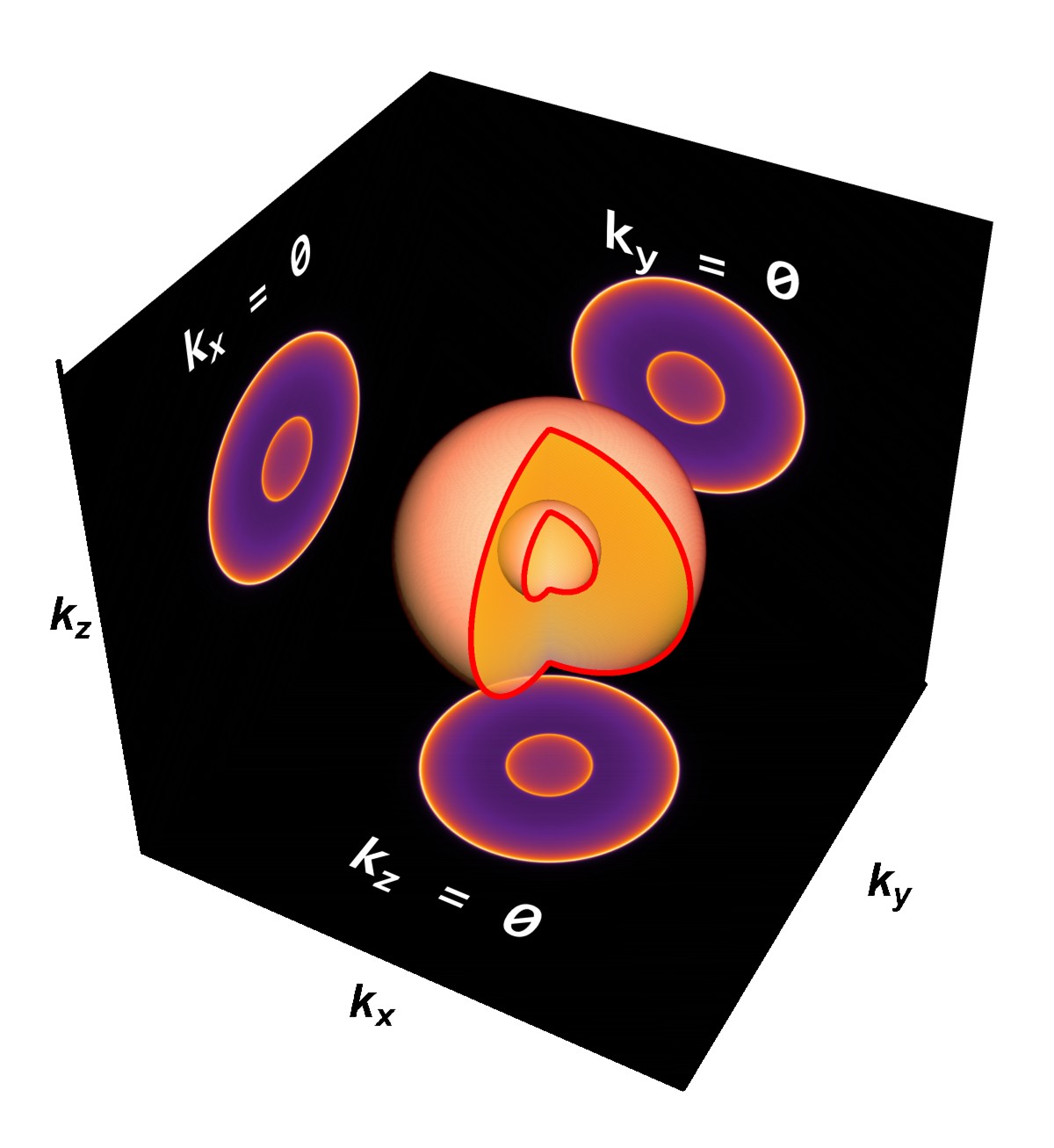}
		\caption{$\omega = 1$}
	\end{subfigure}\\\vspace{0.4cm}
	\begin{subfigure}{0.22\textwidth}
		\centering
		\includegraphics[width=3.0cm]{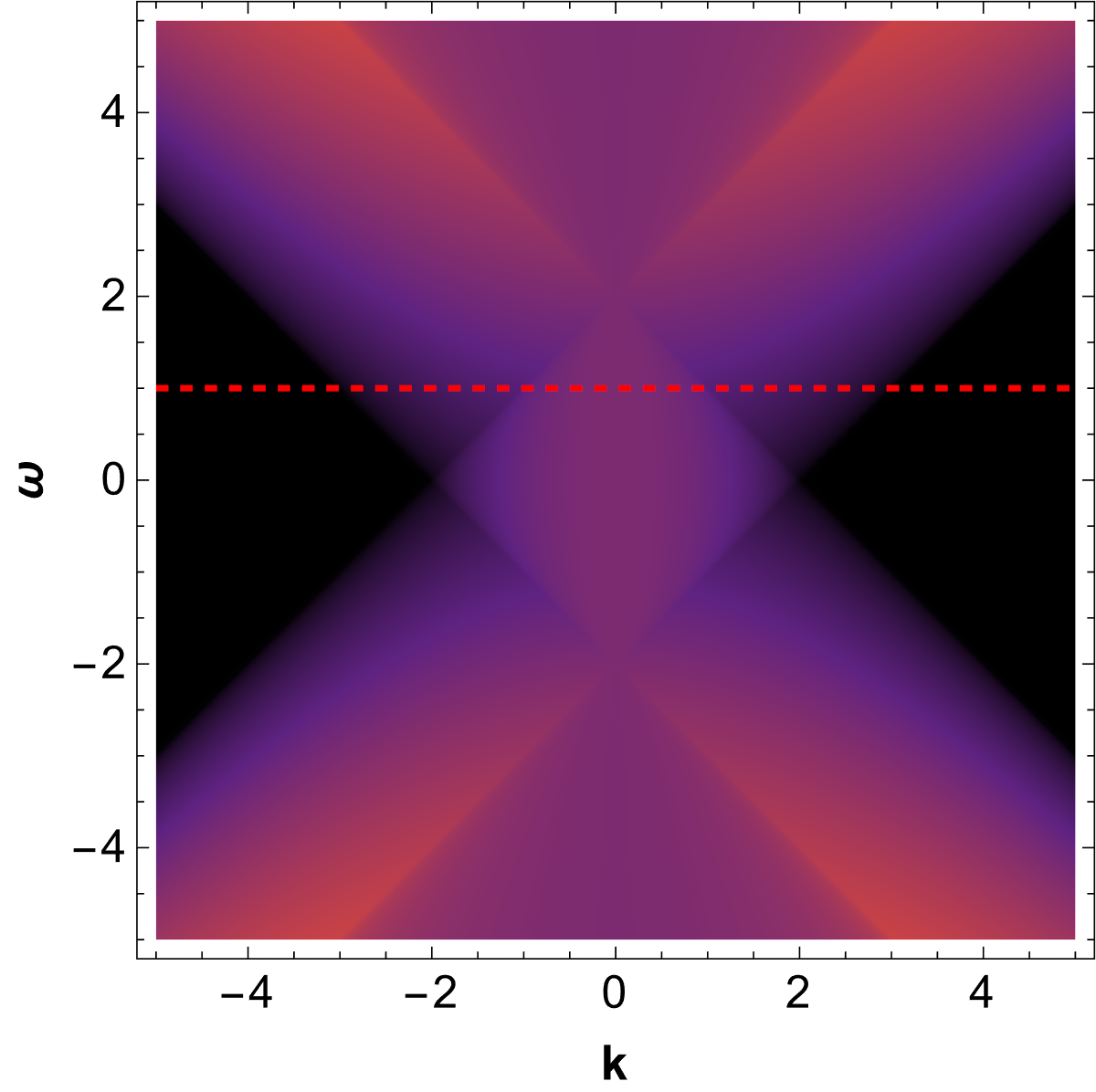}
		\caption{$B^{(0)(SA)}_{t},\omega\text{-}\boldsymbol{k}$}
	\end{subfigure}
	\begin{subfigure}{0.22\textwidth}
		\centering
		\vspace{-0.39cm}
		\includegraphics[width=3.0cm]{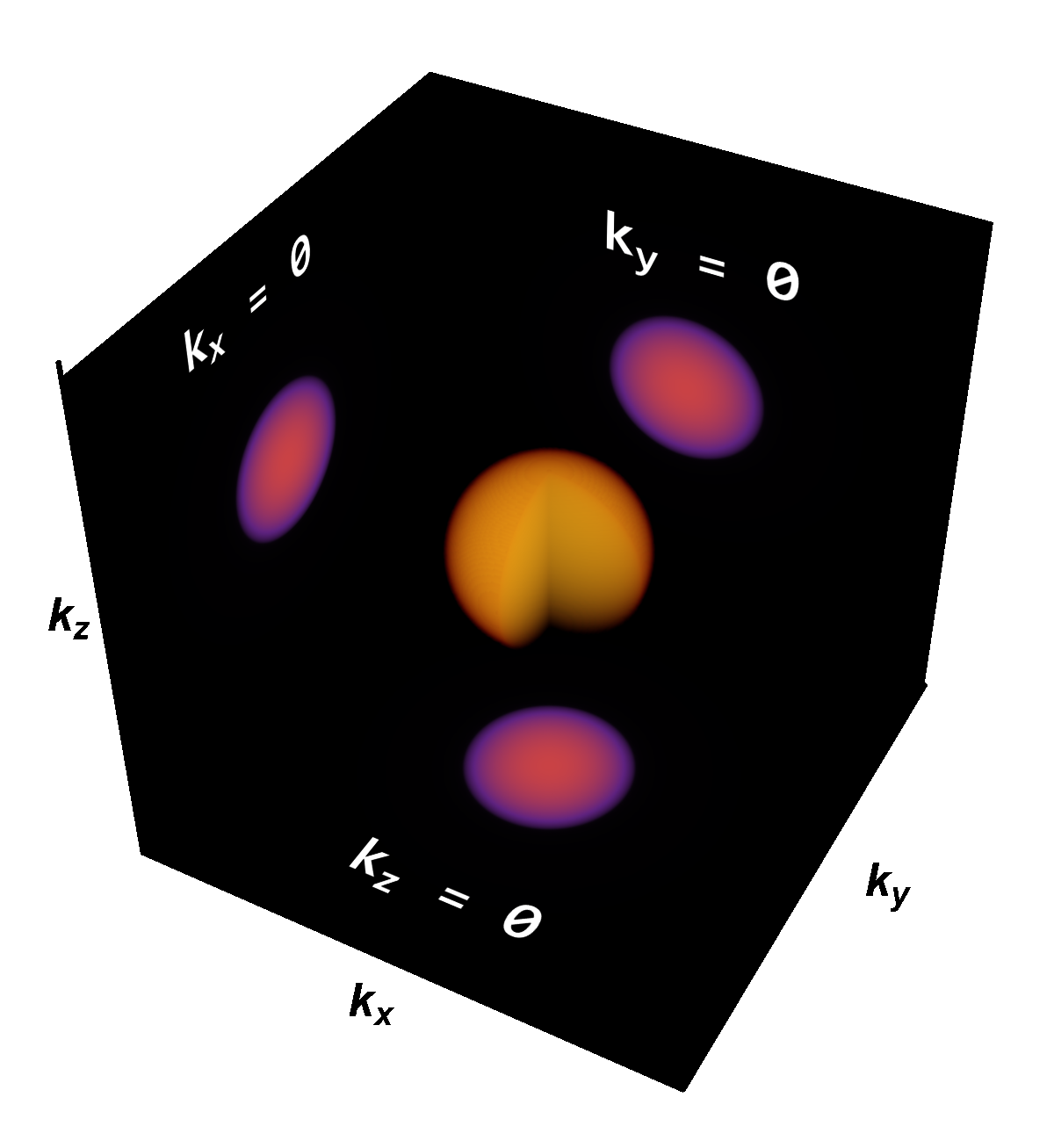}
		\caption{$\omega = 0$}
	\end{subfigure}
	\begin{subfigure}{0.22\textwidth}
		\centering
		\vspace{-0.39cm}
		\includegraphics[width=3.0cm]{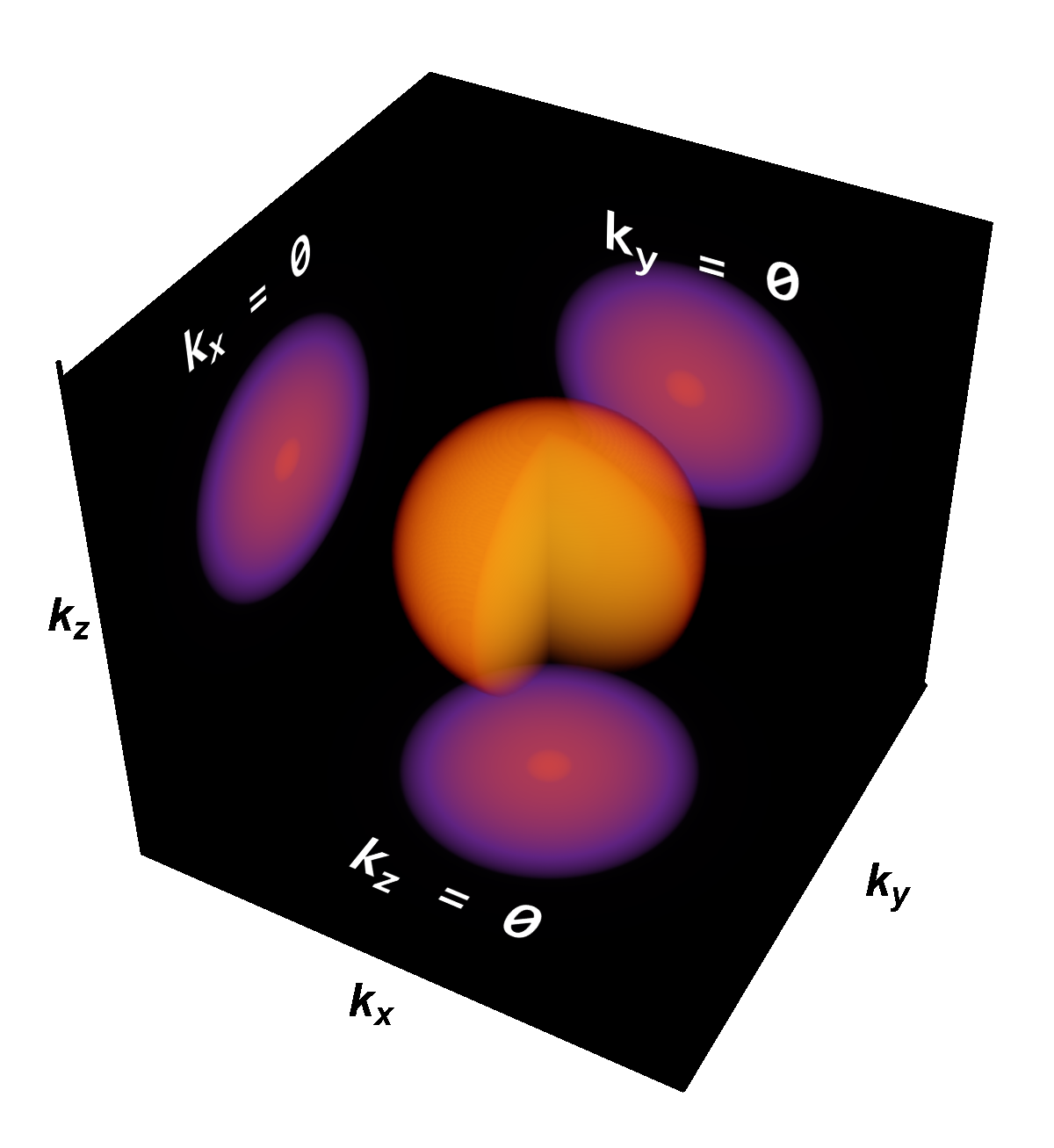}
		\caption{$\omega = 1$}
	\end{subfigure}
	\caption{\small Spectral functions (SFs) of $B_{t}$ source for both quantization choices. (a,d) SFs in $\omega\text{-}\boldsymbol{k}$ plane. (b,e) SFs in $\omega = 0$ plane, and (c,f) for  $\omega = 1$ plane corresponding to the dashed red line, in which the background represents the certain slices at each momentum is zero. The bare orange bulk without the surface shows the spectral function without singularity.}
	\label{fig:Bt}
\end{figure}

\myparagraph{SS}
The trace of the Green's matrix (\ref{G:SSbx}), by choosing $\mu = t$ is given by
\begin{align}
\Tr\mathbb{G}_{B_{t}^{(0)}}^{(SS)} &= 2\Big(\frac{b+\omega}{\sqrt{\boldsymbol{k}^2-(b+\omega)^2}}-\frac{b-\omega}{\sqrt{\boldsymbol{k}^2-(b-\omega)^2}}\Big), \label{TrGBtSS}
\end{align}
where $\boldsymbol{k}^2 = \sum_i^{d-1} k^2_i$. In this case, above result shows two Dirac cones,  shifted  along $\pm\omega$ directions, which are not interacting with each other.  The singularities are located at each cone. Notice that there are spherical symmetry   in $k_x\text{-}k_y\text{-}k_z$. See figure \ref{fig:Bt}(a,b,c)

\myparagraph{SA}
The trace of the Green's matrix (\ref{G:SAbx})   is given by
\begin{align}
\Tr \mathbb{G}_{B_{t}^{(0)}}^{(SA)} &=   \frac{2}{b}\left[ \sqrt{\boldsymbol{k}^2 - (b - \omega)^2} -\sqrt{\boldsymbol{k}^2 - (b + \omega)^2} \right],\label{TrGBtSA}
\end{align}
In this case, the symmetry are the same with SS case. However, there is no  singularity in the Green's function.  Therefore the entire SF is described by a branch-cut without singularity. See figure \ref{fig:Bt}(d,e,f)

\subsubsection{Time-like radial vector $B_{tu}$}
\begin{figure}[t!]
	\centering
	\begin{subfigure}{0.22\textwidth}
		\centering
		\includegraphics[width=3.0cm]{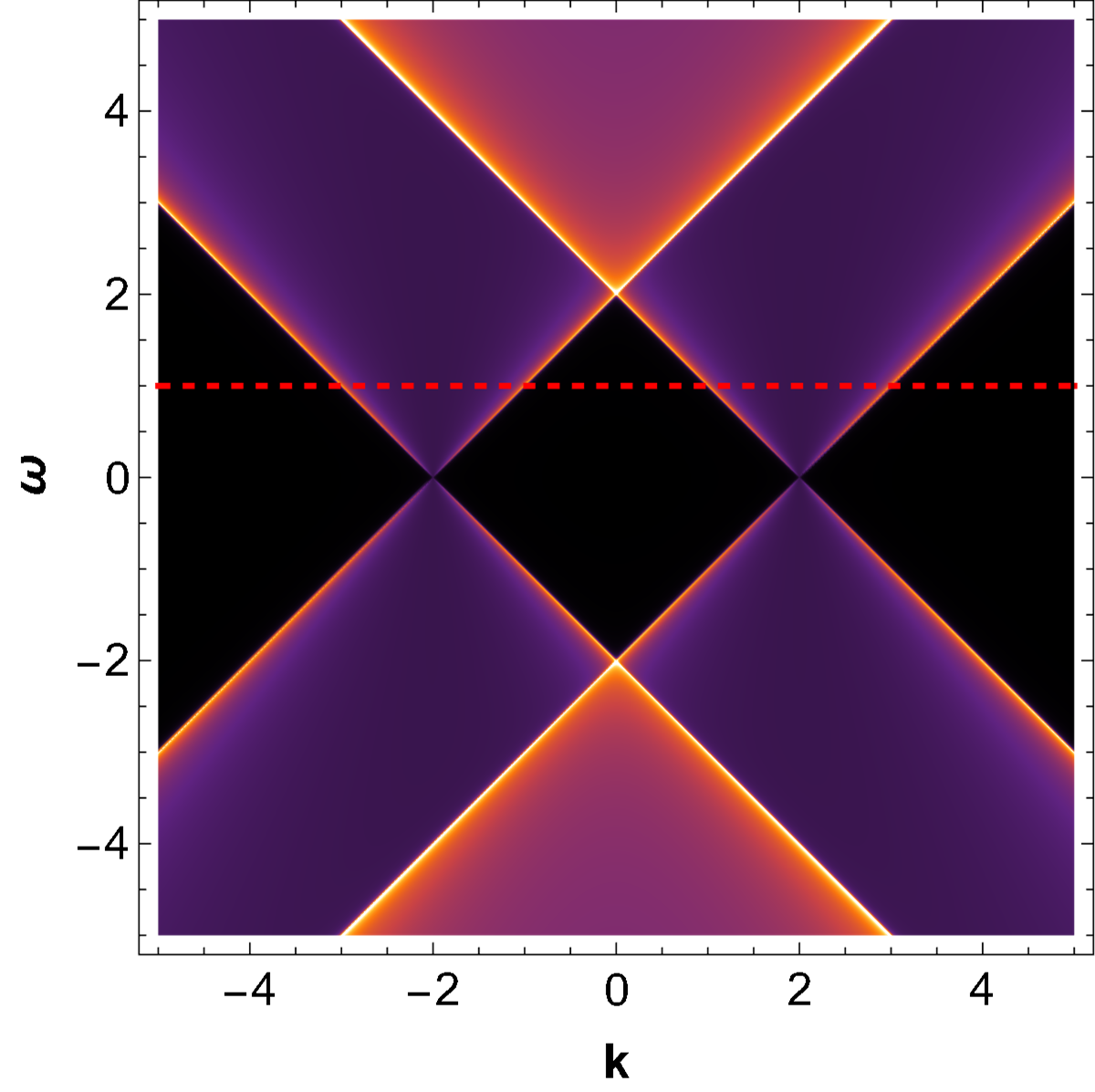}
		\caption{$B^{(-1)(SS)}_{tu},\omega\text{-}\boldsymbol{k}$}
	\end{subfigure}
	\begin{subfigure}{0.22\textwidth}
		\centering
		\vspace{-0.39cm}
		\includegraphics[width=3.0cm]{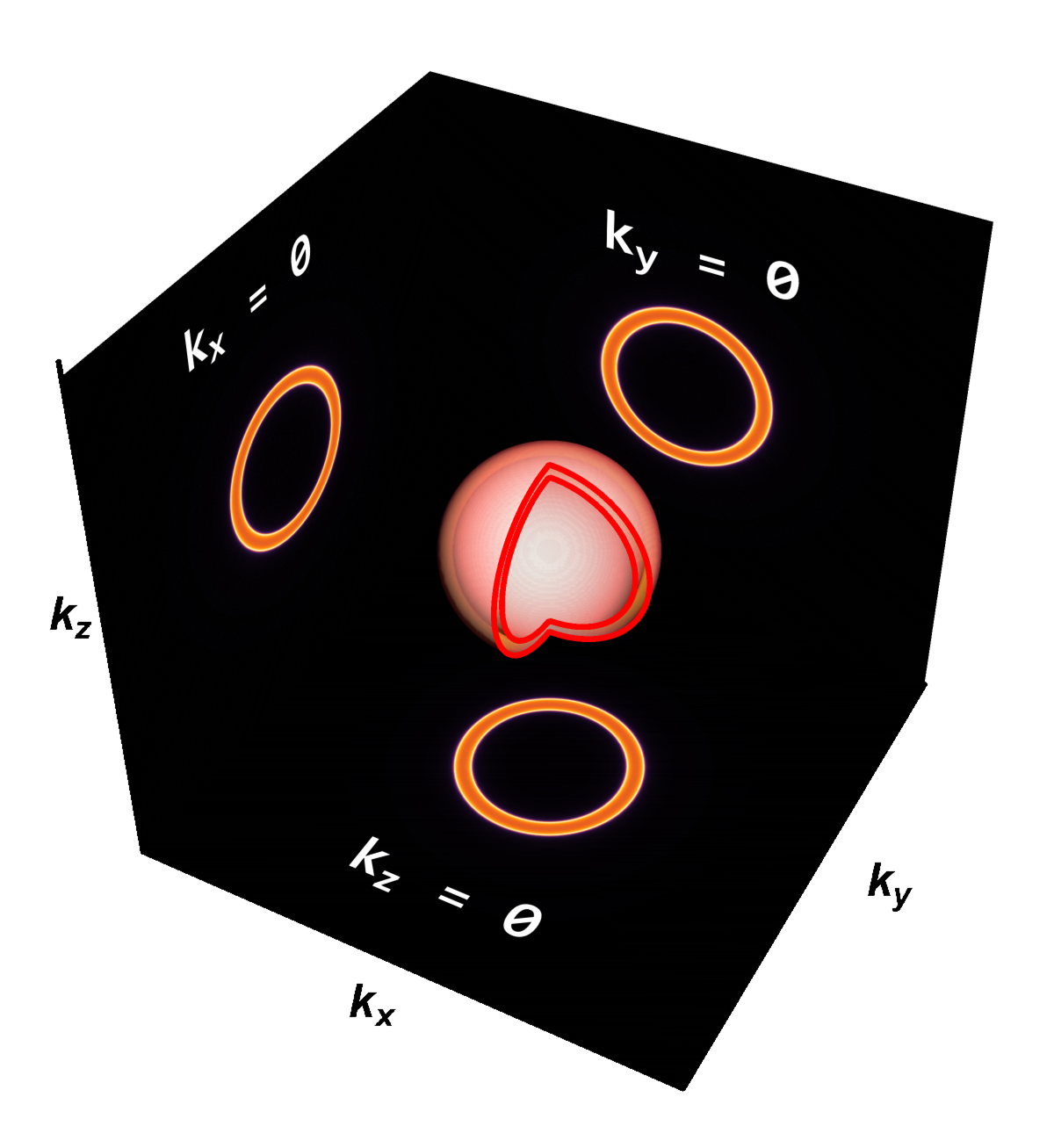}
		\caption{$\omega \simeq 0$}
	\end{subfigure}
	\begin{subfigure}{0.22\textwidth}
		\centering
		\vspace{-0.39cm}
		\includegraphics[width=3.0cm]{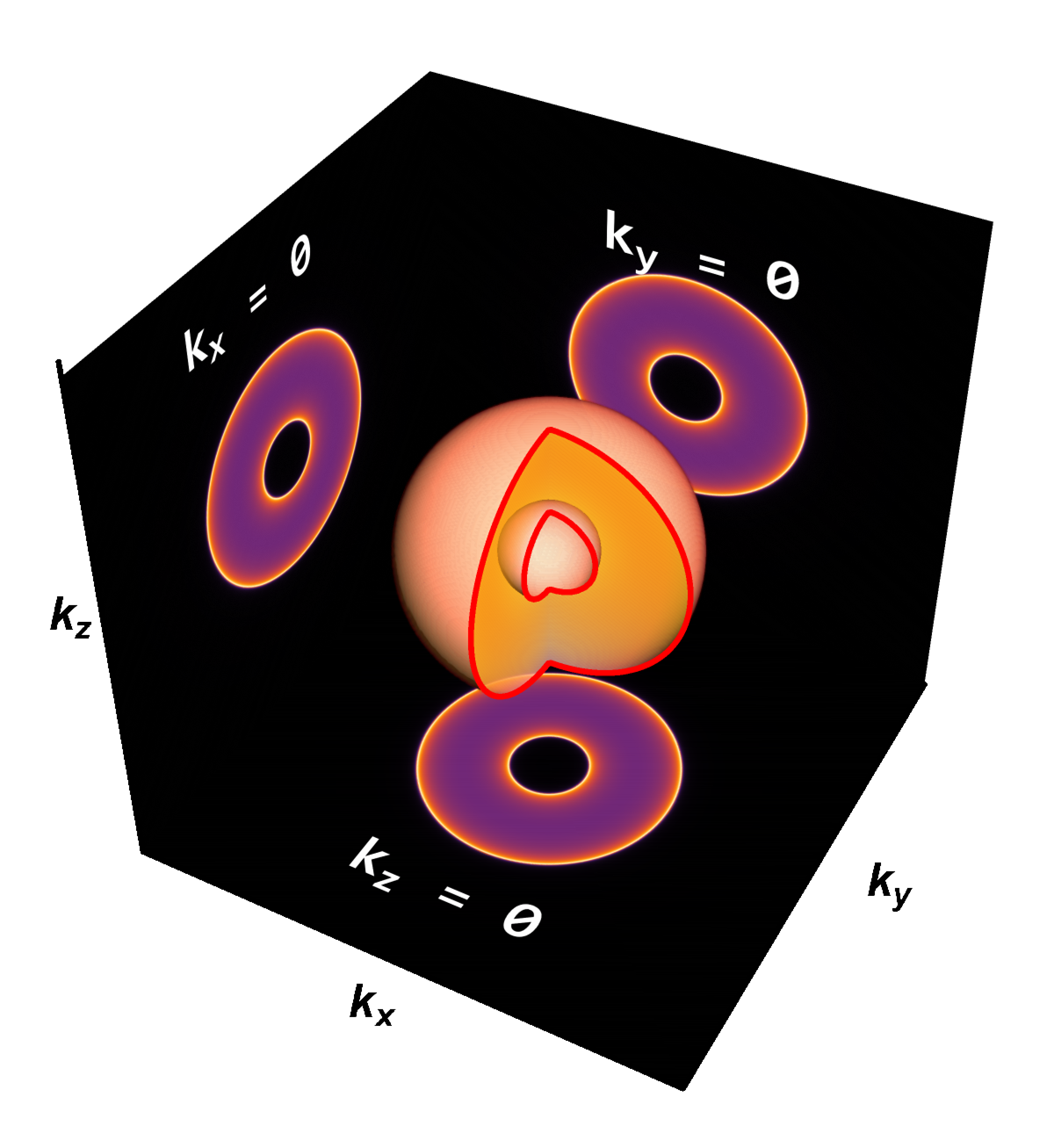}
		\caption{$\omega = 1$}
	\end{subfigure}\\\vspace{0.4cm}
	\begin{subfigure}{0.22\textwidth}
		\centering
		\includegraphics[width=3.0cm]{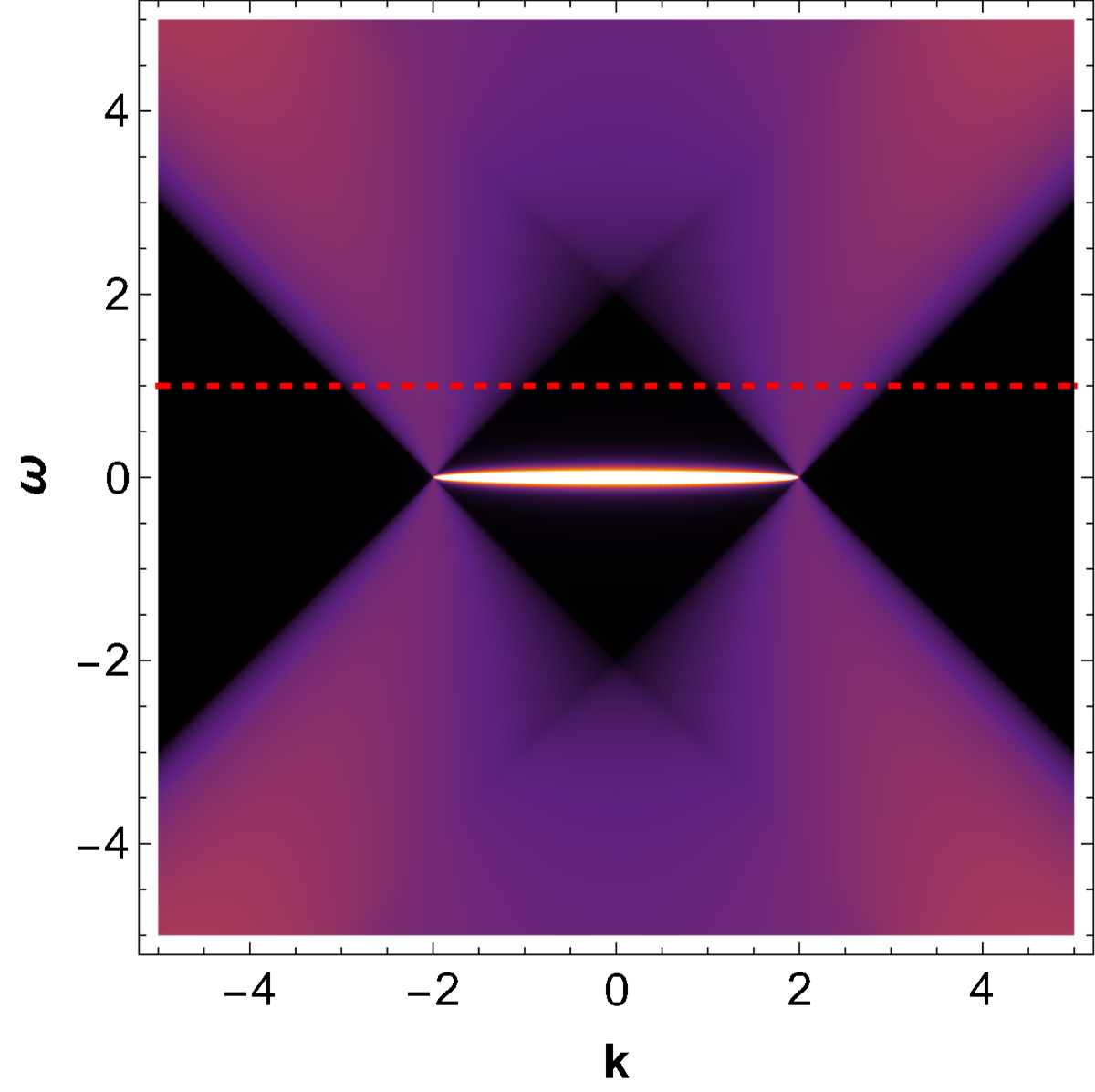}
		\caption{$B^{(-1)(SA)}_{tu},\omega\text{-}\boldsymbol{k}$}
	\end{subfigure}
	\begin{subfigure}{0.22\textwidth}
		\centering
		\vspace{-0.39cm}
		\includegraphics[width=3.0cm]{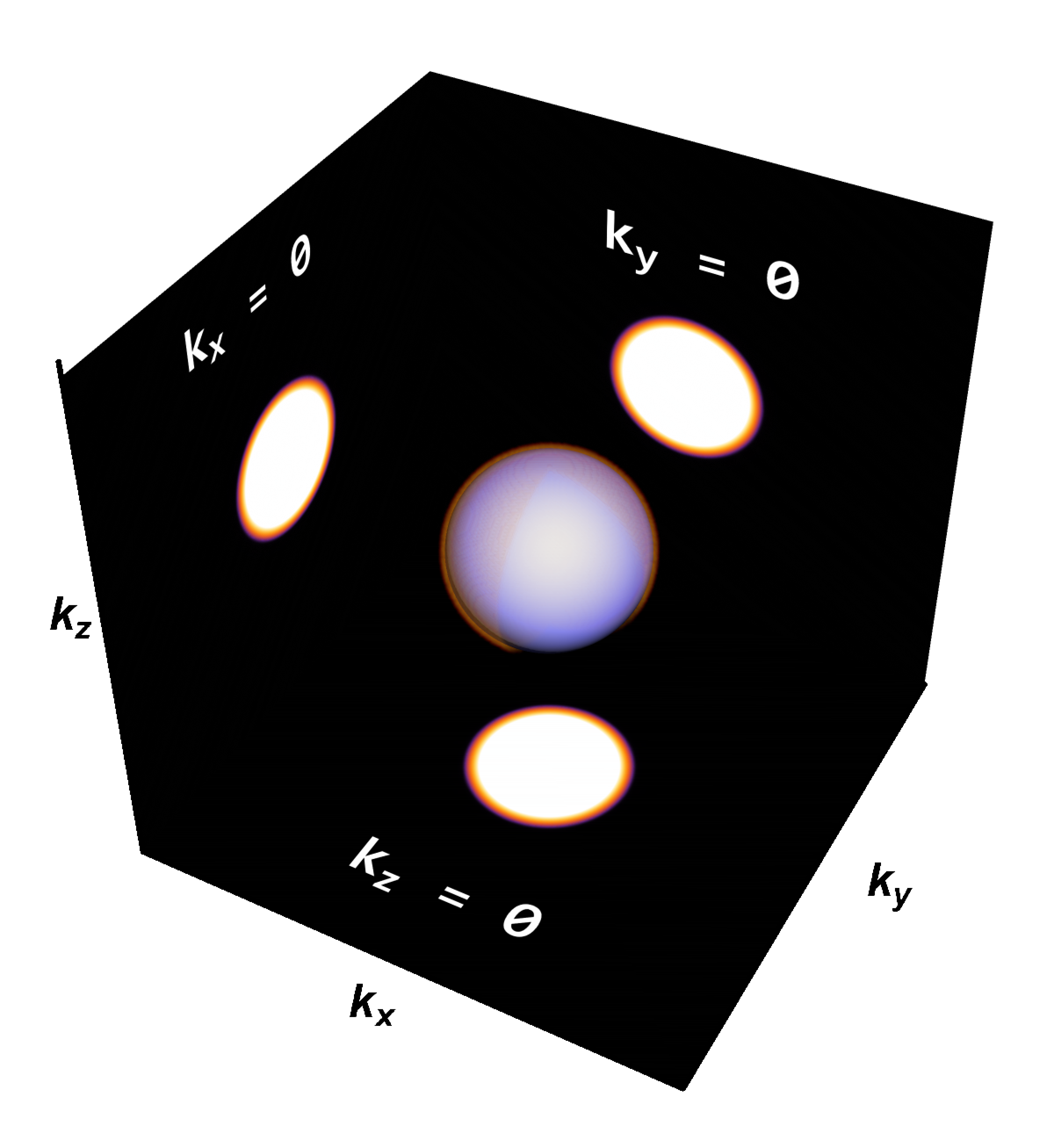}
		\caption{$\omega \simeq 0$}
	\end{subfigure}
	\begin{subfigure}{0.22\textwidth}
		\centering
		\vspace{-0.39cm}
		\includegraphics[width=3.0cm]{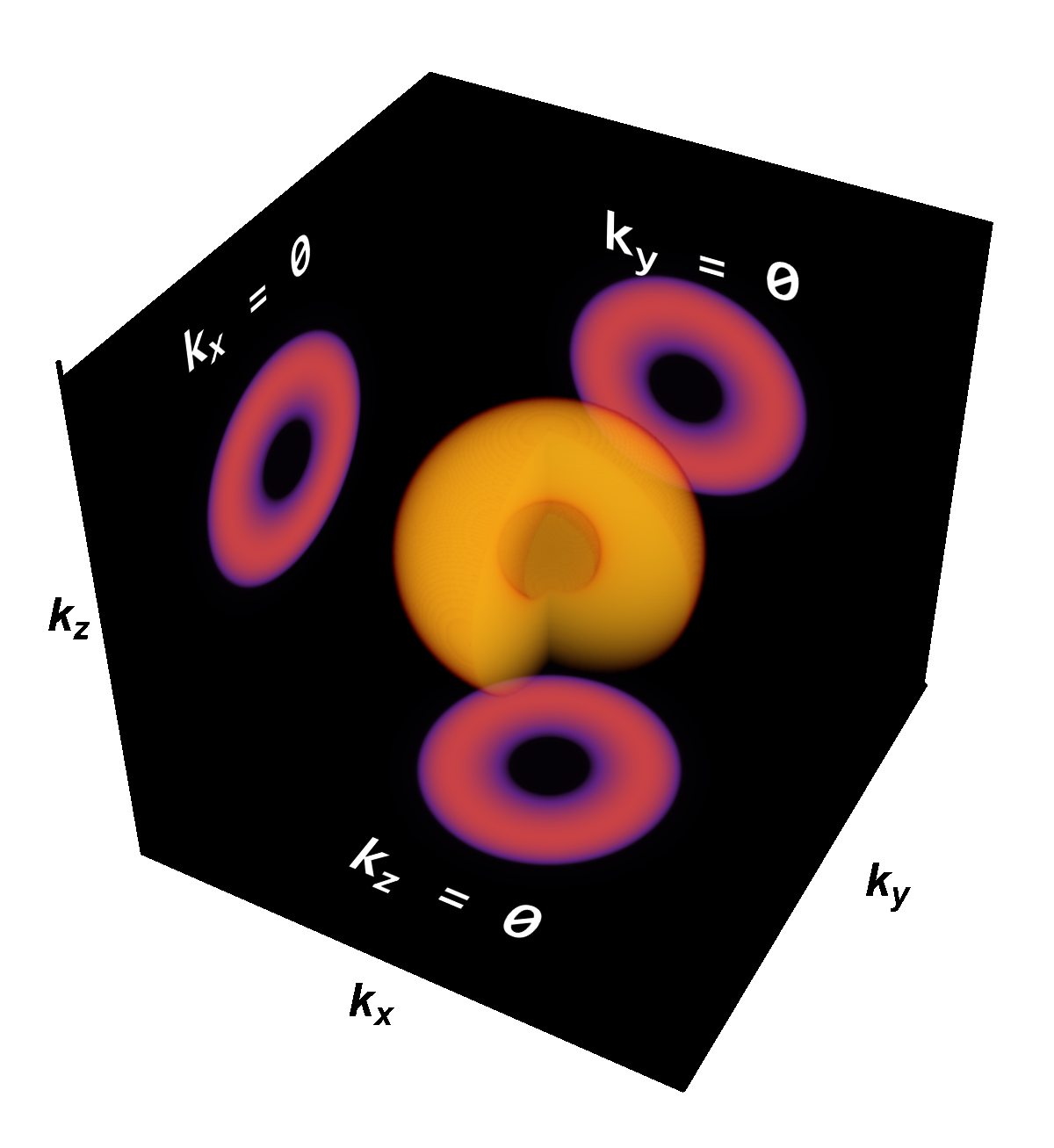}
		\caption{$\omega = 1$}
	\end{subfigure}
	\caption{\small Spectral functions (SFs) for $B_{tu}$ source for both quantization choices. (a,c) SFs in $\omega\text{-}\boldsymbol{k}$ plane. (b,c),(e,f) SFs in $k_x\text{-}k_y\text{-}k_z$  at $\omega = 0,1$ slices, respectively.}
	\label{fig:Btu}
\end{figure}

\myparagraph{SS}
 The analytic expression is given by
\begin{align}
\Tr \mathbb{G}_{B_{tu}^{(-1)}}^{(SS)} &= \frac{2\omega}{\sqrt{(b-|\boldsymbol{k}|)^2-\omega^2}}+\frac{2\omega}{\sqrt{(b+|\boldsymbol{k}|)^2-\omega^2}}. \label{TrGBtuSS}
\end{align}
In this case, the spectrum is isotropic  for each  Dirac cone shifted along the entire $\boldsymbol{k}$-space, and that  is why we cannot distinguish the spectrum of $B_{xy(SA)}$ and $B_{tu(SA)}$ in $AdS_4$. However, in $AdS_5$, the SF has spherical symmetry, while $B_{xy}$ spectrum has  planar  rotational symmetry in $k_x\text{-}k_y$ plane. See figure \ref{fig:Btu}(a,b,c).

\myparagraph{SA}
The analytic expression is given by
\begin{align}
\Tr \mathbb{G}_{B_{tu}^{(-1)}}^{(SA)} &= -\frac{2}{b}\Big[\frac{ (b+|\boldsymbol{k}|)\sqrt{(b-|\boldsymbol{k}|)^2-\omega^2}+(b-|\boldsymbol{k}|)\sqrt{(b+|\boldsymbol{k}|)^2-\omega^2} }{\omega+ i\epsilon}\Big].  \label{TrGBtuSA}
\end{align}
The pole-type singularity appears in this case as a flat band. In AdS$_5$, It is  a 3D  flat band  in the solid sphere with radius $b$. See figure \ref{fig:Btu}(d,e). However, the flat band immediately disappears if  move  to $\omega \neq 0$ slice. See figure \ref{fig:Btu}(f). 

\subsubsection{Space-like polar vector, $B_{x}$}
\begin{figure}[t!]
	\centering
	\begin{subfigure}{0.22\textwidth}
		\centering
		\includegraphics[width=3.0cm]{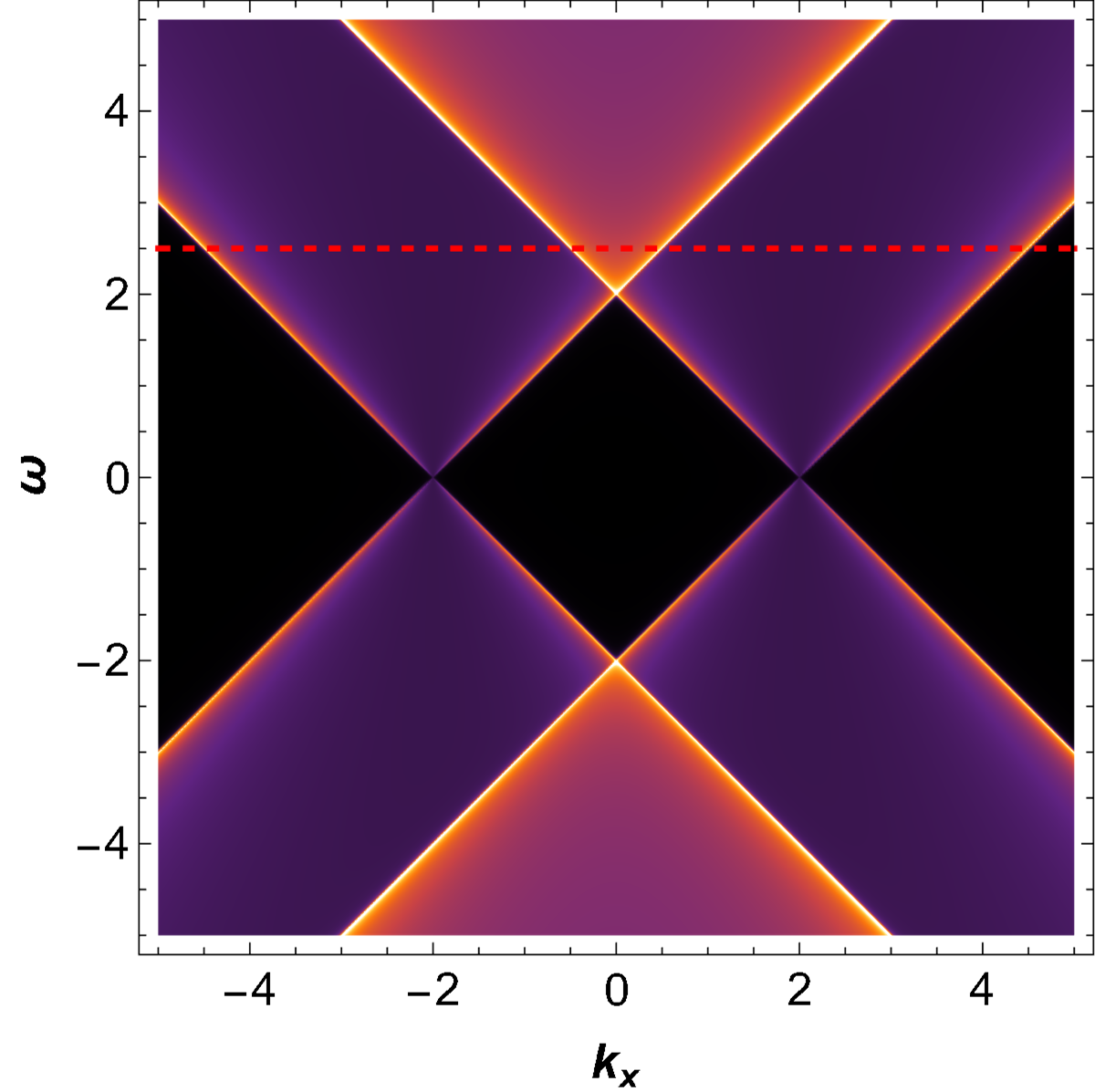}
		\caption{$B^{(0)(SS)}_{x},\omega\text{-}k_x$}
	\end{subfigure}
\begin{subfigure}{0.22\textwidth}
	\centering
	\includegraphics[width=3.0cm]{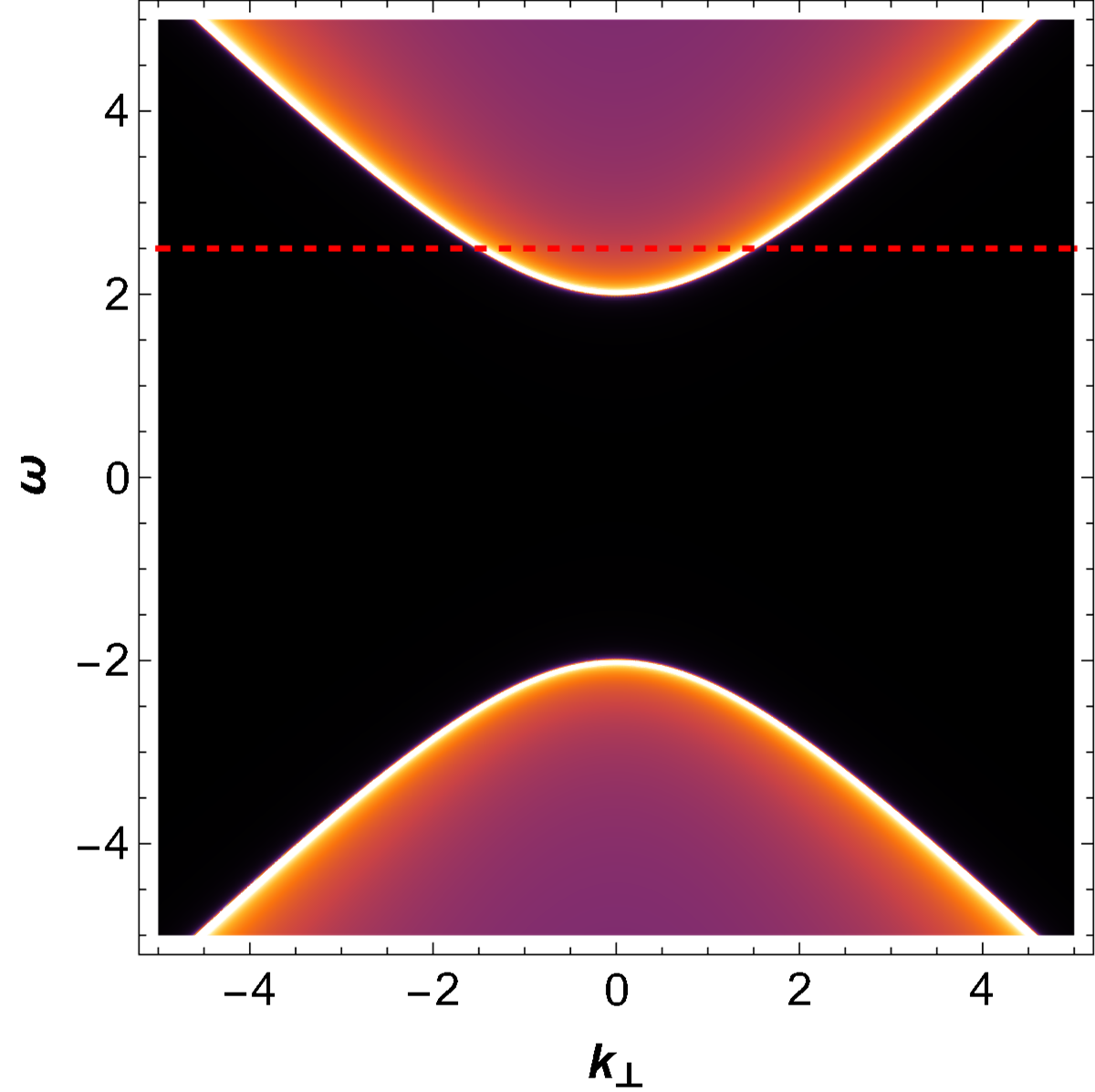}
	\caption{$B^{(0)(SS)}_{x},\omega\text{-}\kp$}
\end{subfigure}
	\begin{subfigure}{0.22\textwidth}
		\centering
		\vspace{-0.39cm}
		\includegraphics[width=3.0cm]{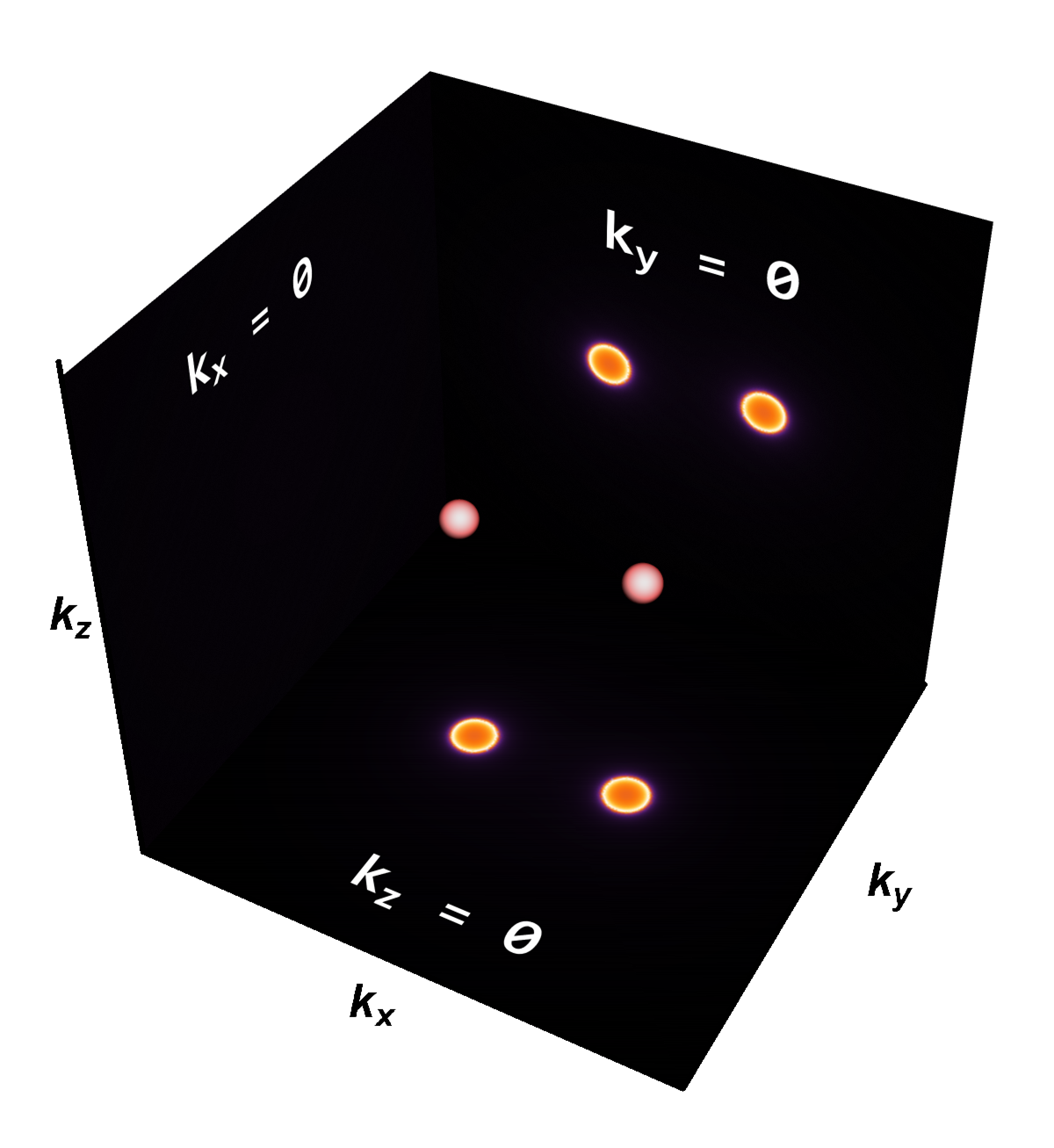}
		\caption{$\omega \simeq 0$}
	\end{subfigure}
	\begin{subfigure}{0.22\textwidth}
		\centering
		\vspace{-0.39cm}
		\includegraphics[width=3.0cm]{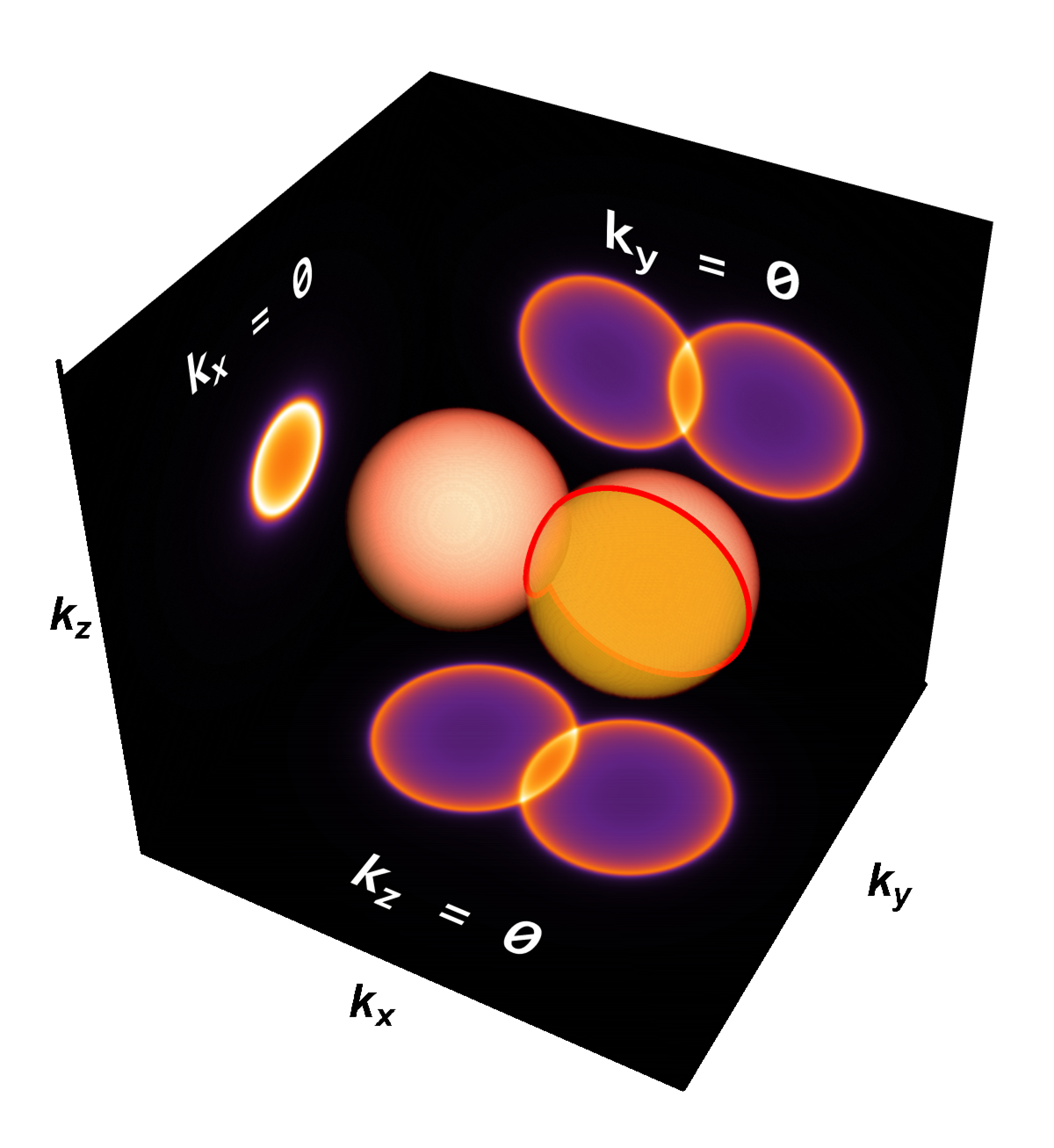}
		\caption{$\omega = 2.5$}
	\end{subfigure}\\\vspace{0.4cm}
		\begin{subfigure}{0.22\textwidth}
		\centering
		\includegraphics[width=3.0cm]{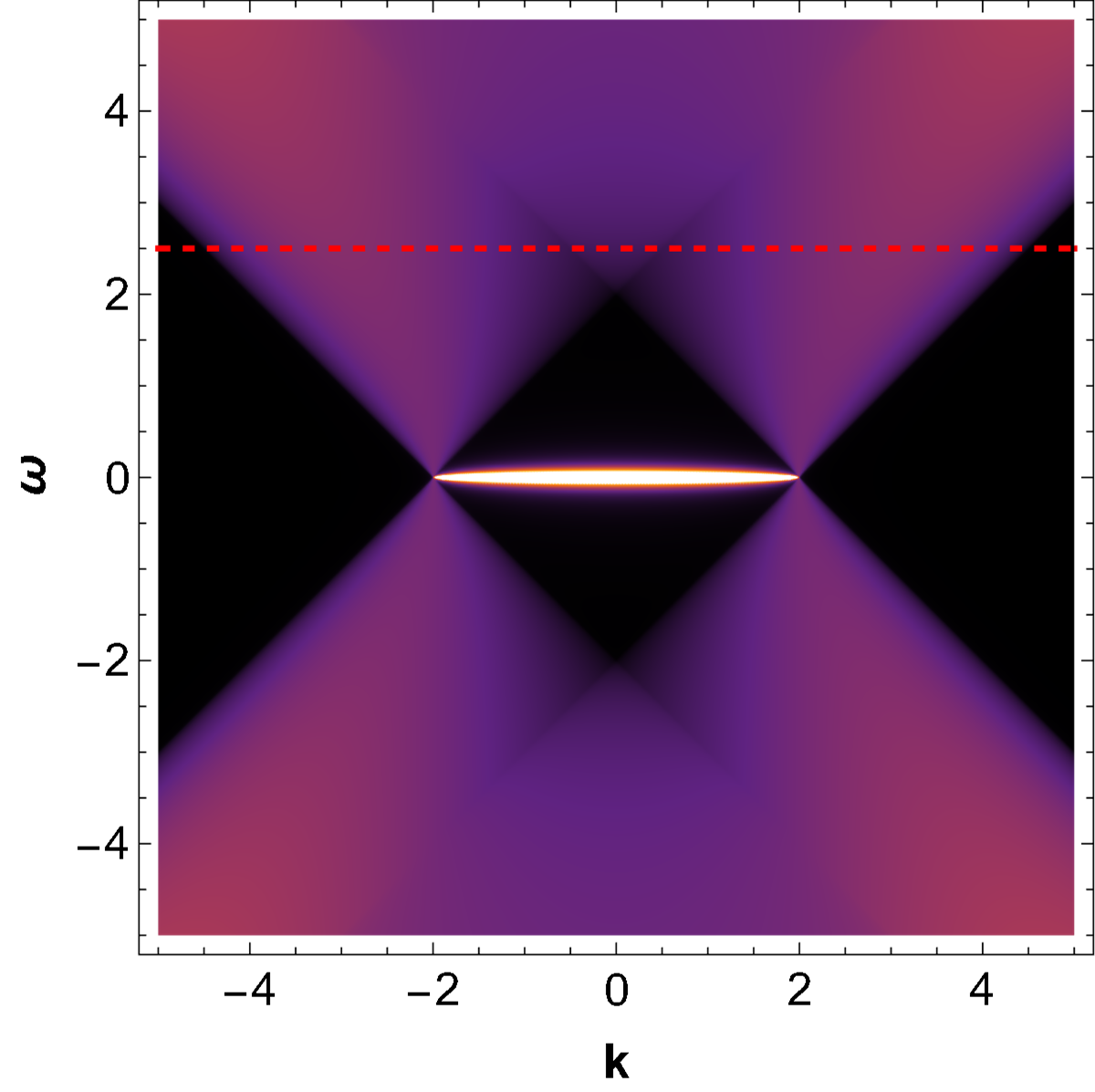}
		\caption{$B^{(0)(SA)}_{x},\omega\text{-}k_x$}
	\end{subfigure}
	\begin{subfigure}{0.22\textwidth}
	\centering
	\includegraphics[width=3.0cm]{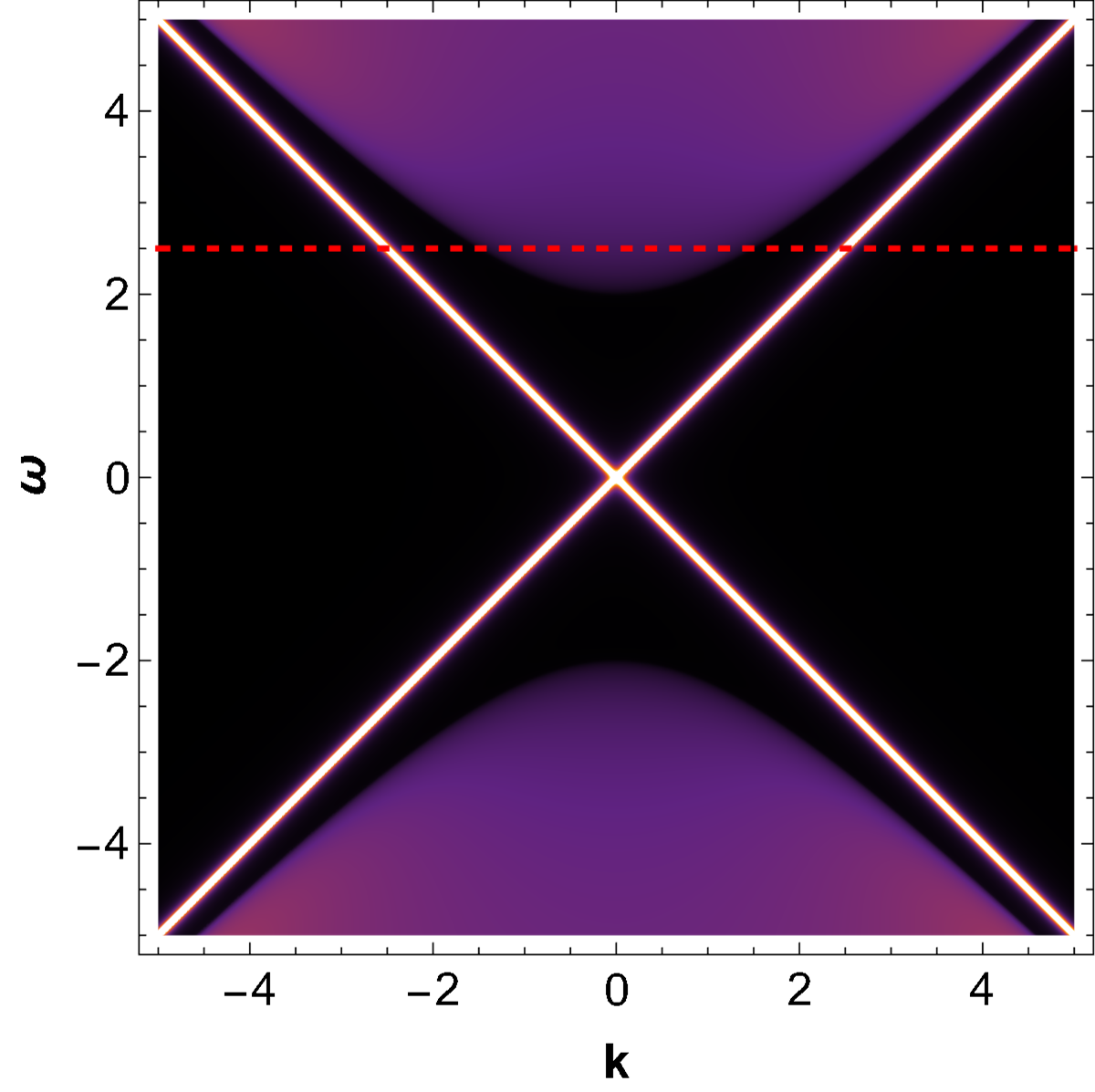}
	\caption{$B^{(0)(SA)}_{x},\omega\text{-}\kp$}
	\end{subfigure}
	\begin{subfigure}{0.22\textwidth}
		\centering
		\vspace{-0.39cm}
		\includegraphics[width=3.0cm]{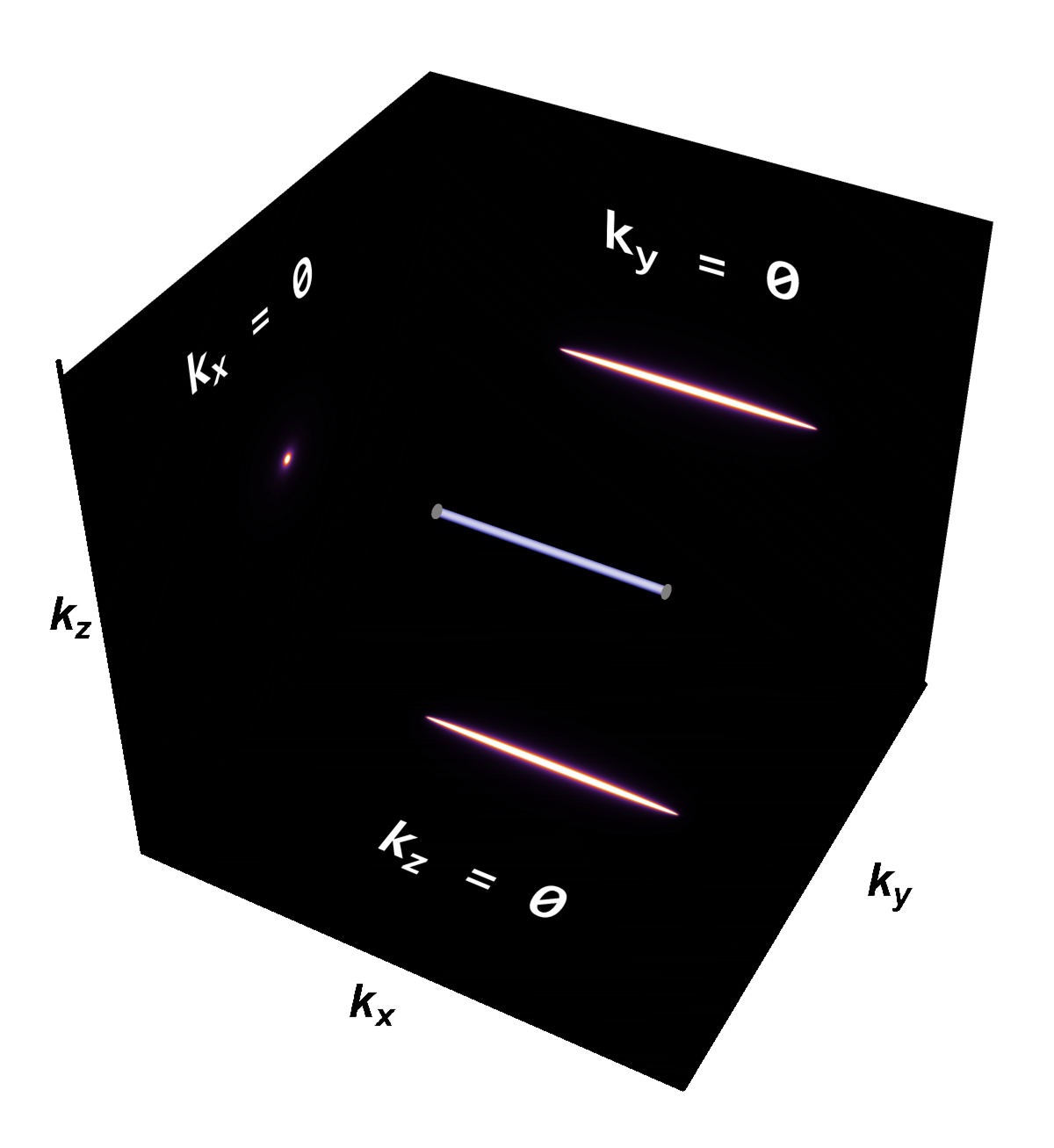}
		\caption{$\omega = 0$}
	\end{subfigure}
	\begin{subfigure}{0.22\textwidth}
		\centering
		\vspace{-0.39cm}
		\includegraphics[width=3.0cm]{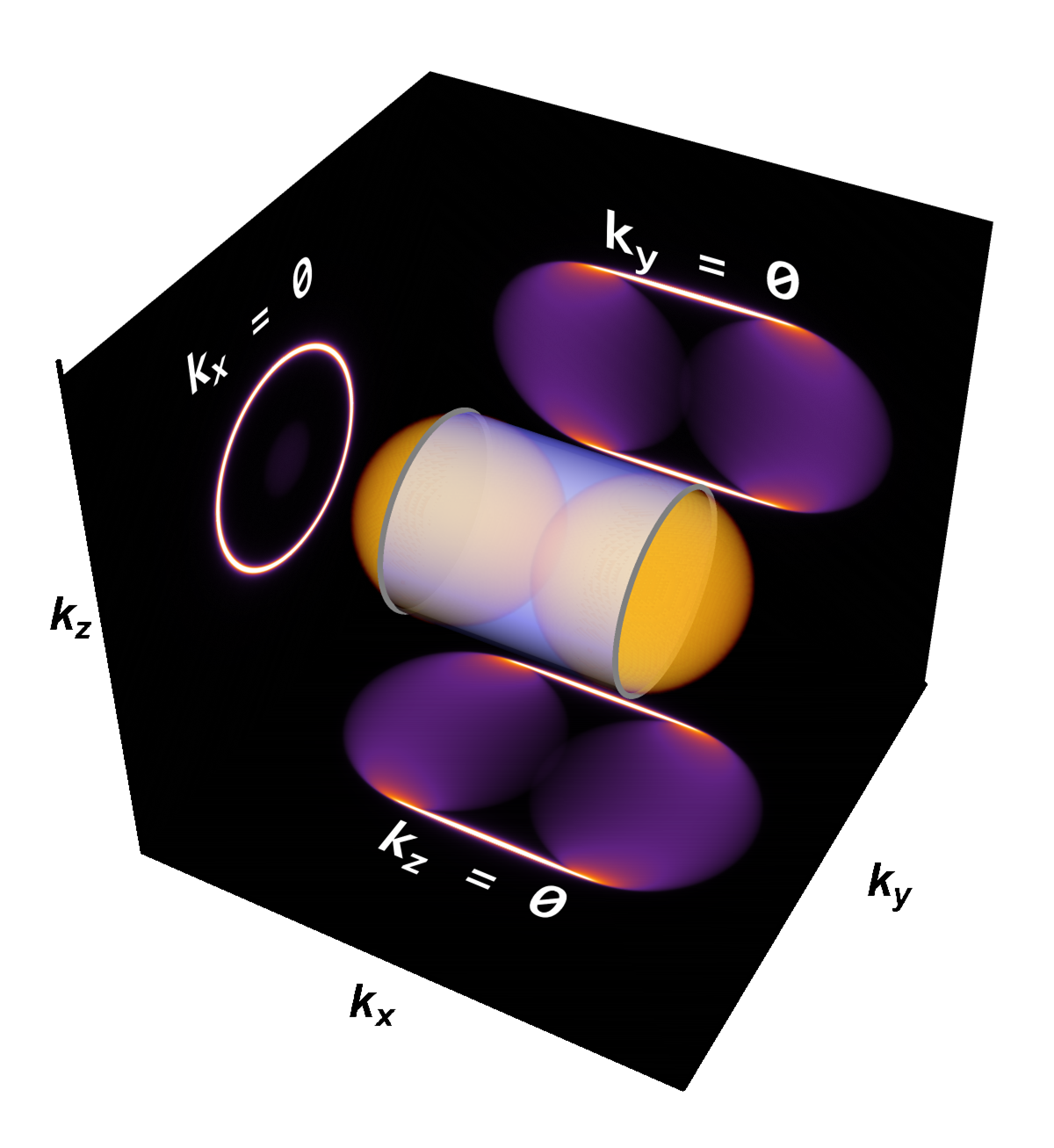}
		\caption{$\omega= 2.5$}
	\end{subfigure}
	\caption{\small Spectral functions (SFs) of $B_x$ source for both quantization choices. (a,b,e,f) SFs in $\omega\text{-}k_x,\omega\text{-}\kp$ plane. (c,g) SFs in $\omega = 0$, and (d,h) SFs for $\omega = 2.5$ correspondingly to the dashed red lines. In (g,h), if $k_x \notin [-b,b]$ the pole type singularity disappears so that only $k_x \notin [-b,b]$ which the arc lines visible. The box's background represents the certain slices at each momentum is zero. }
	\label{fig:Bx}
\end{figure}

\myparagraph{SS}
The trace of the Green's matrix (\ref{G:SSbx}), by choosing $\mu = x$
\begin{align}
\Tr \mathbb{G}_{B_{x}^{(0)}}^{(SS)} &=  \frac{2\omega}{\sqrt{(b-k_x)^2+\boldsymbol{k}_\perp^2 -\omega^2}}+\frac{2\omega}{\sqrt{(b+k_x)^2+\boldsymbol{k}_\perp^2 -\omega^2}} , \label{TrGBxSS}
\end{align}
where $\boldsymbol{k}^2_{\perp} = k_y^2+k_z^2$. The SF shows the superposition of two Dirac cones shifted  along the $k_x$ direction, which are non-interacting with each other.(\ref{TrGBxSS}). The distance  between the Dirac points is $2b$ and the surface of the cones are branch-cut type singularity. Notably, the SF in the $\omega\text{-}k_x$ plane exhibits a shifting of 2dimensional Dirac cones, see fig \ref{fig:Bx}(a).  In the section of $\omega\text{-} k_\perp $ plane; it shows a gap, see figure \ref{fig:Bx}(b).

\myparagraph{SA}
The trace of the Green's function matrix (\ref{G:SAbx})  for  $\mu = x$ is given by 
\begin{align}
\Tr\mathbb{G}_{B_{x}^{(0)}}^{(SA)}  &= \frac{2\omega}{b}\Big[\frac{ (b+k_x)\sqrt{(b - k_x)^2 +\boldsymbol{k}_\perp^2-\omega^2} +(b-k_x)\sqrt{(b + k_x)^2 +\boldsymbol{k}_\perp^2-\omega^2} }{\boldsymbol{k}_\perp^2 - \omega^2-i\epsilon}\Big]. \label{TrGBxSA}
\end{align}
The main feature of the spectrum is shifted Dirac cones in $\pm k_x$ direction: two Dirac points is connected  by flat band of  1-dimensional pole  singularity $(\omega^2-\boldsymbol{k}_\perp^2)^{-1} $ along $k_x \in [-b,b]$. See figure \ref{fig:Bx}(e,f). It is important to note that the residue is zero for $k_x \notin [-b,b]$,  so there is no singularity  outside the interval. See figure \ref{fig:Bx}(g,h).

\subsubsection{Space-like radial vector, $B_{ux}$}
\begin{figure}[t!]
	\centering
	\begin{subfigure}{0.22\textwidth}
		\centering
		\includegraphics[width=3.0cm]{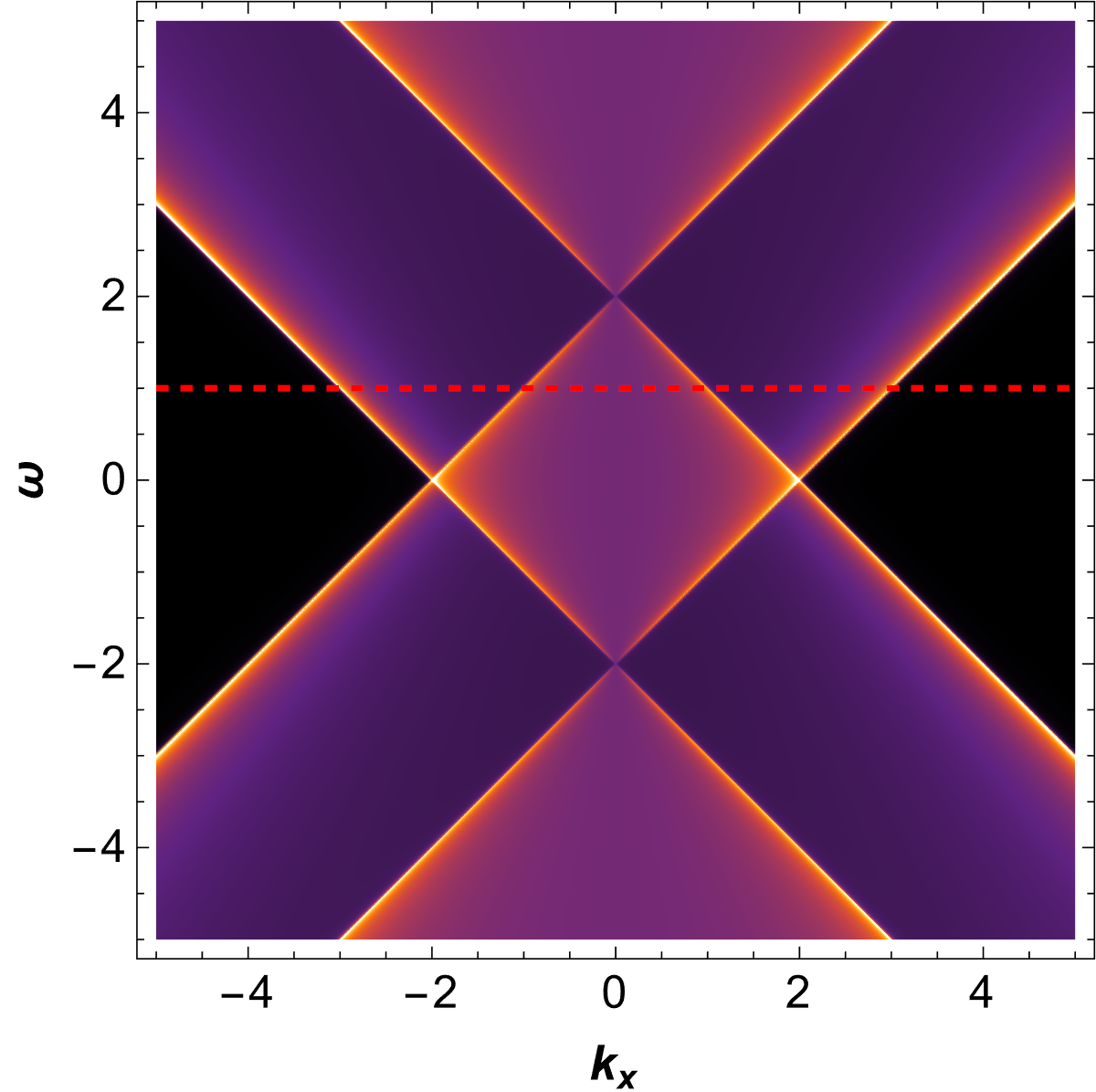}
		\caption{$B^{(-1)(SS)}_{ux},\omega\text{-}k_x$}
	\end{subfigure}
	\begin{subfigure}{0.22\textwidth}
		\centering
		\includegraphics[width=3.0cm]{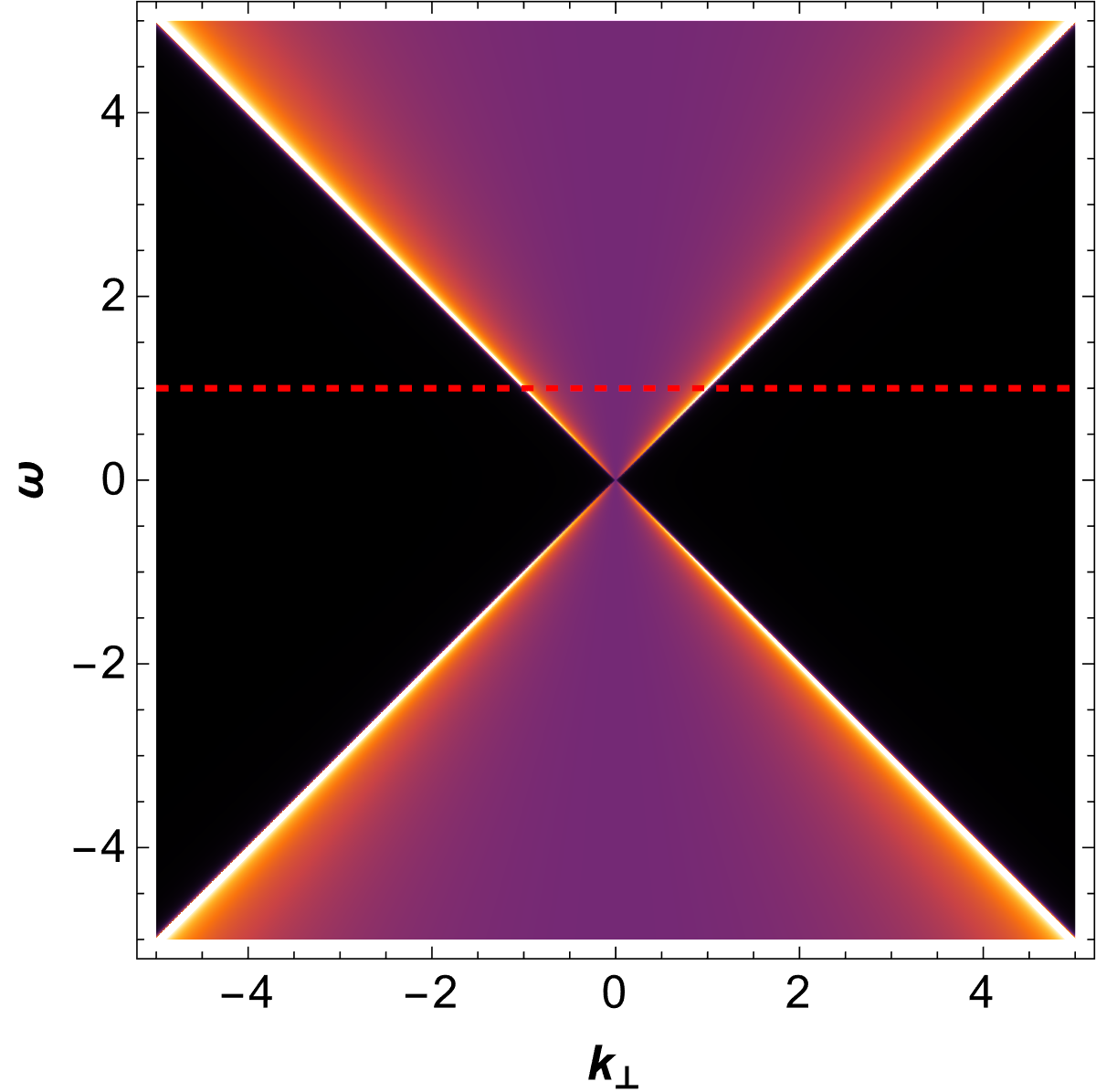}
		\caption{$B^{(-1)(SS)}_{ux},\omega\text{-}\kp$}
	\end{subfigure}
	\begin{subfigure}{0.22\textwidth}
		\centering
		\vspace{-0.39cm}
		\includegraphics[width=3.0cm]{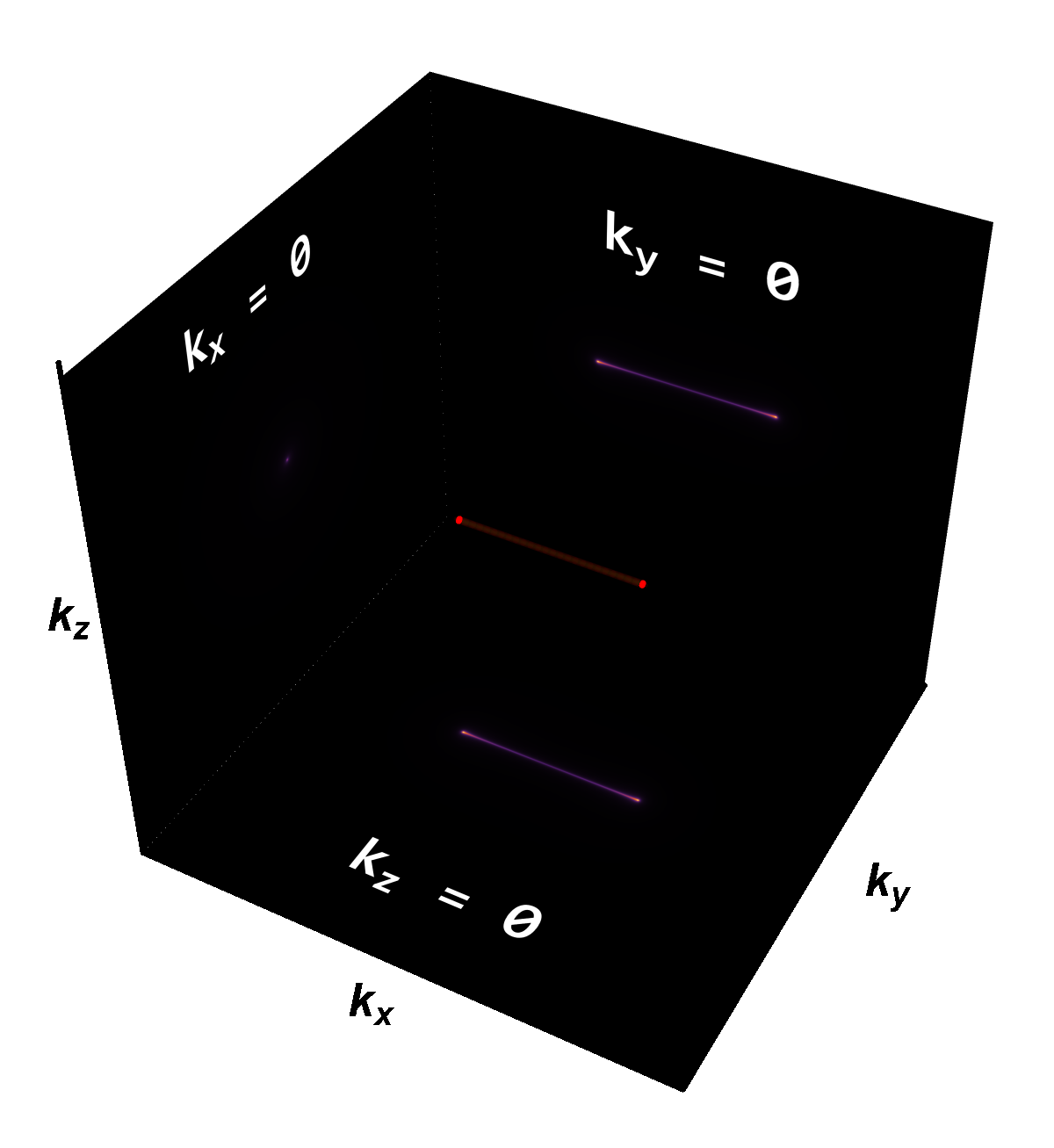}
		\caption{$\omega = 0$}
	\end{subfigure}
	\begin{subfigure}{0.22\textwidth}
		\centering
		\vspace{-0.39cm}
		\includegraphics[width=3.0cm]{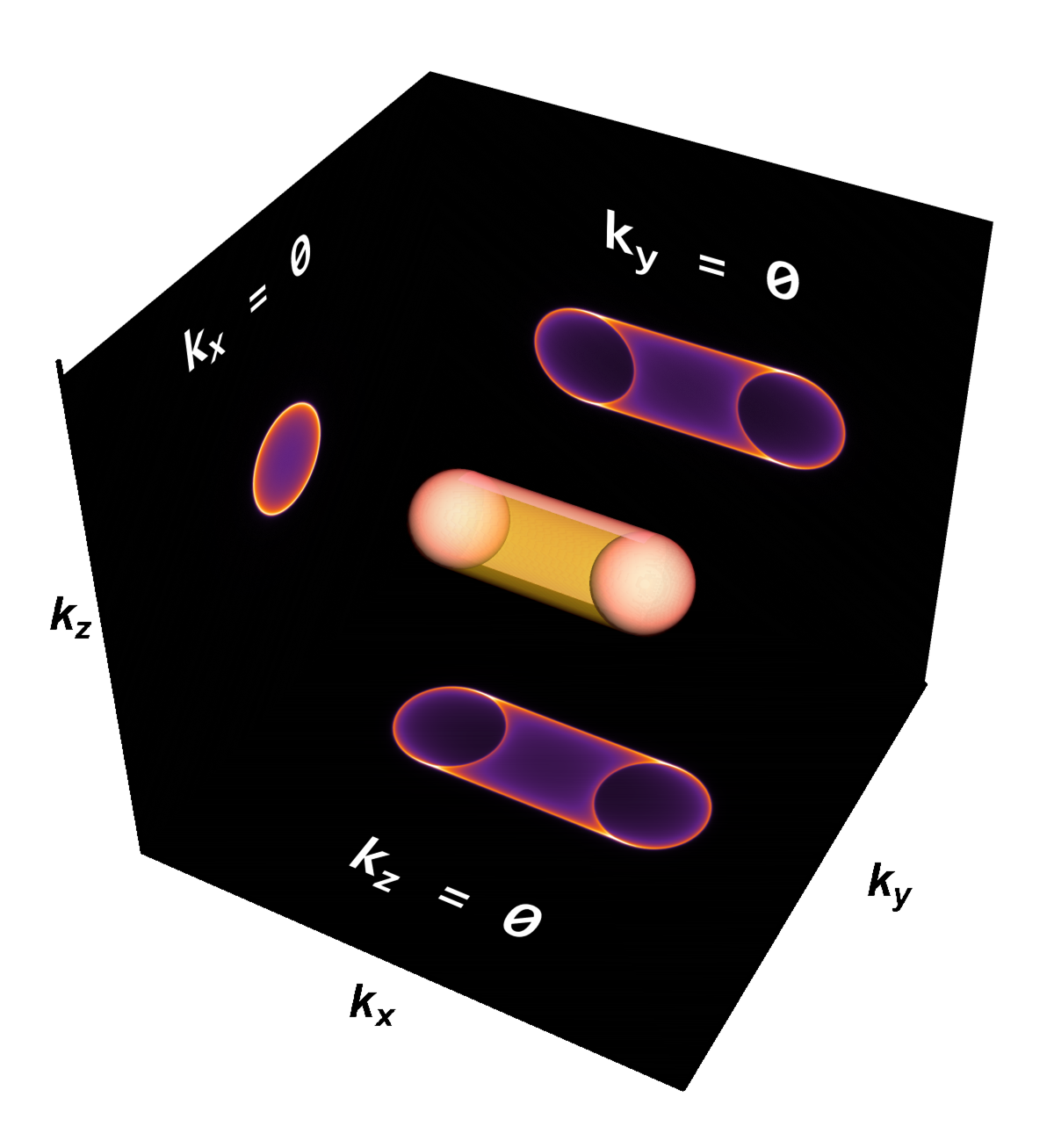}
		\caption{$\omega = 1$}
	\end{subfigure}\\\vspace{0.4cm}
	\begin{subfigure}{0.22\textwidth}
		\centering
		\includegraphics[width=3.0cm]{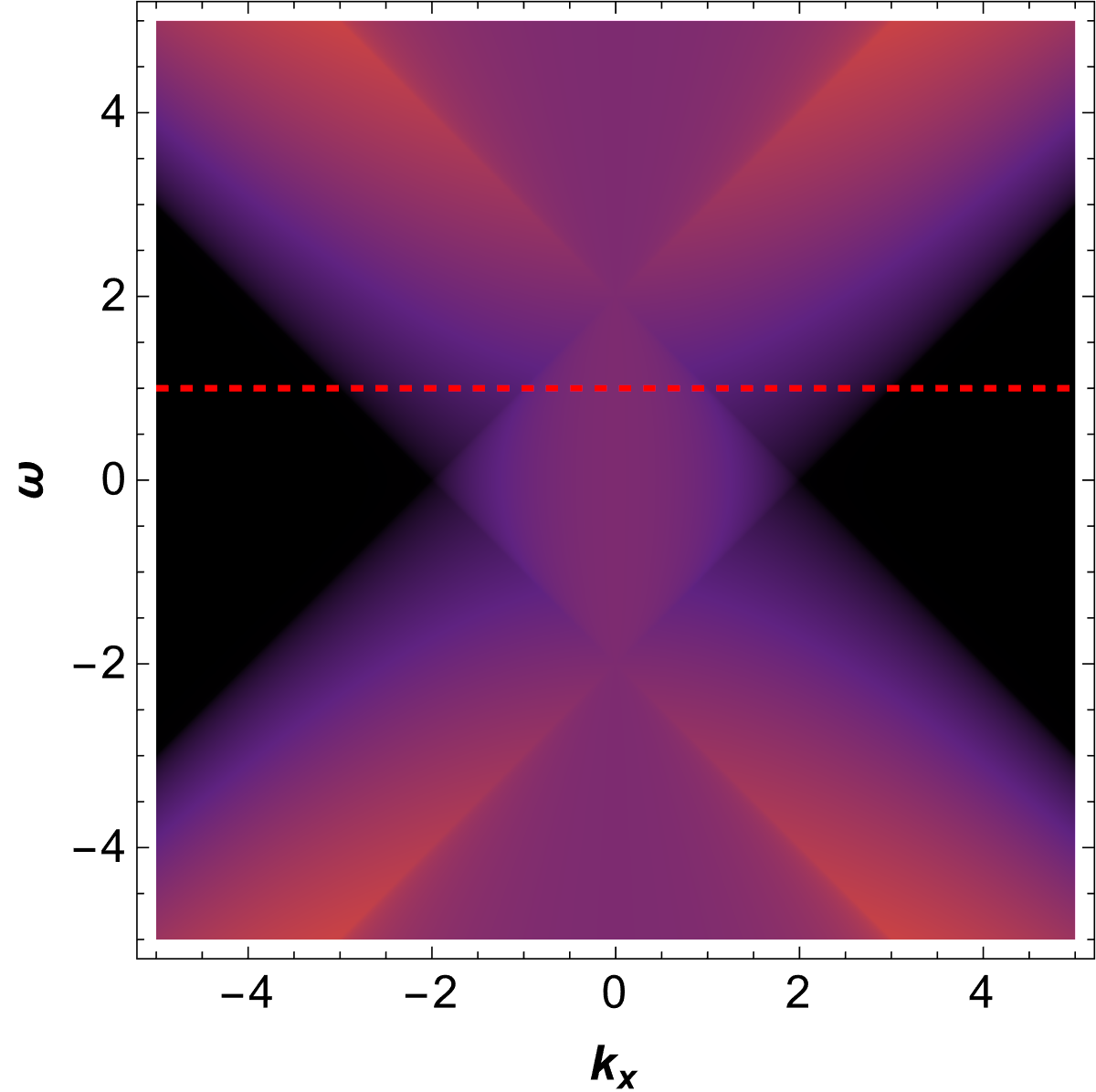}
		\caption{$B^{(-1)(SA)}_{ux},\omega\text{-}k_x$}
	\end{subfigure}
	\begin{subfigure}{0.22\textwidth}
		\centering
		\includegraphics[width=3.0cm]{./fig/ScaledBux.png}
		\caption{$B^{(-1)(SA)}_{ux},\omega\text{-}\kp$}
	\end{subfigure}
	\begin{subfigure}{0.22\textwidth}
		\centering
		\vspace{-0.39cm}
		\includegraphics[width=3.0cm]{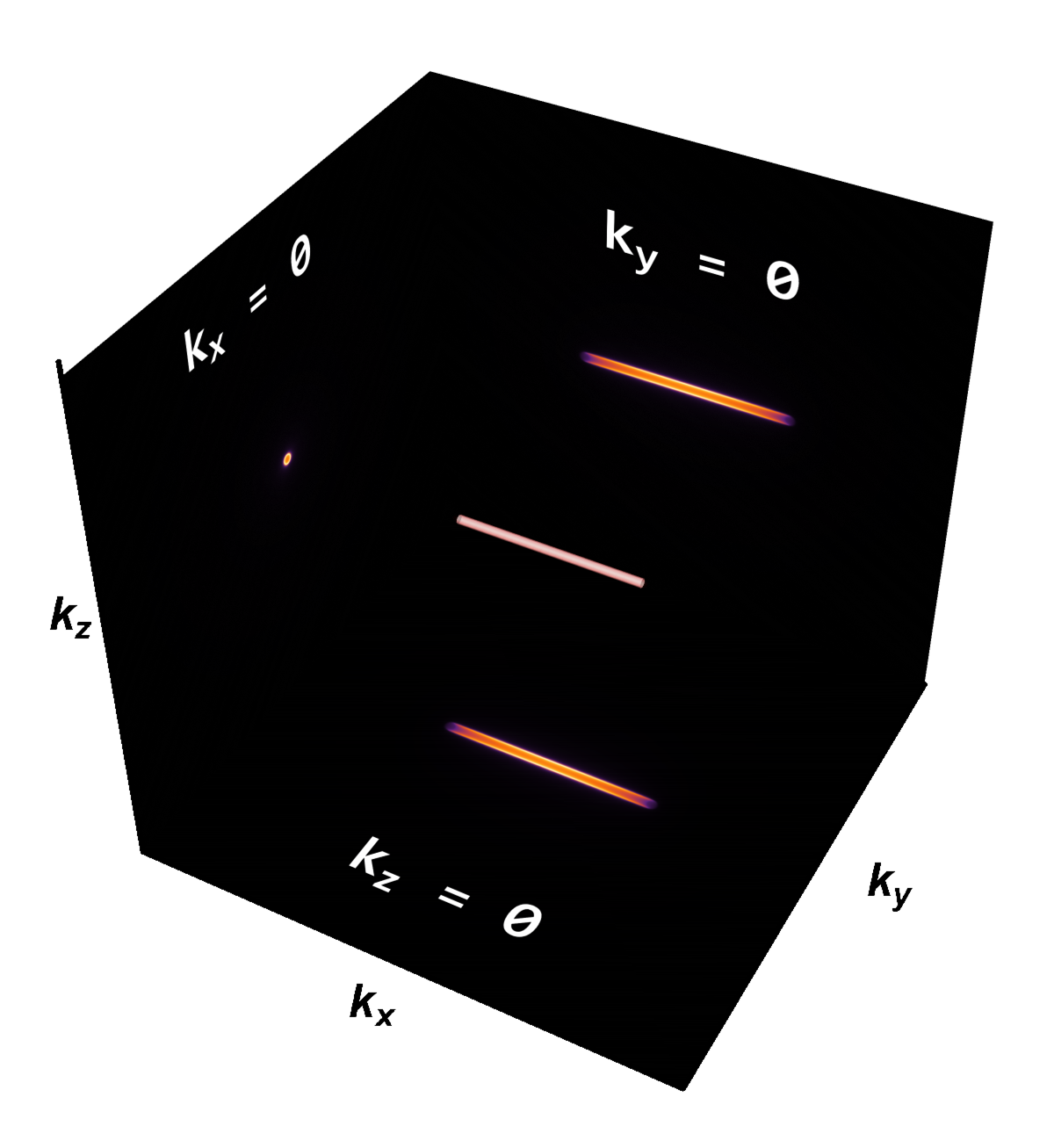}
		\caption{$\omega = 0$}
	\end{subfigure}
	\begin{subfigure}{0.22\textwidth}
		\centering
		\vspace{-0.39cm}
		\includegraphics[width=3.0cm]{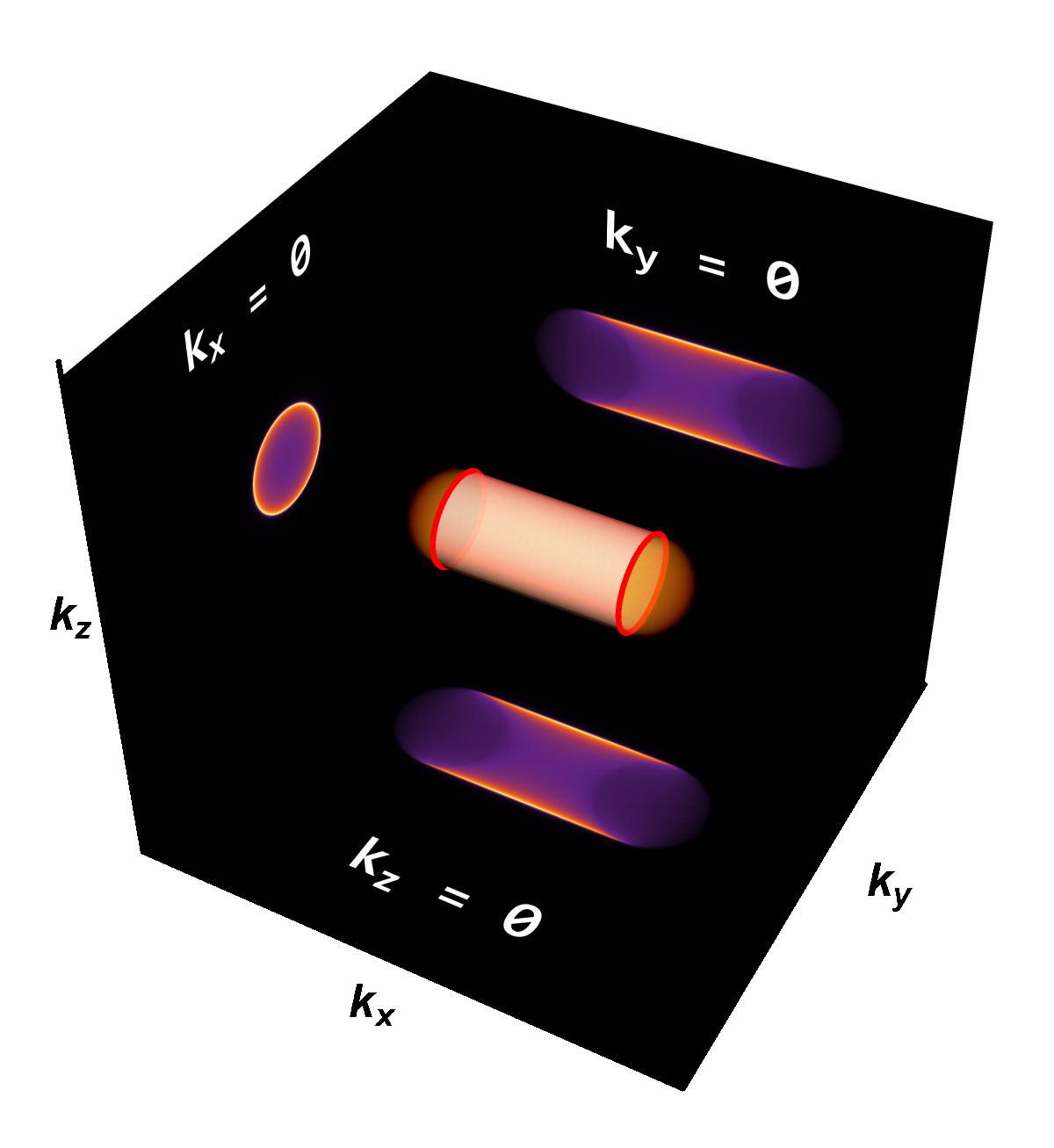}
		\caption{$\omega = 1$}
	\end{subfigure}
	
	\caption{\small Spectral functions (SFs) for $B_{ux}$ source for both quantization choices. (a,b,e,f) SFs  in $\omega\text{-}\boldsymbol{k}$ plane. (c,d,g,h) SFs in $k_x\text{-}k_y\text{-}k_z$  at $\omega = 0,1$ slices, respectively. The spectral features are analogous to $B_{x}$ case, with extra branch-cut singularity pieces. Notice that the spectrum shown in (g) is just the nonsingular branch-cut.}
	\label{fig:Bux}
\end{figure}

The analytic expression is given by
\begin{align}
\Tr \mathbb{G}_{B_{ux}^{(-1)}}^{(SS)} &=  4\omega \frac{b^2+\boldsymbol{k}^2 -\omega^2+f_+f_-}{f_+f_-(f_+ + f_-)},\\
\Tr \mathbb{G}_{B_{ux}^{(-1)}}^{(SA)} &=  4\omega \frac{(f_{+}+f_{-})\sqrt{\omega^2-\kp^2}- b(f_{+}-f_{-})}{\sqrt{\omega^2-\kp^2}(b^2+\boldsymbol{k}^2 -\omega^2+f_+f_-)}.
\end{align}
where $f_\pm = \sqrt{k^2_x - \Big(b \pm \sqrt{\omega^2-\kp^2}\Big)^2}$. The structure of the $f_\pm$ is nothing but shifting of $|\omega|$-radius semispheres in $k_x$ direction. It is useful to realize that $f_-f_+$ is shifting of two $|\omega|$-radius spheres in $k_x$ direction. See figure \ref{fig:Bux}.

\subsection{Antisymmetric 2-tensors}
\subsubsection{Space-like tensor $B_{xy}$}
\begin{figure}[t!]
	\centering
	\begin{subfigure}{0.22\textwidth}
		\centering
		\includegraphics[width=3.0cm]{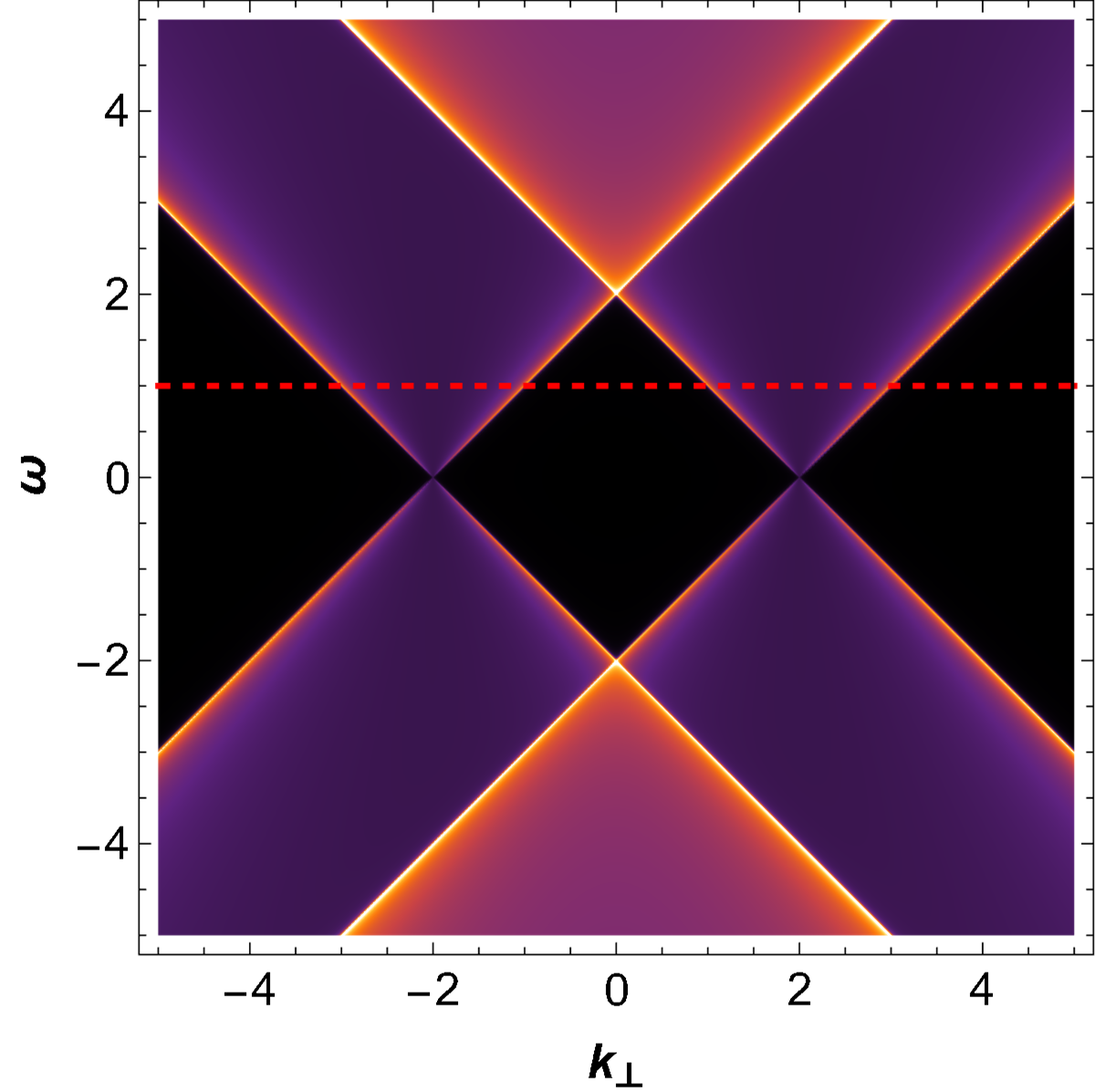}
		\caption{$B^{(-1)(SA)}_{xy},\omega\text{-}\kp$}
	\end{subfigure}
	\begin{subfigure}{0.22\textwidth}
		\centering
		\includegraphics[width=3.0cm]{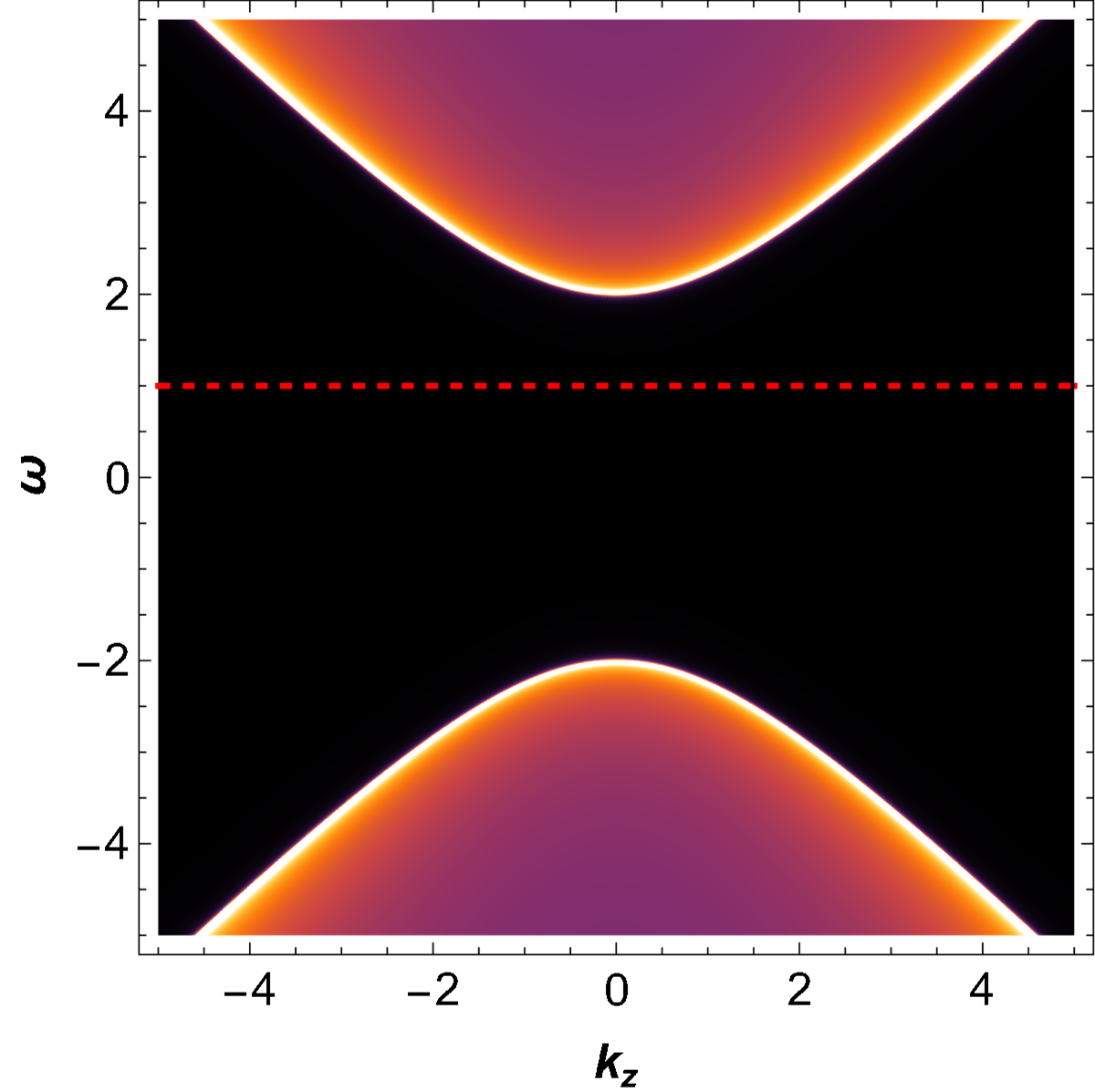}
		\caption{$B^{(-1)(SA)}_{xy},\omega\text{-}k_z$}
	\end{subfigure}
	\begin{subfigure}{0.22\textwidth}
		\centering
		\vspace{-0.39cm}
		\includegraphics[width=3.0cm]{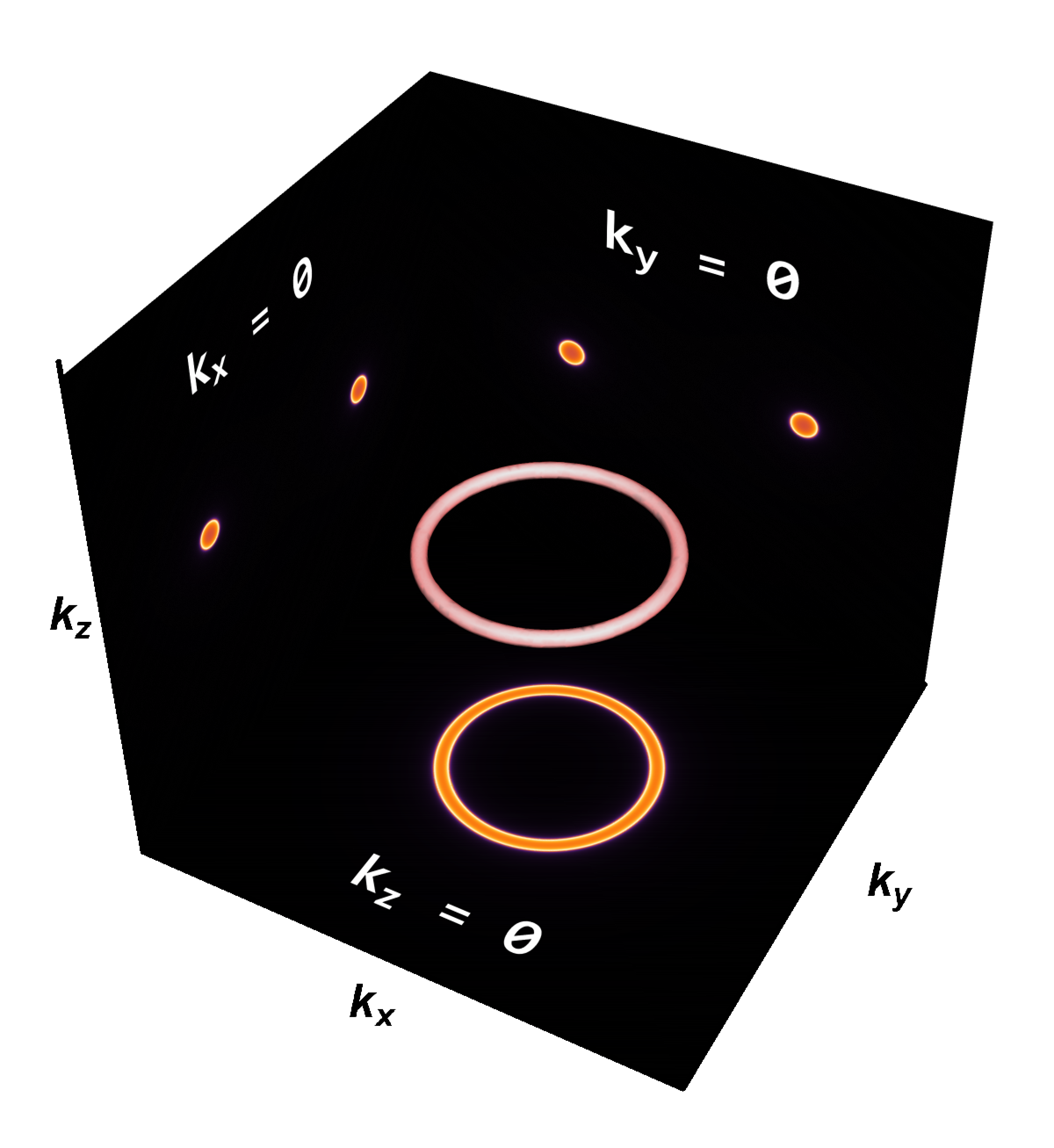}
		\caption{$\omega \simeq 0$}
	\end{subfigure}
	\begin{subfigure}{0.22\textwidth}
		\centering
		\vspace{-0.39cm}
		\includegraphics[width=3.0cm]{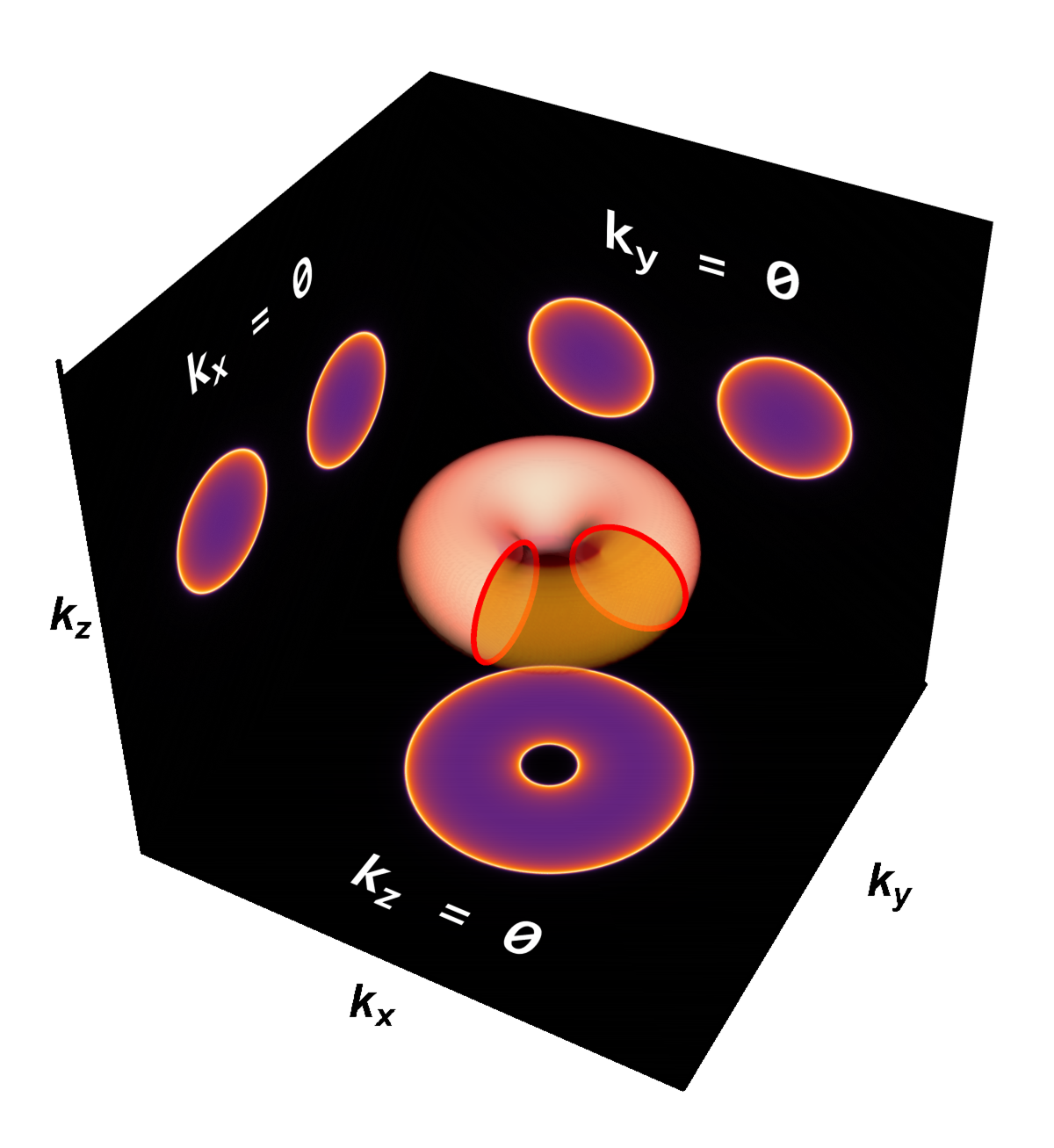}
		\caption{$\omega = 1$}
	\end{subfigure}\\\vspace{0.4cm}
	\begin{subfigure}{0.22\textwidth}
		\centering
		\includegraphics[width=3.0cm]{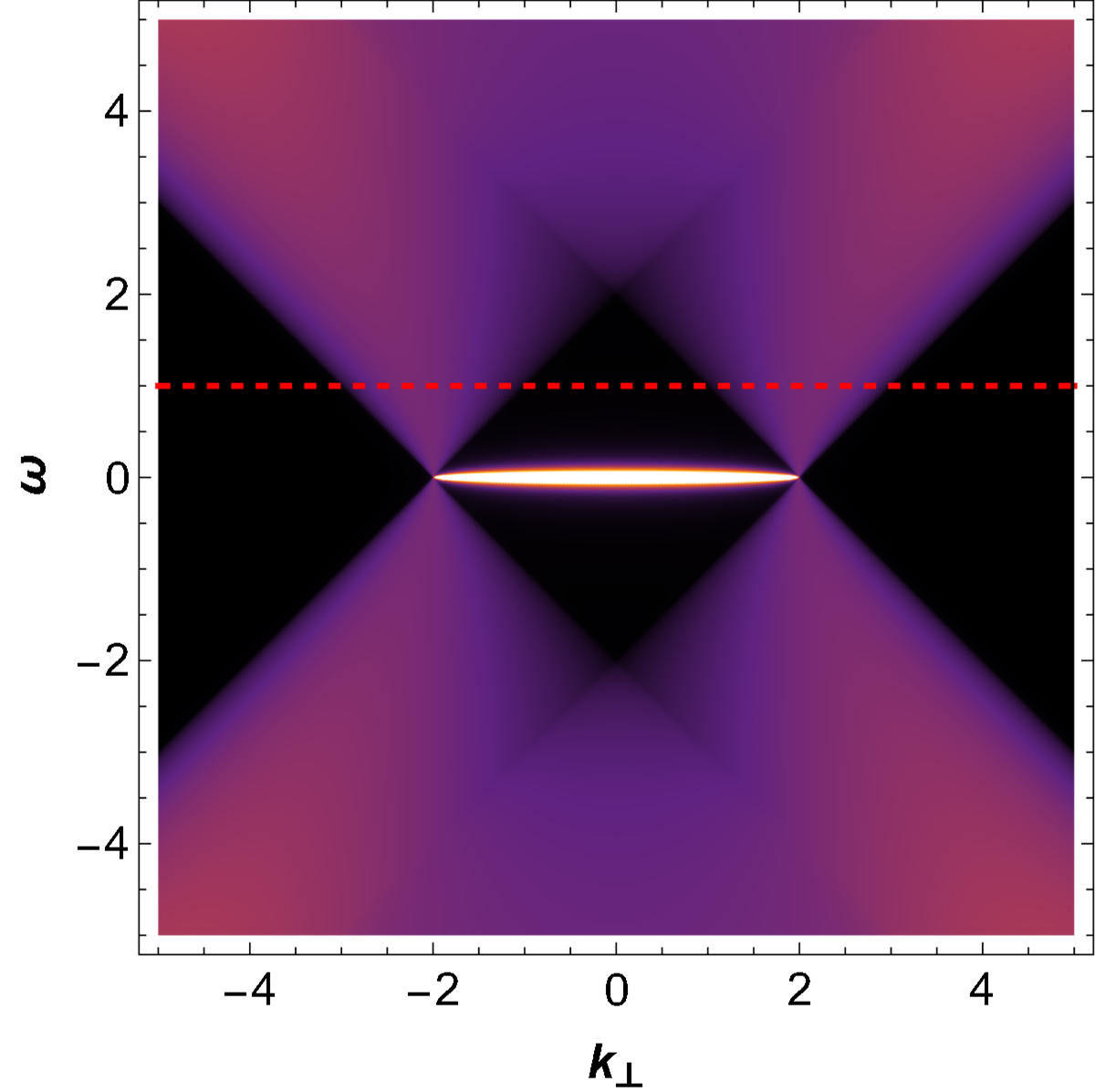}
		\caption{$B^{(-1)(SS)}_{xy},\omega\text{-}\kp$}
	\end{subfigure}
	\begin{subfigure}{0.22\textwidth}
		\centering
		\includegraphics[width=3.0cm]{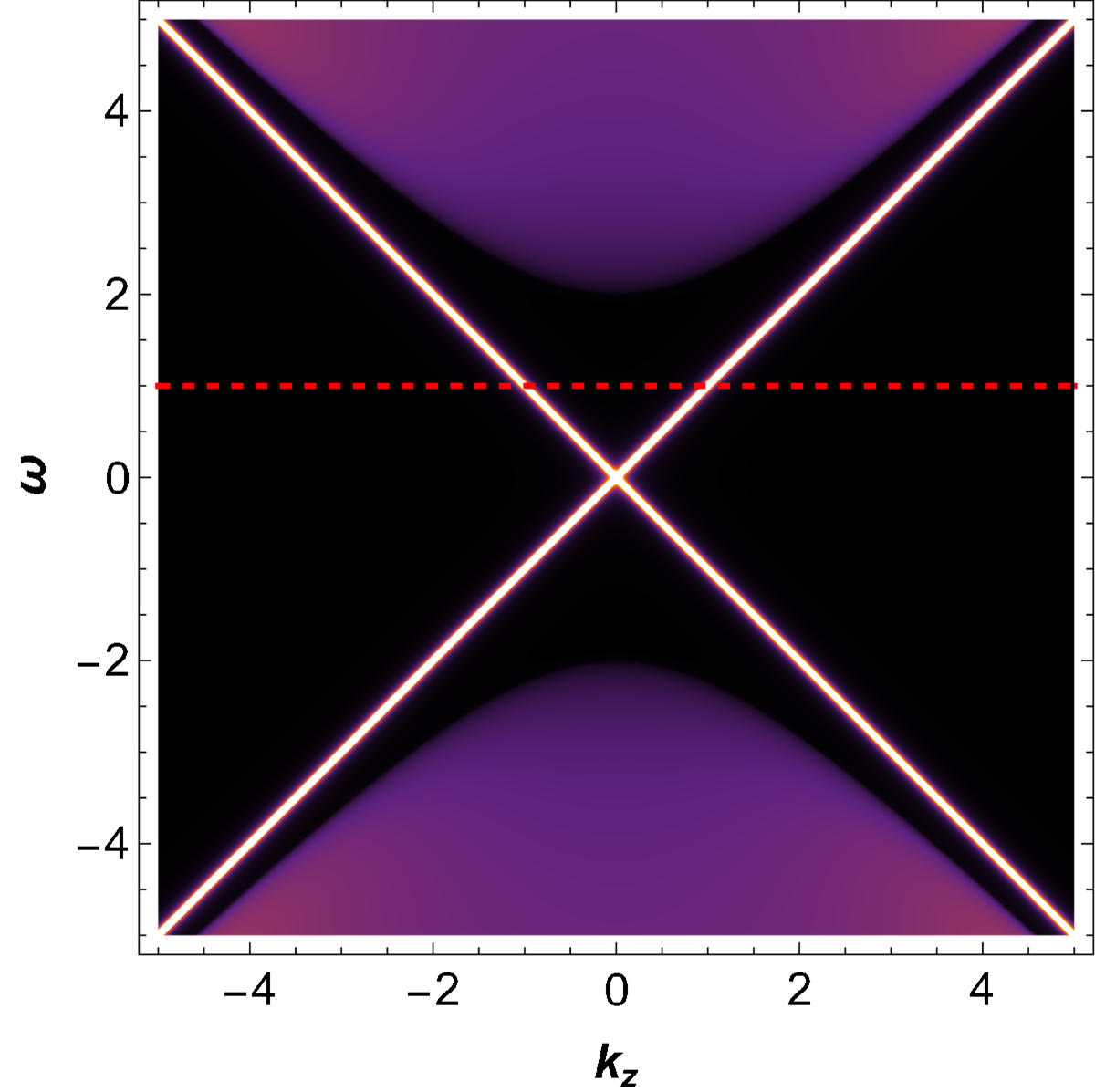}
		\caption{$B^{(-1)(SS)}_{xy},\omega\text{-}k_z$}
	\end{subfigure}
	\begin{subfigure}{0.22\textwidth}
		\centering
		\vspace{-0.39cm}
		\includegraphics[width=3.0cm]{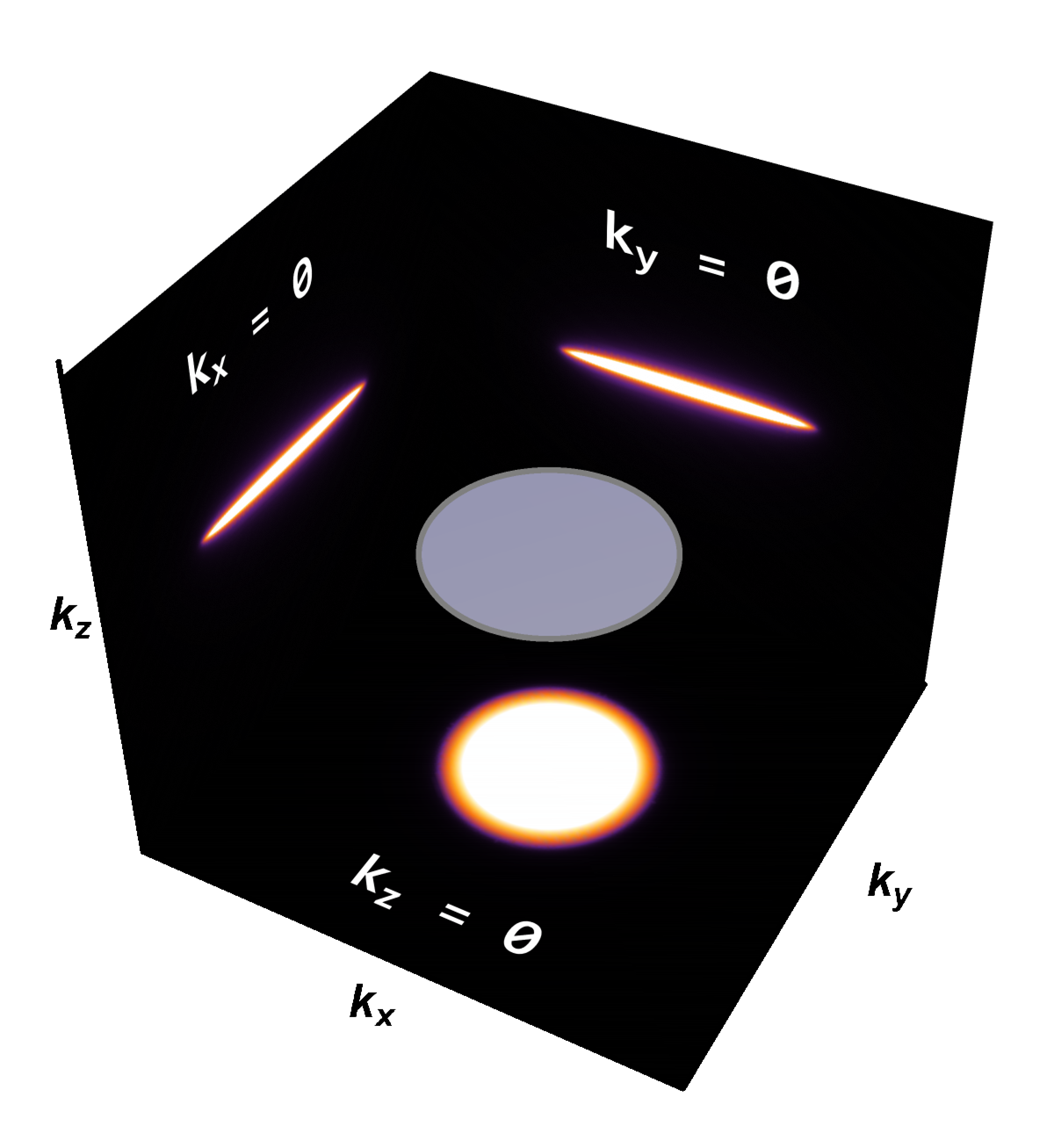}
		\caption{$\omega = 0$}
	\end{subfigure}
	\begin{subfigure}{0.22\textwidth}
		\centering
		\vspace{-0.39cm}
		\includegraphics[width=3.0cm]{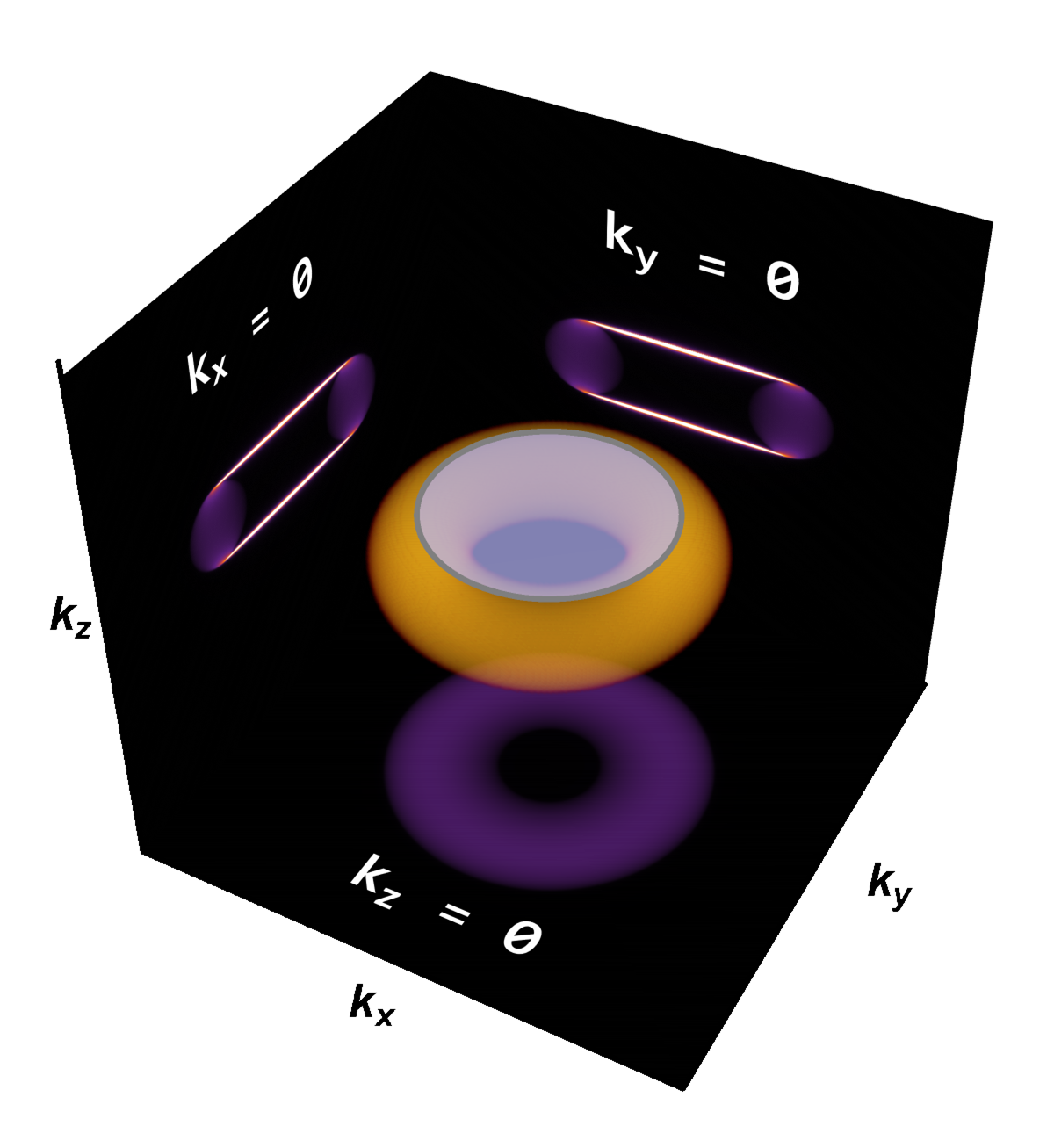}
		\caption{$\omega = 1$}
	\end{subfigure}
	\caption{\small Spectral function (SFs) of $B_{xy}$ source for both quantization choices. (a,b,e,f) SFs in $\omega\text{-}k_x,\omega\text{-}\kp$ planes. (c,g,) SFs in $\omega = 0$, and (d,h) $\omega = 1$ correspondingly to the dashed red lines. The spectral functions have rotational symmetry for each fixing $k_z$. The background of the box represents the certain slices at each momentum is zero.}
	\label{fig:Bxy}
\end{figure}

\myparagraph{SS}
The polar spatial tensor source of SS-quantization yields Green's functions with the rotational symmetry in $k_x\text{-}k_y$ plane. The trace of the Green's function matrix 
(\ref{G:SSbxy}) yields
\begin{align}
\Tr\mathbb{G}_{B_{xy}^{(-1)}}^{(SS)}  &= \frac{2\omega}{b}\Big[\frac{ (b+|\kp|)\sqrt{(b - |\kp|)^2 +k_z^2-\omega^2} +(b-|\kp|)\sqrt{(b + |\kp|)^2 +k_z-\omega^2} }{k_z^2 - \omega^2-i\epsilon}\Big]. \label{TrGBxySS}
\end{align}
Where $\kp^2 = k_x^2+k_y^2$, which is perpendicular to $k_z$. The structure of SF is different to $B_x$ case due to rotational symmetry in k-space. In this case, the cone shifts along $\kp$ directions, which makes the nodal line instead of separated two-Dirac points. Meanwhile, an infinite 1-dimensional pole-type singularity exists on a disk $\kp \in [-b,b]$. See figure \ref{fig:Bxy}(e,f)  In $k_x\text{-}k_y\text{-}k_z$ space, if $\omega$ slightly increases from 0, the singularity splits in $k_z$ direction and connects the torus's center; see figure \ref{fig:Bxy}(g,h). For AdS4, we lost the third momentum, so that no cone appears and    flat band remains only  in $k_x\text{-}k_y$ plane \cite{ABC}.

\myparagraph{SA}
The spectrum exhibits a notable characteristic of rotational symmetry in the $k_{x}\text{-}k_{y}$ plane (\ref{TrGBxySA}), so the nodal line is this case's main feature. The radius of the nodal line is 2b, and the surface of the SF appears as the branch-cut type singularity. See figure \ref{fig:Bxy}(a,b,c,d)
\begin{align}
\Tr G_{B_{xy}^{(-1)}}^{(SA)} &= \frac{2\omega}{\sqrt{(b-|\kp|)^2+k_z^2 -\omega^2}}+\frac{2\omega}{\sqrt{(b+|\kp|)^2+k_z^2 -\omega^2}}. \label{TrGBxySA}
\end{align}

\subsubsection{Time-space-like tensor $B_{tz}$}
\begin{figure}[t!]
	\centering
	\begin{subfigure}{0.22\textwidth}
		\centering
		\includegraphics[width=3.0cm]{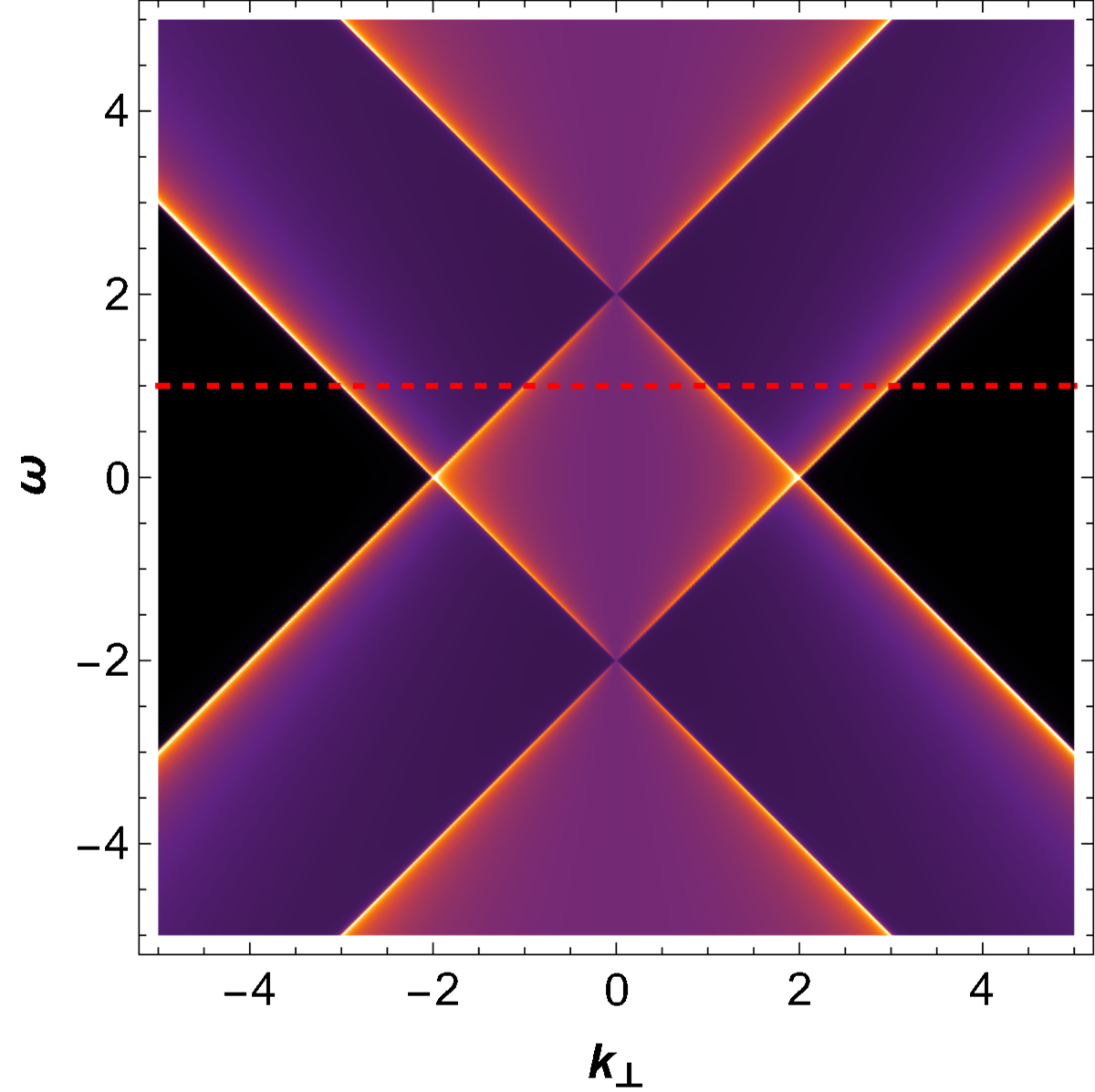}
		\caption{$B^{(-1)(SA)}_{tz},\omega\text{-}\kp$}
	\end{subfigure}
	\begin{subfigure}{0.22\textwidth}
		\centering
		\includegraphics[width=3.0cm]{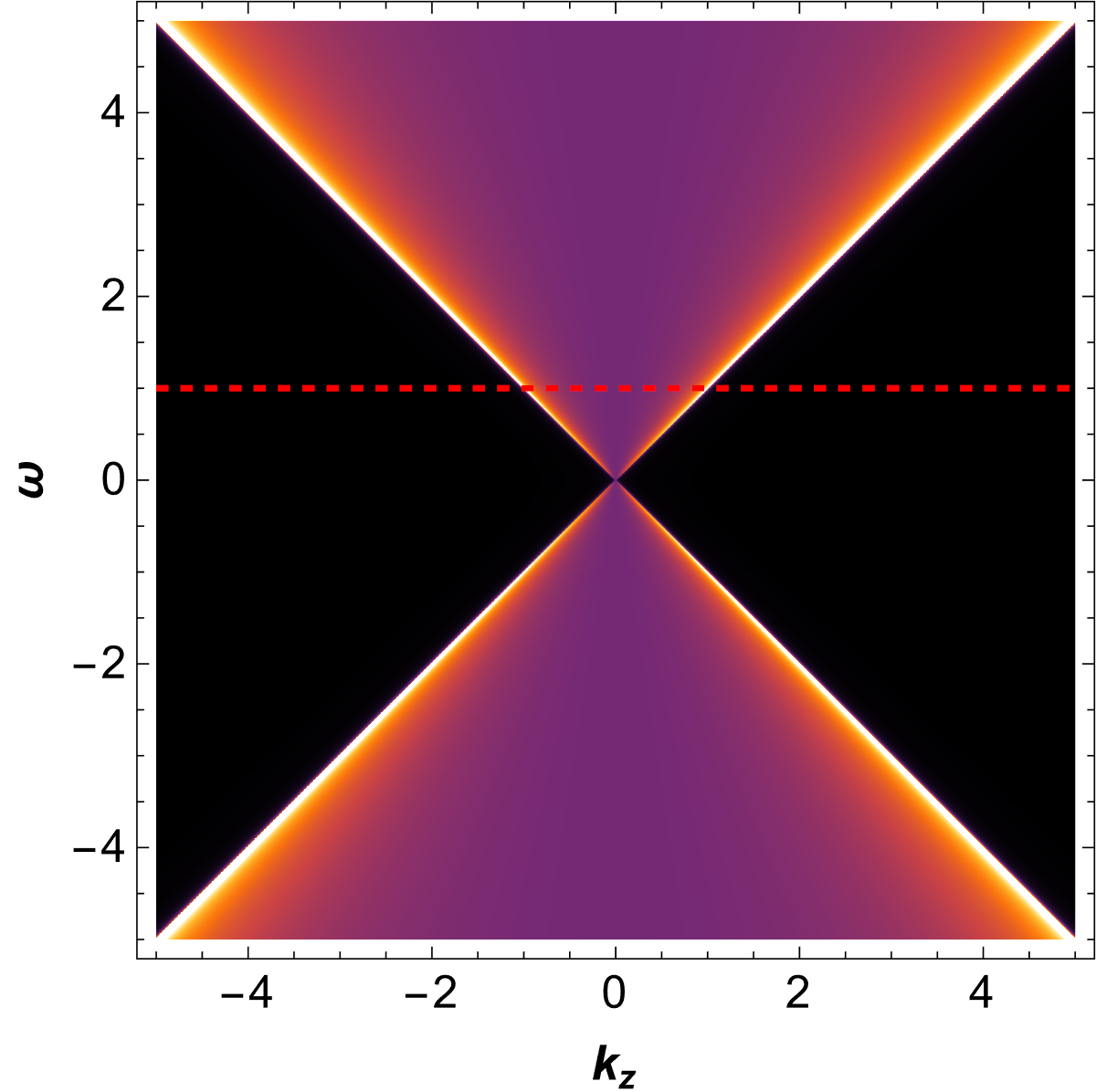}
		\caption{$B^{(-1)(SA)}_{tz},\omega\text{-}k_z$}
	\end{subfigure}
	\begin{subfigure}{0.22\textwidth}
		\centering
		\vspace{-0.39cm}
		\includegraphics[width=3.0cm]{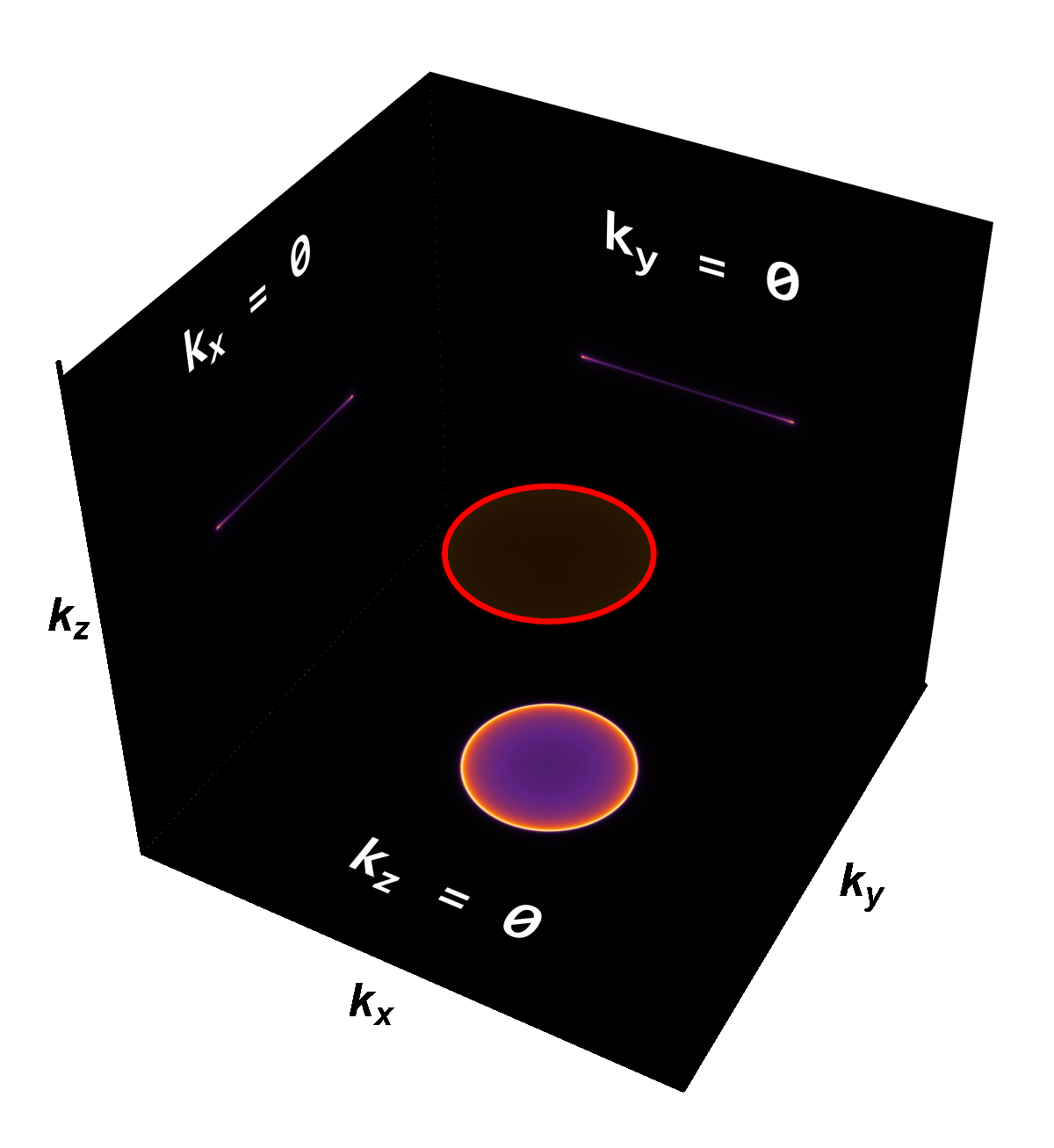}
		\caption{$\omega \simeq 0$}
	\end{subfigure}
	\begin{subfigure}{0.22\textwidth}
		\centering
		\vspace{-0.39cm}
		\includegraphics[width=3.0cm]{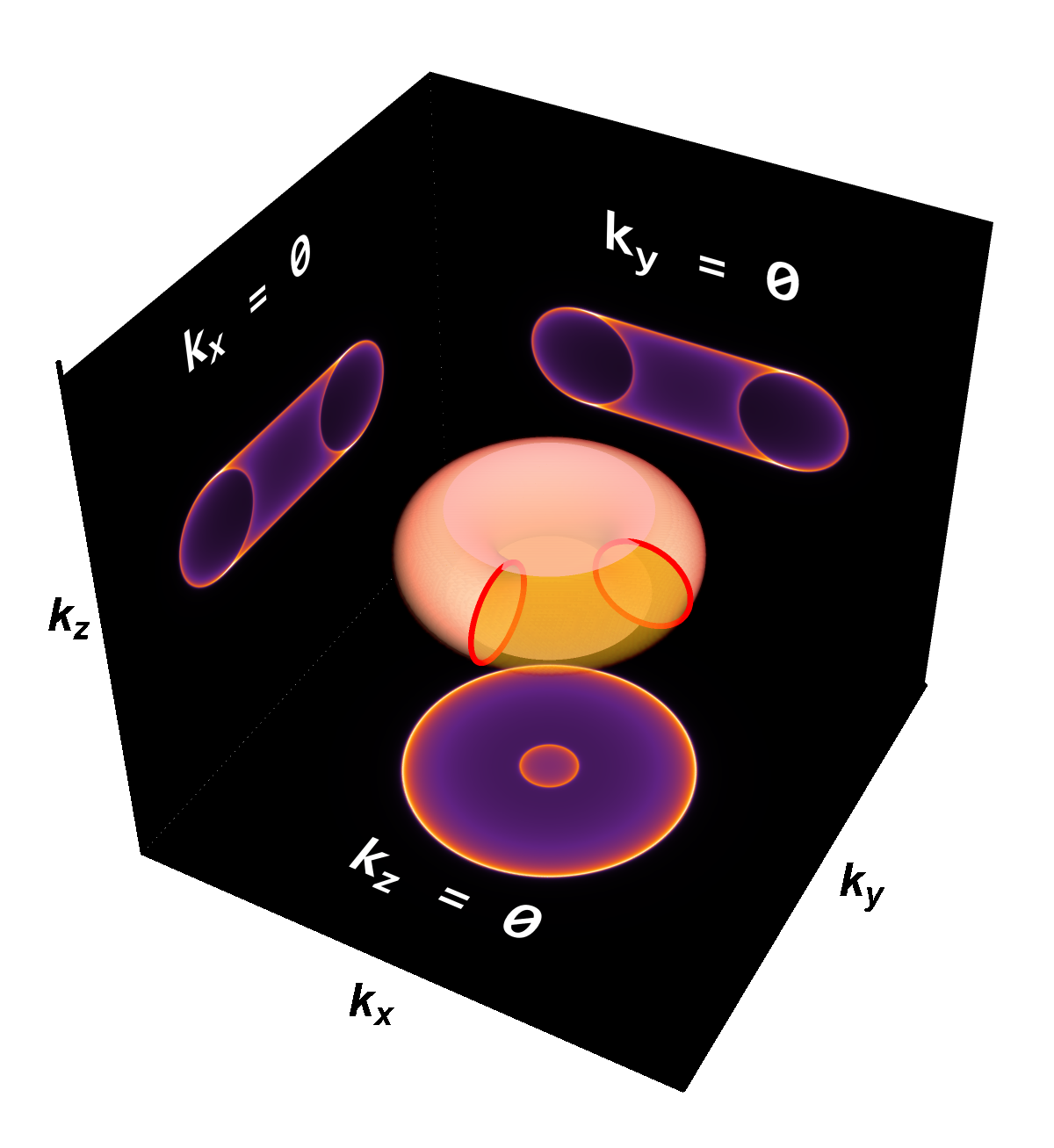}
		\caption{$\omega = 1$}
	\end{subfigure}\\\vspace{0.4cm}
	\begin{subfigure}{0.22\textwidth}
		\centering
		\includegraphics[width=3.0cm]{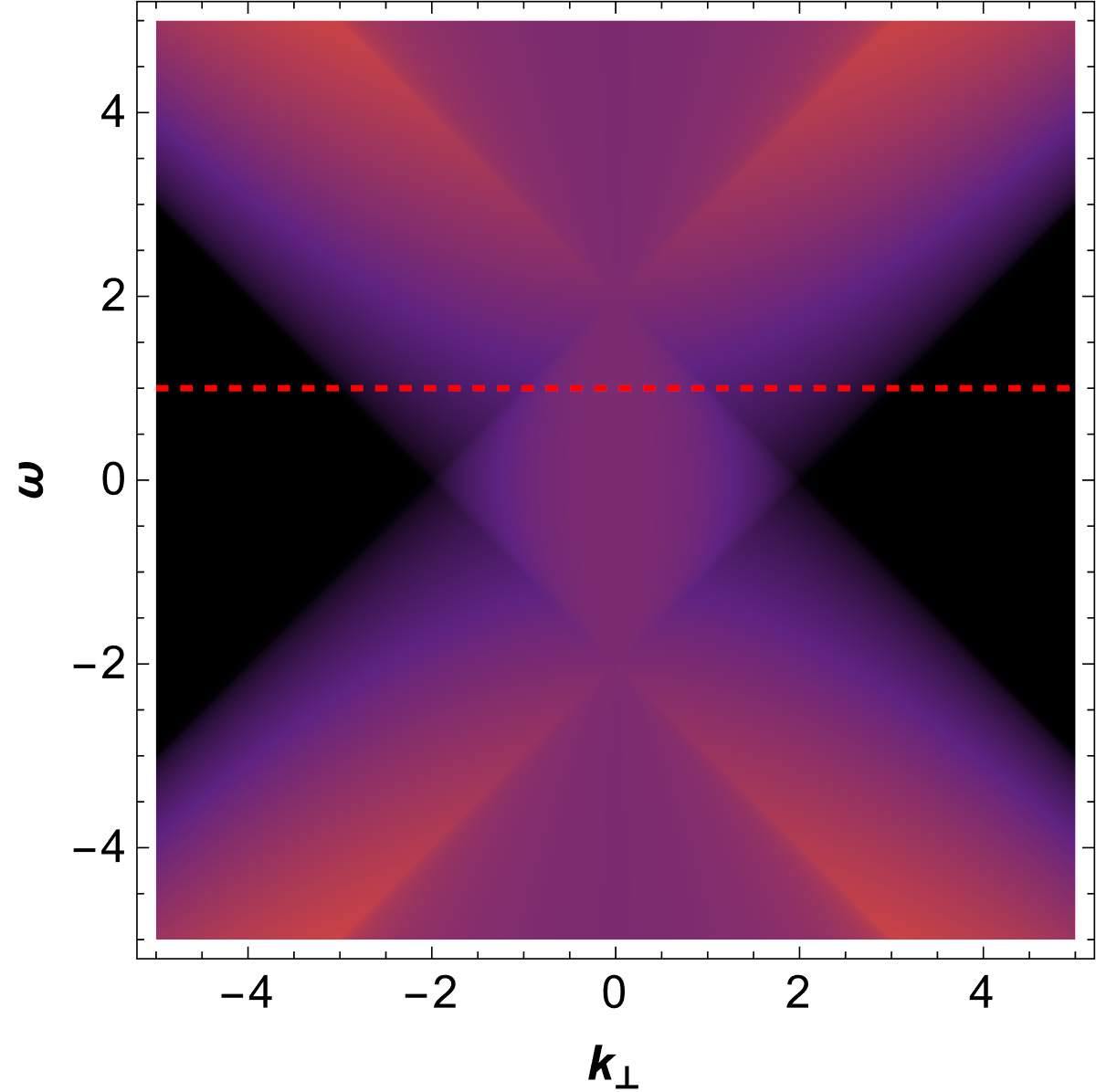}
		\caption{$B^{(-1)(SS)}_{tz},\omega\text{-}\kp$}
	\end{subfigure}
	\begin{subfigure}{0.22\textwidth}
		\centering
		\includegraphics[width=3.0cm]{./fig/ScaledBtz.png}
		\caption{$B^{(-1)(SS)}_{tz},\omega\text{-}k_z$}
	\end{subfigure}
	\begin{subfigure}{0.22\textwidth}
		\centering
		\vspace{-0.39cm}
		\includegraphics[width=3.0cm]{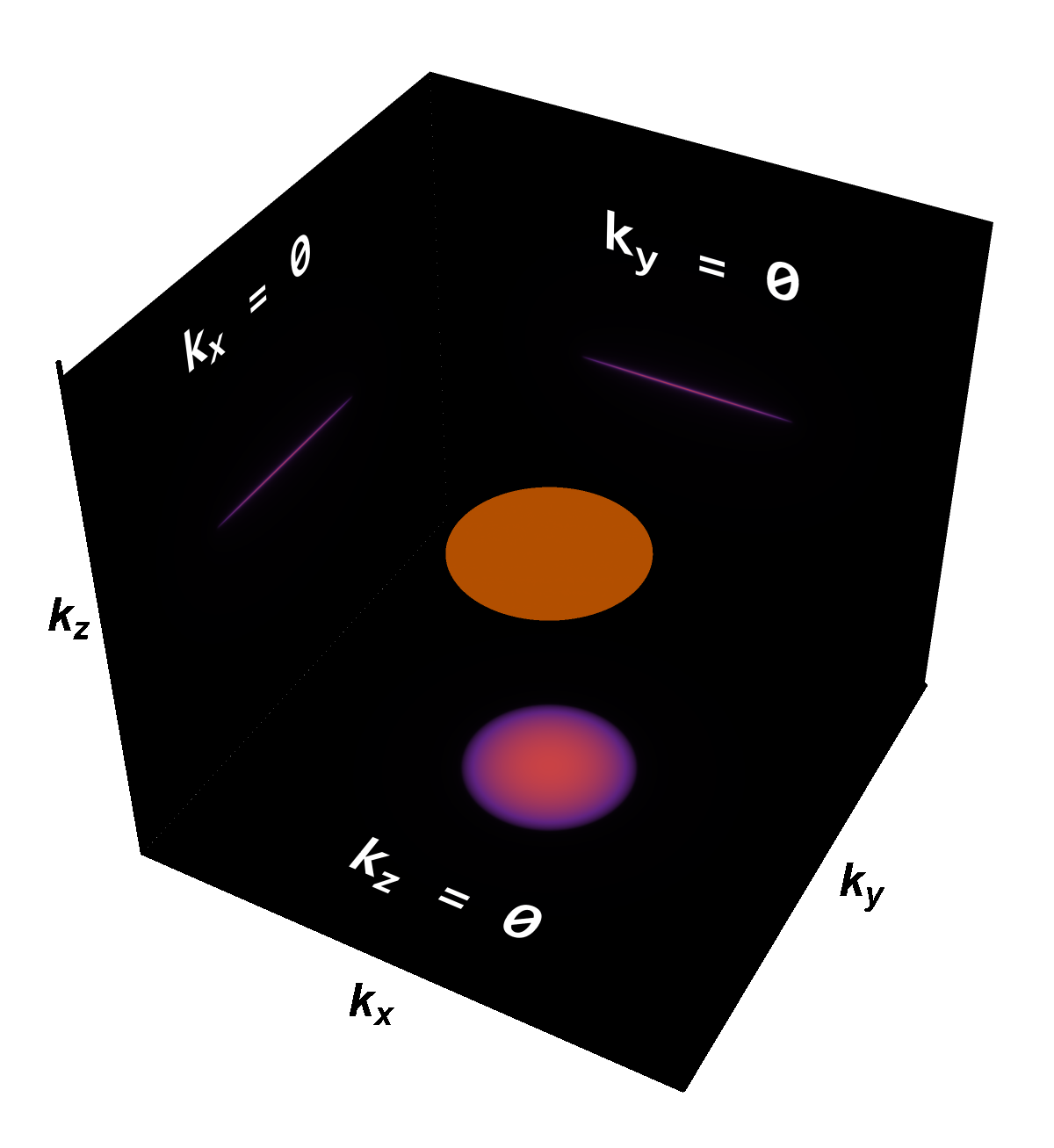}
		\caption{$\omega = 0$}
	\end{subfigure}
	\begin{subfigure}{0.22\textwidth}
		\centering
		\vspace{-0.39cm}
		\includegraphics[width=3.0cm]{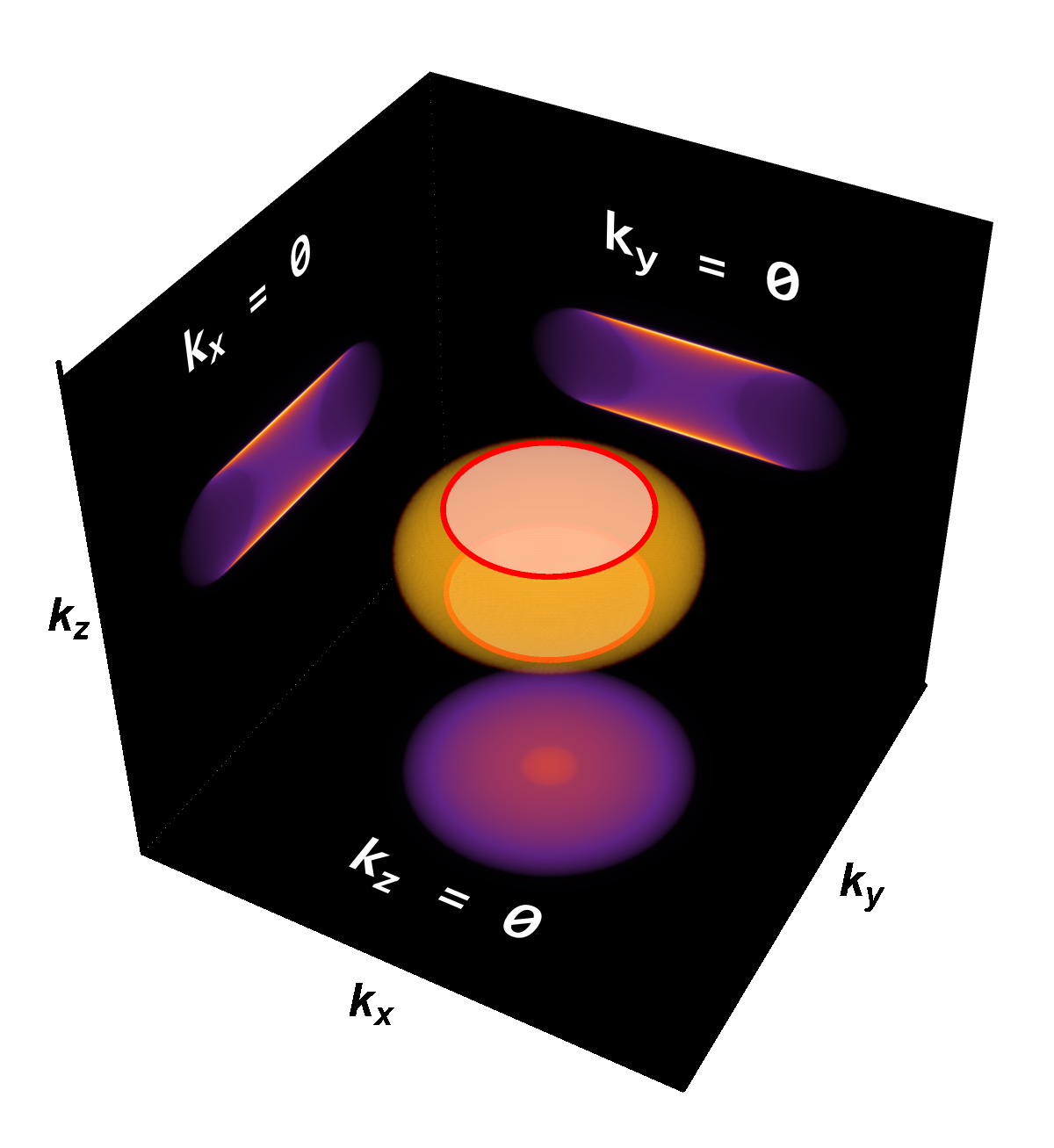}
		\caption{$\omega = 1$}
	\end{subfigure}
	
	\caption{\small Spectral functions (SFs) for $B_{tx}$ source for both quantization choices. (a,b,e,f) SFs  in $\omega\text{-}\boldsymbol{k}$ plane. (c,d,g,h) SFs in $k_x\text{-}k_y\text{-}k_z$  at $\omega = 0,1$ slices, respectively. The spectral features are analogous to $B_{xy}$. Notice that the disk appearing in (g) is just the nonsingular branch-cut.}
	\label{fig:Btz}
\end{figure}

The trace of the Green's function is given by
\begin{align}
 \Tr\mathbb{G}_{B_{tz}^{(-1)}}^{(SA)} &= 4\omega \frac{b^2+\boldsymbol{k}^2 -\omega^2+h_+h_-}{h_+h_-(h_+ + h_-)},\\
 \Tr\mathbb{G}_{B_{tz}^{(-1)}}^{(SS)} &= 4\omega \frac{(h_{+}+h_{-})\sqrt{\omega^2-\kp^2}- b(h_{+}-h_{-})}{\sqrt{\omega^2-\kp^2}(b^2+\boldsymbol{k}^2 -\omega^2+h_+h_-)}.
\end{align}
where $h_\pm = \sqrt{\kp^2 - \Big(b \pm \sqrt{\omega^2-k_z^2}\Big)^2}$,  $h_{\pm} $ has a semi-torus structure. Here, $h_-h_+ = \sqrt{\big((b-|\kp|)^2+k_z^2-\omega^2\big)\big((b+|\kp|)^2+k_z^2-\omega^2\big)}$, which is nothing other than a torus. This structure analogous to $B_{xy}$ with extra branch-cut singularity pieces. For SA case at nonzero $\omega$, we observe a torus with connecting planes. See figure \ref{fig:Btz}(d). For SS case at nonzero $\omega$, it is branch-cut version of $B_{xy}^{(SS)}$. See figure \ref{fig:Btz}(h). The crucial difference is that there is no singularity at $\omega = 0$ in SS case.

\begin{figure}[t!]
	\centering
	\includegraphics[width=\textwidth]{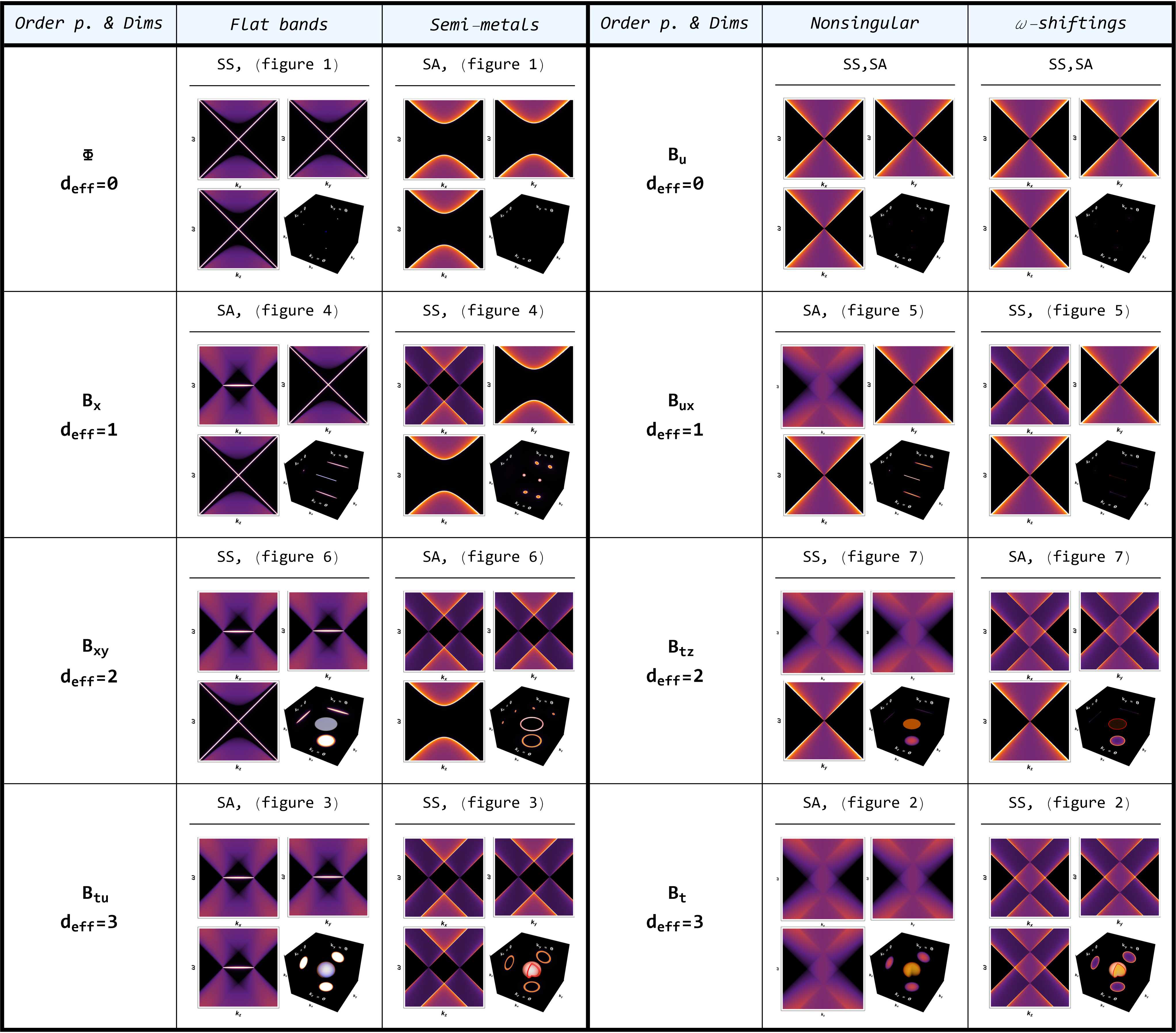}
	\caption{\small The classifications of spectral features for all interaction and quantization types. The table consists of spectral functions in $\omega\text{-}k_{x,y,z}$ and $k_x\text{-}k_y\text{-}k_z$ at $\omega \simeq 0$. The spectra have identical symmetry on the horizontal alignment and have the same spectral feature on the vertical alignment. $d_{eff}$ is the number of the flat band, semi-metals, nonsingular, and $\omega$-shifting spectra appearing in each k-space section.}
	\label{fig:classification}
\end{figure}

We have successfully obtained spectral functions for all interaction and quantization types where $\langle \mathcal{O}_{\Phi} \rangle = 0$. From their spectral features, we classify them according to their spectral features: various dimensional of flat bands, semi-metals,  singularity types, and $\omega$-shiftings. See figure \ref{fig:classification}.

\subsection{Spectrum in the presence of the order parameter's condensation}	
We have exhibited the SFs corresponding to the condensation in the alternative quantization   by employing the analytic Green's functions.  Here we   mention that the flow equation   can numerically compute the case where  order parameter fields  condensation in the standard quantization \cite{Yuk:2022lof,ABC,Lieb} although the analytic expressions are not possible.  
The spectral features in the probe limit can be generated by employing seven spectral features and modifying the symmetry in $\boldsymbol{k}$-space.

\begin{figure}[h!]
\centering
	\begin{subfigure}{0.3\textwidth}
		\centering
		\includegraphics[width=3.0cm]{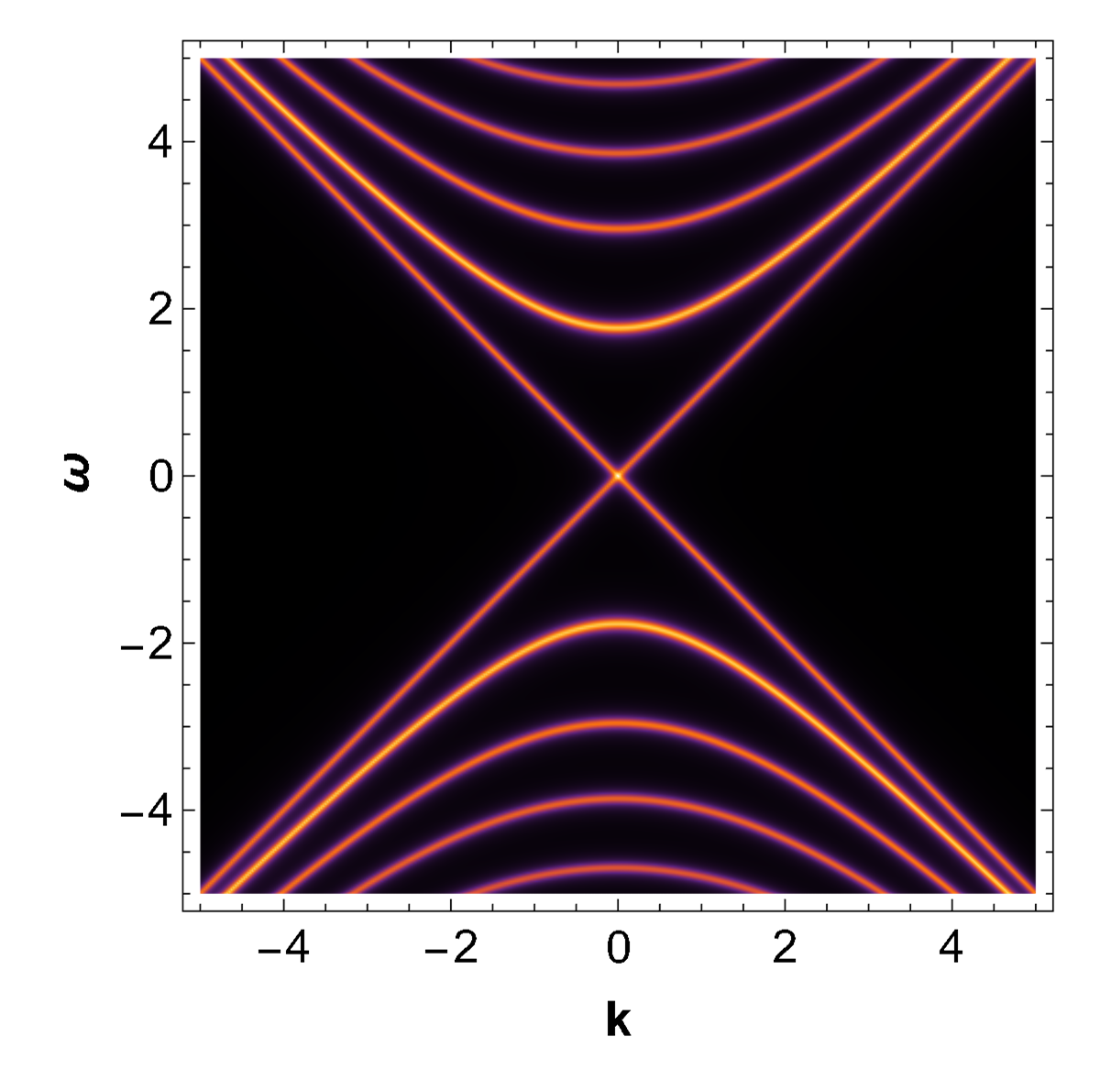}
		\caption{$M^{(SS)}$}
	\end{subfigure}
	\hspace{0.1cm}
	\begin{subfigure}{0.3\textwidth}
		\centering
		\includegraphics[width=3.0cm]{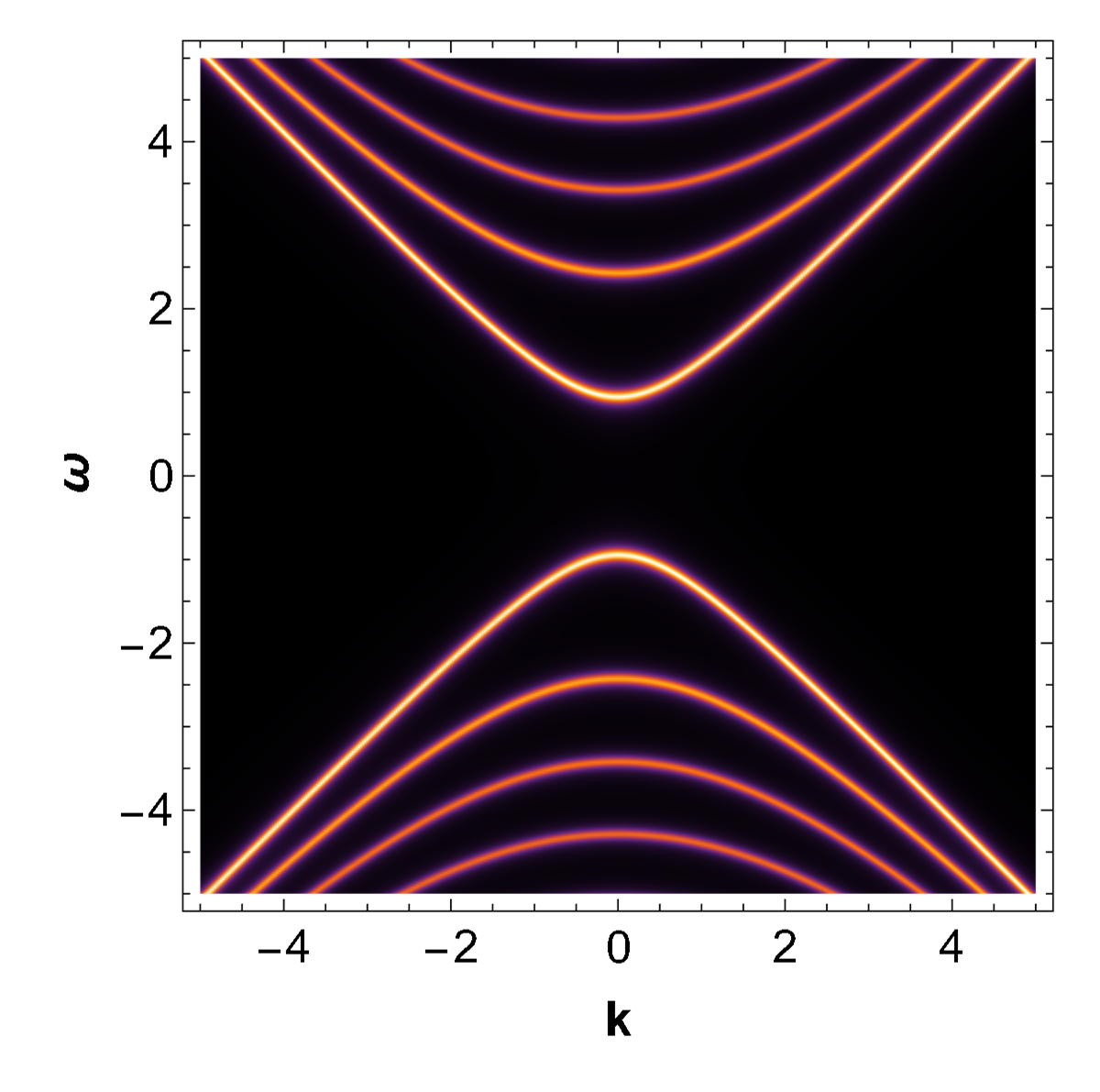}
		\caption{$M^{(SA)}$}
	\end{subfigure}
	\begin{subfigure}{0.3\textwidth}
		\centering
		\includegraphics[width=3.0cm]{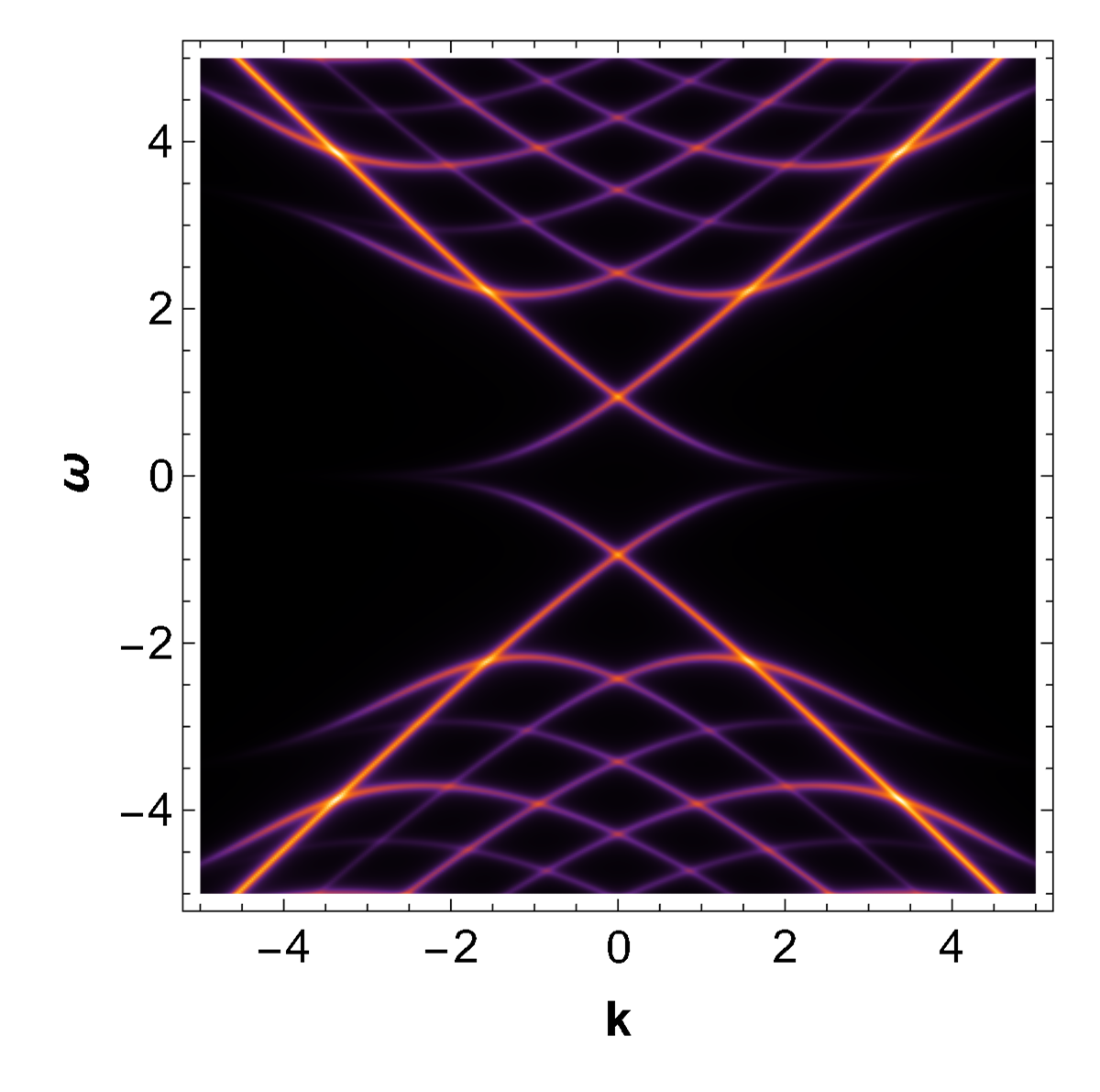}
		\caption{$B^{(1)(SS)}_{tu}$}
	\end{subfigure}\\
	\begin{subfigure}{0.3\textwidth}
		\centering
		\includegraphics[width=3.0cm]{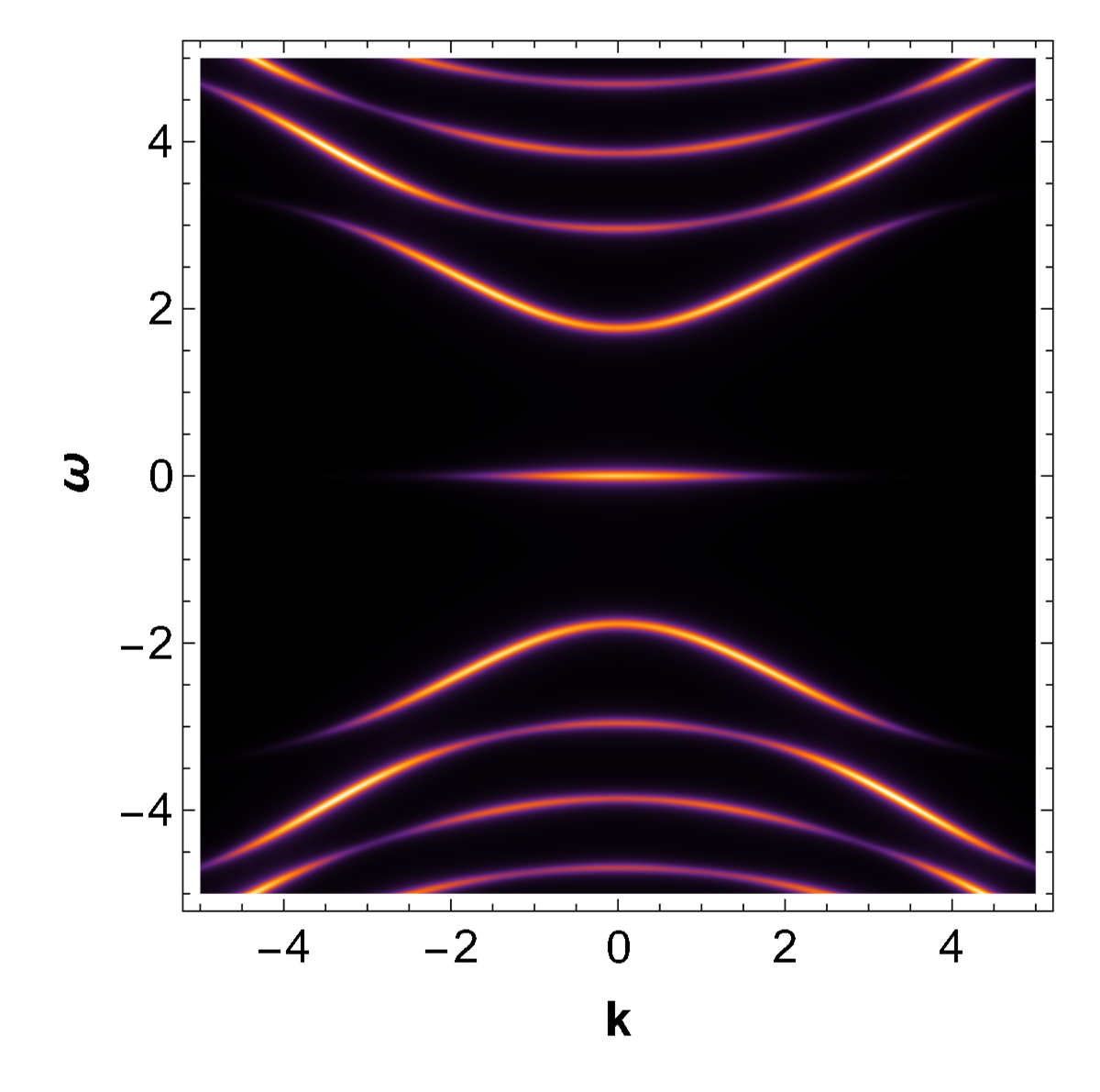}
		\caption{$B^{(1)(SA)}_{tu}$}
	\end{subfigure}
	\begin{subfigure}{0.3\textwidth}
		\centering
		\includegraphics[width=3.0cm]{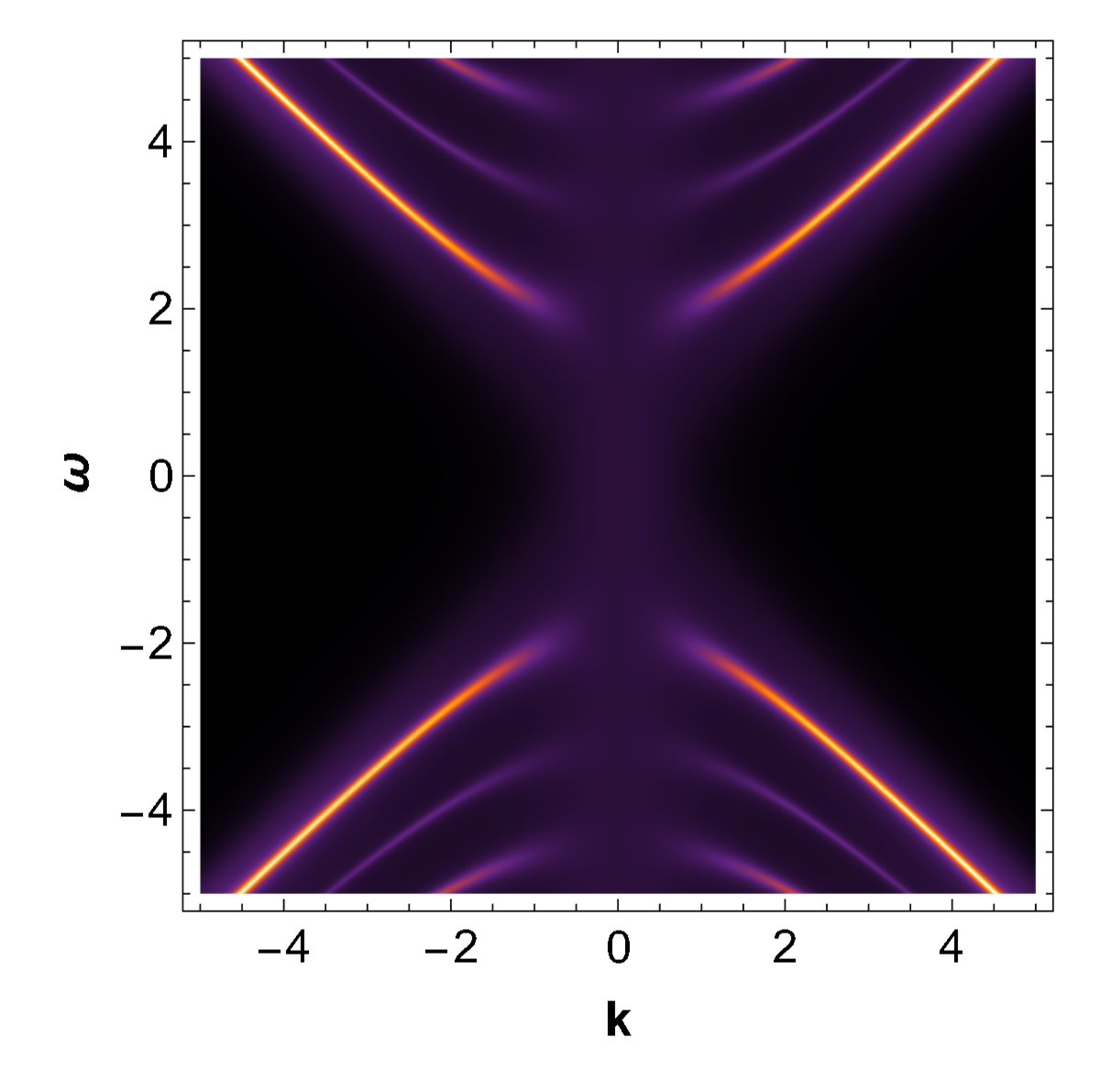}
		\caption{$B^{(2)(SS)}_{t}$}
	\end{subfigure}
	\begin{subfigure}{0.3\textwidth}
		\centering
		\includegraphics[width=3.0cm]{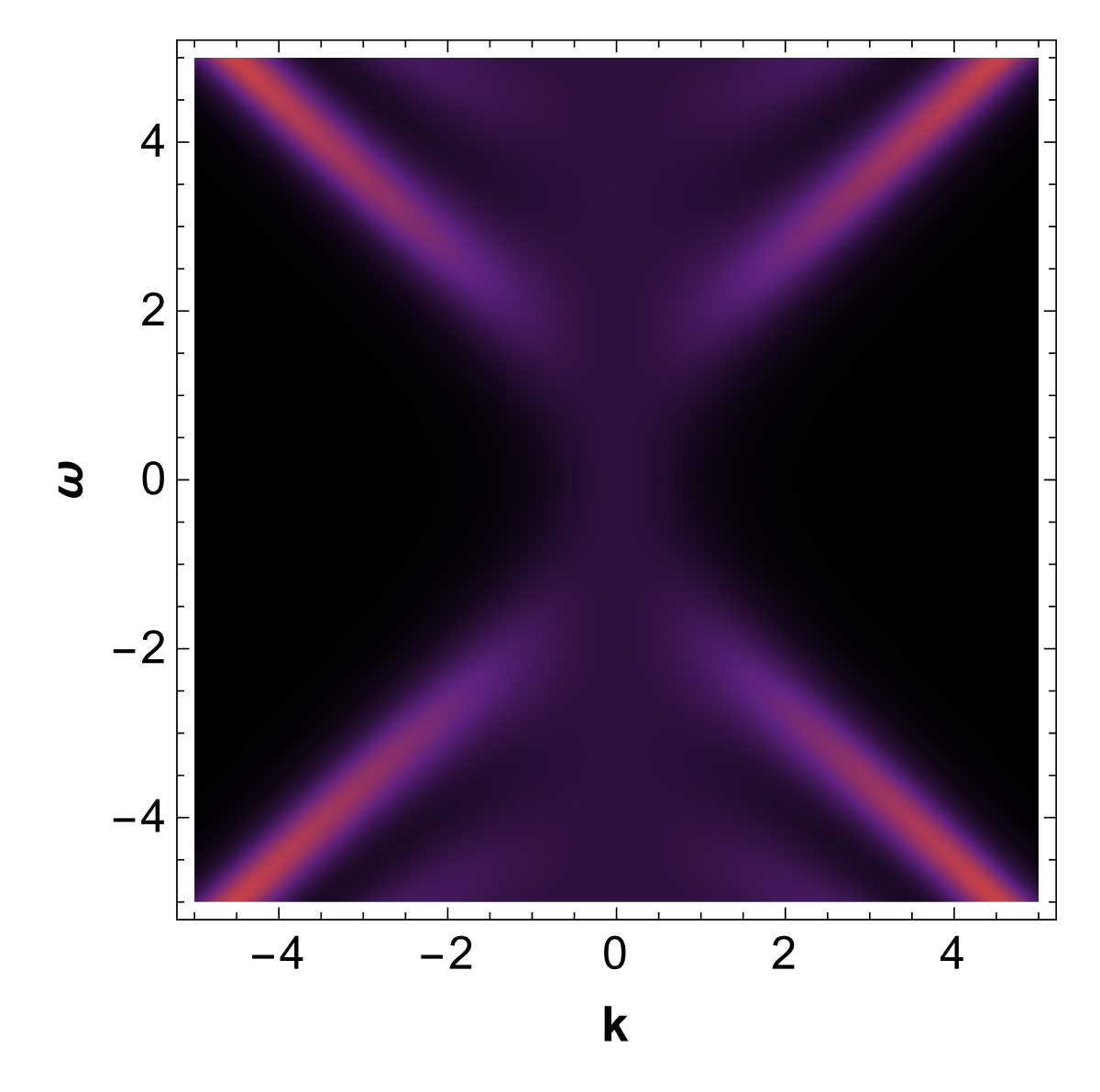}
		\caption{$B^{(2)(SA)}_{t}$}
	\end{subfigure}
	\caption{\small Spectral functions (SFs) which are associated with the condensation order parameter fields. (a,b) and (c,d) scalar and $B_{tu}$ condensation SFs, lead to the complete disappearance of the fuzzy spectra, resulting in the emergence of only the Kaluza-Klein (KK) modes. It is worth noting that even in the source case of these interactions, both simple pole and branch-cut singularities can occur. However, only the simple pole type is observed in the condensation cases. (e,f) $B_t$ condensation, the types of Green's functions appear similar to those observed in the source cases. Keep in mind that combining these SFs and introducing variations makes it possible to generate all 16 types of interaction SFs.}
	\label{fig:condensation}
\end{figure}

Our calculation shows that the interactions leading to the simple pole type in the alternative quantization, which are given by $\Phi,B_x,B_{xy},B_{tu}$, also yield the simple pole types singularity in the standard quantizated case as well.  But not vice versa.  For example, the scalar interaction with standard quantization have poles  in SA fermion,    but  in alternative quantization has branch-cut in the same SA fermion.  See figure 9(b). 
 In contrast, the remaining interactions $B_u,B_{ux},B_{tz},B_t$   yield branch-cut type Green's functions alternative quantization.  See figure \ref{fig:condensation}. 

In this section, we found three types of Green's function: pole type, branch-cut type singularity and branch-cut  non-singularity. In the upcoming section, we will compute the full back-reaction for each order parameter field. It is noteworthy that we have observed a remarkably stable pole-type singularity. This observation agrees with our previous research, wherein we interpreted the pole type, referred to as the zero mode, as a topological mode. The stability of this mode is assured by the boundary conditions of the fermions, which makes it topological \cite{Strangemetal}.

\section{Backreacted spectral functions}
So far, our calculations were done in the probe limit, where  back-reactions to the metric by the order parameter fields were neglected. Consequently,  the reliability of these approximate analytic expressions can be asked and the only way to answer is to actually carry out the full back-reacted solution. In this section,  we carry out this program and compare it with the essential probe limit analytic results.
 Since any of the back reaction calculation in volve full scale numerical work, and preliminary calculation with low numerical grid  gave us the tentative result that 
 the Green function with pole type singularity is stable under perturbation while 
 the branch cut type singularity is not. Therefore we also expect that the full scale back reacktion should be similar. In this section, we perform only for the space-like antisymmetric tensor type order parameter, postponing other cases to the future work. We use special lagrangian which permit the non-zero source and zero condensation. In other words, we use the theory that allowed the alternative quantization of the order parameter field. 
 Our result will show that indeed our expectation is correct. 
 
%
%
%
  We follow the fundamental action model which is given in \cite{PhysRevD.92.046001,Altschul:2009ae}. Additionally, to quantize real $B_{xy}$ alternatively, where only the leading term is nonvanishing, we introduce $A_t$ and set the highest order coupling term between $A_t$ and $B_{xy}$ as the source of spontaneous symmetry breaking. The model is then given by
\begin{align}
	\mathcal{S}_{g,B,A} &= \int d^5x\sqrt{-g}(R-2\Lambda-\frac{1}{4}F^2(1+8\gamma B^2)-\frac{1}{3}H^2- m^2 B^2 ),
\end{align}
here $A_\mu = A_t(u) dt$ and the antisymmetric field-strength tensor $H_{\lambda\mu\nu}$, given by
\begin{align}
	H_{\lambda\mu\nu} = \nabla_{\lambda}B_{\mu\nu}+\nabla_{\mu}B_{\nu\lambda}+\nabla_{\nu}B_{\lambda\mu}\quad; \quad \nabla_{\lambda}B_{\mu\nu} = \partial_\lambda B_{\mu\nu}-\Gamma^{\gamma}_{\lambda \mu}B_{\gamma\nu}-\Gamma^{\sigma}_{\lambda \nu}B_{\mu\sigma},
\end{align}
where $B_{\mu\nu} = B_{xy}(u)~dx\wedge dy$. Notice that the choice of our $A_\mu,B_{xy}$,  leading to $F^{\mu\nu}B_{\mu\nu} = 0$, and in this scenario, $F^2B^2$ becomes the highest order term. 

It is important to note that by setting $m^2=1$, the space-like antisymmetric 2-tensor field $B_{xy}(u)$ diverges near the boundary,
\begin{align}
	B_{xy}(u) \approx \frac{B_{xy}^{(-1)}}{u} + B_{xy}^{(1)} u + \cdots.
\end{align}
The presence of the singularity in the expression causes the numerical challenges. Consequently, we define a new variable, $\mathcal{B}_{xy}(u) \coloneqq u B_{xy}(u)$,  significantly enhancing numerical convenience by eliminating the singularity from the expression. With a small value $\epsilon$, we can determine the sub-leading term $B_{xy}^{(1)}$  by $\mathcal{B}_{xy}^{'}(\epsilon)/2\epsilon$ .

\begin{figure}[t!]
	\centering
	\begin{subfigure}{0.4\textwidth}
		\centering
		\includegraphics[width=5cm]{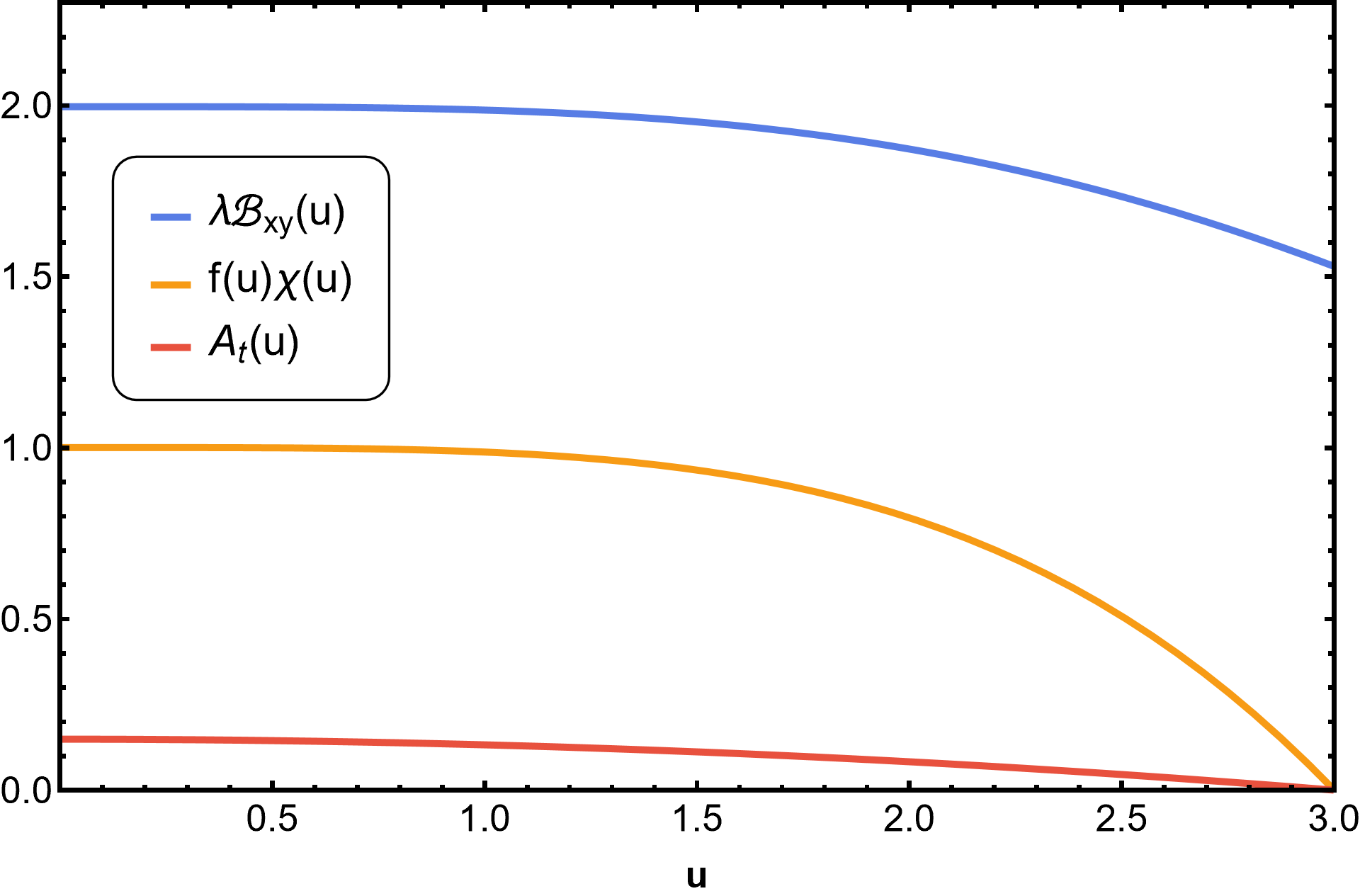}
		\caption{Background Fields Profile}
	\end{subfigure}\\
	\vspace{0.3cm}
	\begin{subfigure}{0.2\textwidth}
		\centering
		\includegraphics[width=2.8cm]{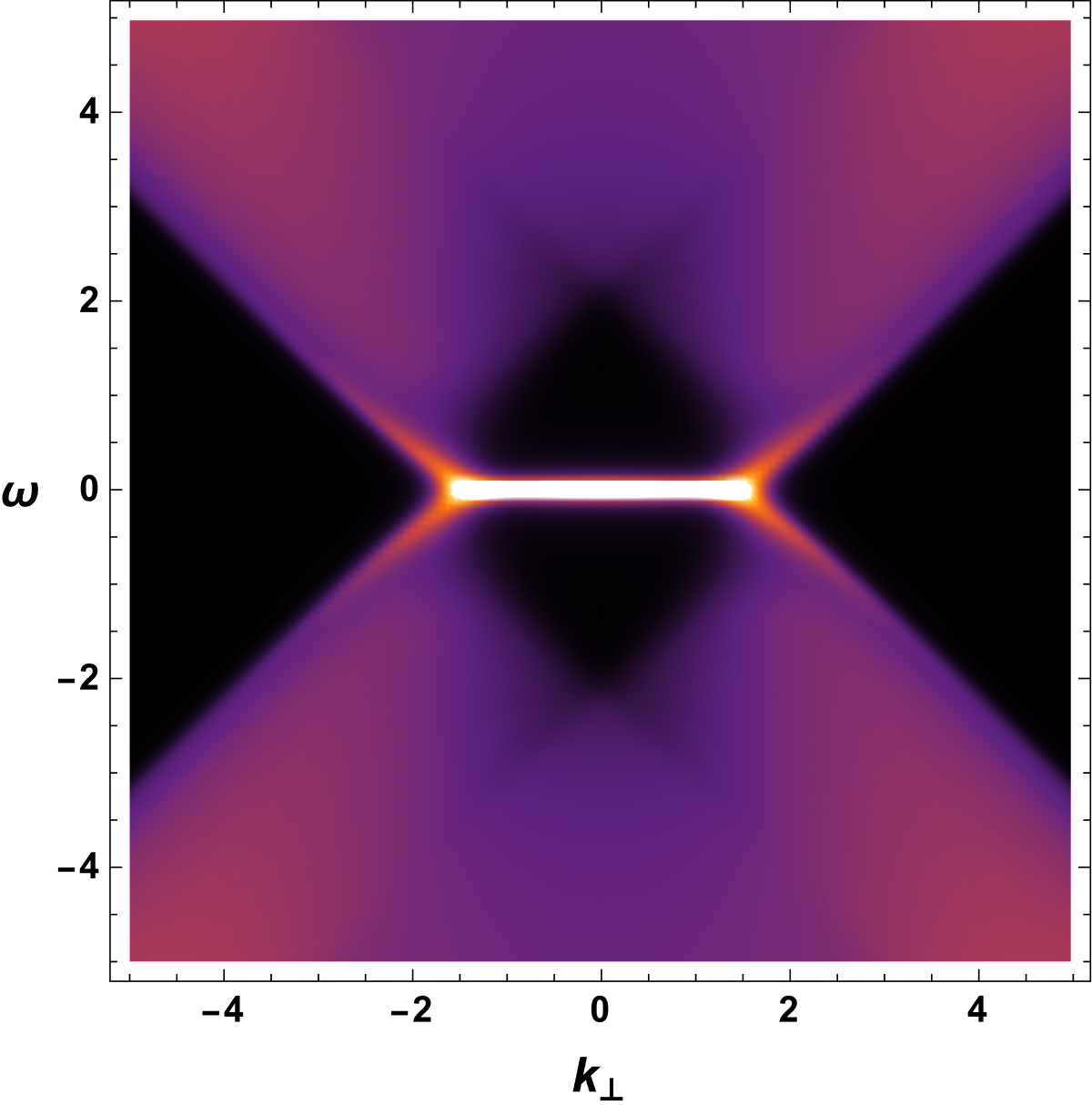}
		\caption{$B_{xy}^{(SS)}$, $\omega\text{-}\kp$}
	\end{subfigure}
	\begin{subfigure}{0.2\textwidth}
		\centering
		\includegraphics[width=2.8cm]{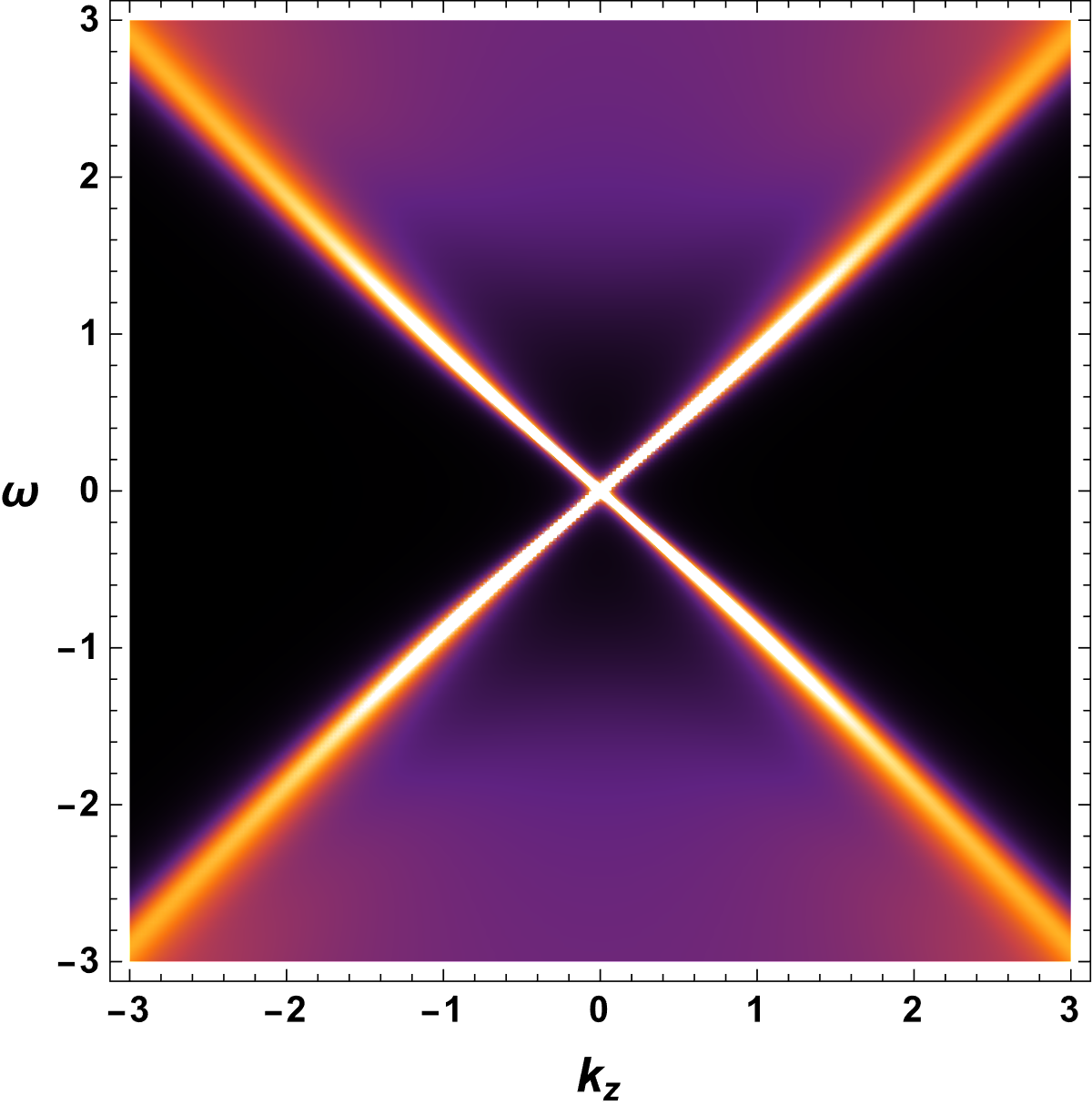}
		\caption{$B_{xy}^{(SS)}$, $\omega\text{-}k_z$}
	\end{subfigure}
	\begin{subfigure}{0.2\textwidth}
		\centering
		\includegraphics[width=2.8cm]{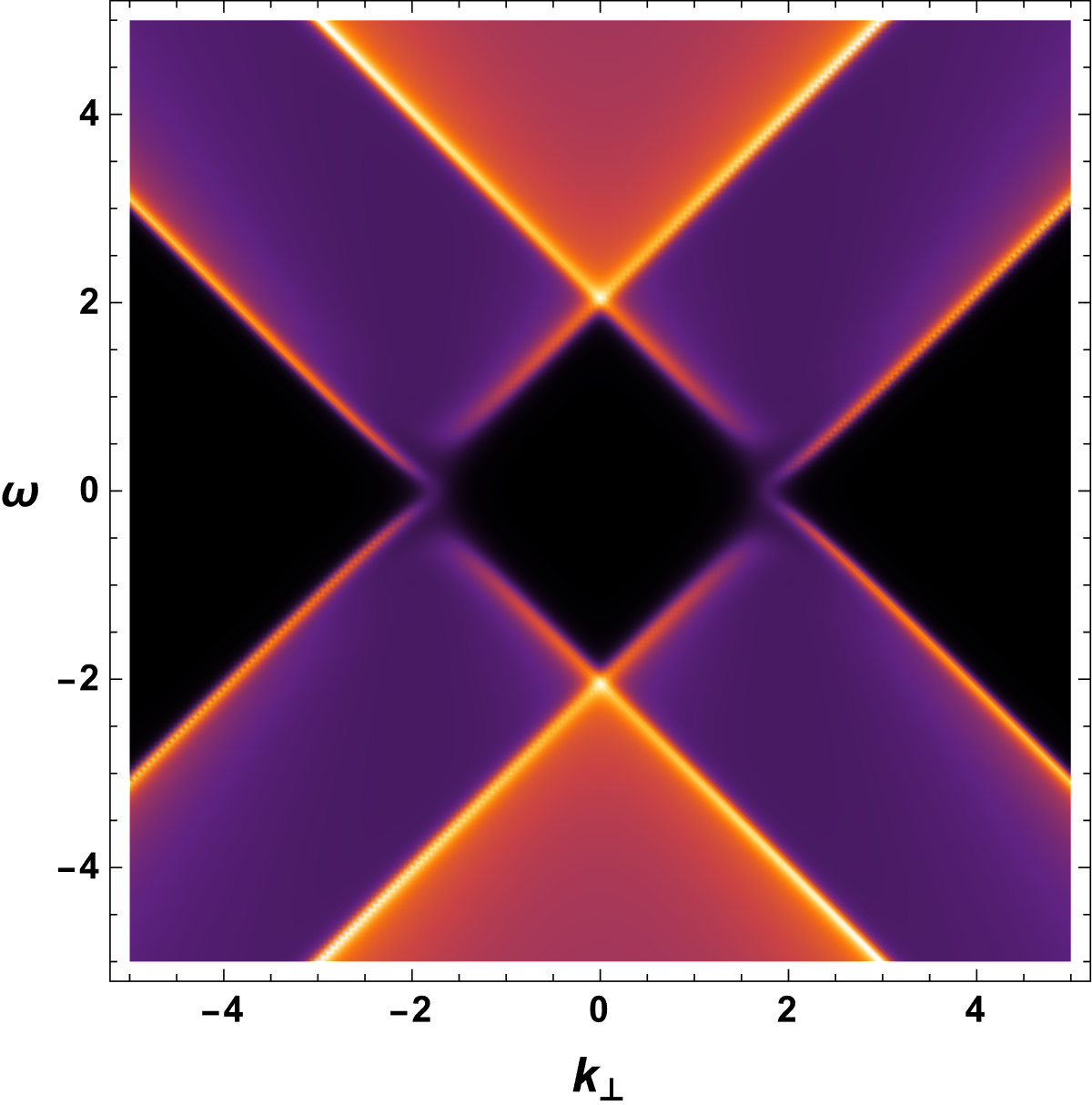}
		\caption{$B_{xy}^{(SA)}$, $\omega\text{-}\kp$}
	\end{subfigure}
	\begin{subfigure}{0.2\textwidth}
		\centering
		\includegraphics[width=2.8cm]{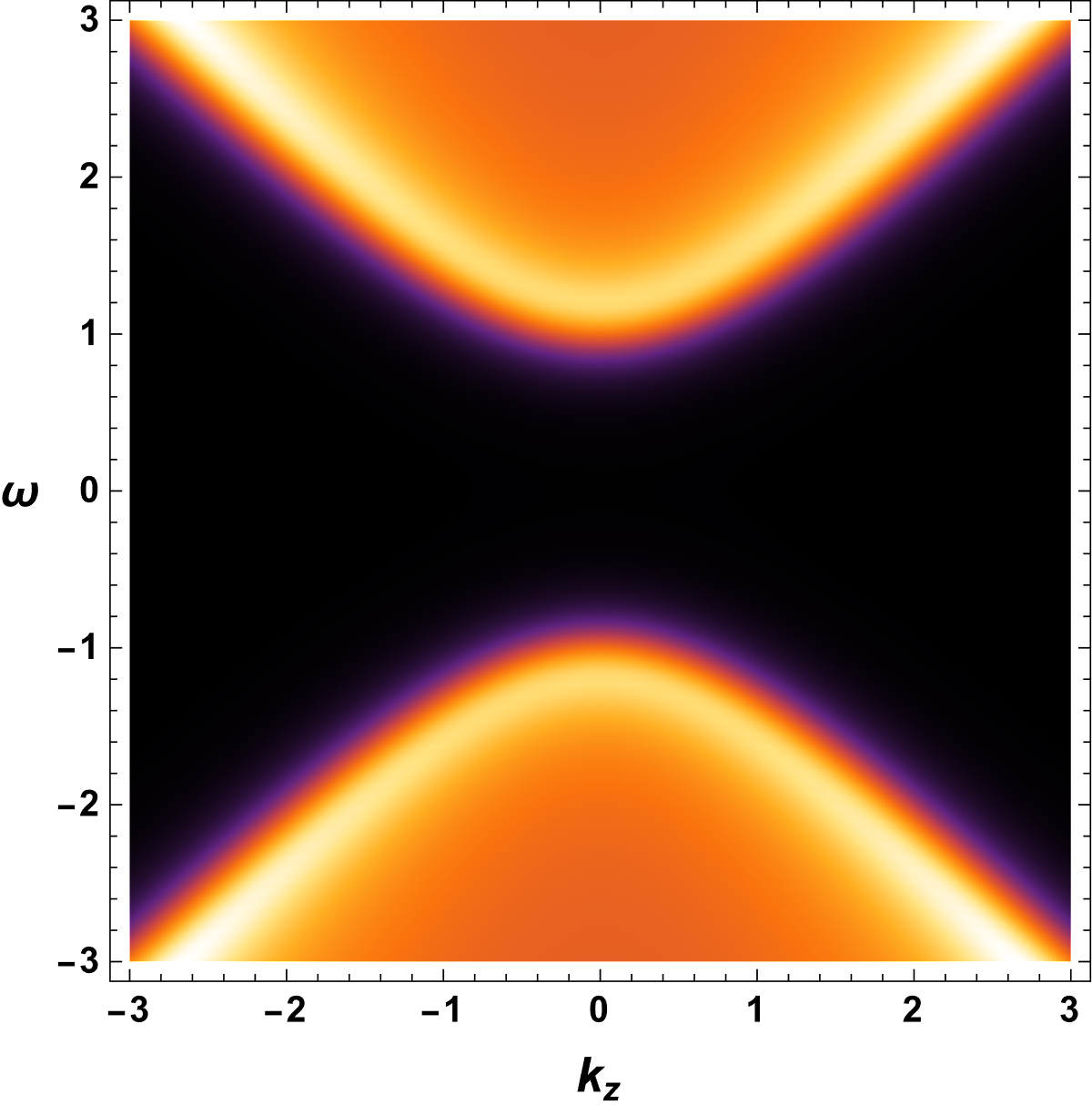}
		\caption{$B_{xy}^{(SA)}$, $\omega\text{-}k_z$}
	\end{subfigure}
	\caption{\small Backreacted fields profile and fermions spectral function (SF) of $B_{xy}$. (a) Backreacted background fields. (b,c) Backreacted fermions SF of $B_{xy}^{(SS)}$ compared to the analytic result, figure \ref{fig:Bxy}(e,f). (d,f) Backreacted fermions SF of $B_{xy}^{(SA)}$ compared to the analytic result, figure \ref{fig:Bxy}(a,b). We set parameters $\gamma = -2.0, Q = 3.2 \times 10^{-2}, u_H = 3.0$, and $\lambda = 153$, which is chosen for comparing with figure \ref{fig:Bxy}. }
	\label{Bxyback}
\end{figure}

In the back-reaction calculation, we take the following ansatz,
\begin{align}
	ds^2  = \frac{1}{u^2} (-f(u)\chi(u) dt^2 +\sum_{i = 1}^{3} dx_i^2 + \frac{du^2}{f(u)}),
\end{align}
the background fields equations of motion are given as follows
\begin{align}
	\mathcal{B}_{xy}^{''} +\mathcal{B}_{xy}^{'}(\frac{f'}{f}+\frac{\chi^{'}}{2\chi}-\frac{1}{u})
	-\frac{\mathcal{B}_{xy}}{u}(\frac{f'}{f}+\frac{\chi^{'}}{2\chi}-\frac{1}{u}-\frac{4u^5Q^2\gamma \mathcal{B}_{xy}}{f(1+8u^2\gamma\mathcal{B}_{xy}^2)^2}+\frac{1}{uf}) &= 0, \label{Beq}\\
	f^{'} - \frac{f}{3u} \big(12+u^2(\mathcal{B}_{xy}-u\mathcal{B}_{xy}^{'})^2 \big) -\frac{1}{3}u\mathcal{B}_{xy}^2- \frac{u^5Q^2}{6(1+8u^2\gamma\mathcal{B}_{xy}^2)}+\frac{4}{u} &= 0,\\
	\chi^{'}+\frac{2}{3}u\chi(\mathcal{B}_{xy}-u\mathcal{B}_{xy}^{'})^2 &= 0,  \\
	A_t^{'}+\frac{uQ\sqrt{\chi}}{1+8u^2\gamma \mathcal{B}_{xy}^2} &= 0. \label{Aeq}
\end{align}
where $Q$ is the effective charge density. We utilize the shooting method to search for the solution which satisfies the following boundary conditions,
\begin{align}
	f(u_H),~A_t(u_H) = 0, \quad \chi(\epsilon) = 1 , \quad \mathcal{B}_{xy}(u_H) = \text{finite}, \quad \mathcal{B}_{xy}^{'}(\epsilon)/2\epsilon = 0, \label{bdycons}
\end{align}
to perform the calculation, we consider the near horizon behaviour of fields which can be obtained by the Taylor   expansion of the fields in the following ways:
\begin{align}
	(\mathcal{B}_{xy}(u),f(u),\chi(u),A_t(u)) \simeq \sum_{i=0}^{n}(b_i,f_i,\chi_i,a_i)\Big(1-\frac{u}{u_H}\Big)^i.
\end{align}
By plugging in the above expansion into the fields equations (\ref{Beq})-(\ref{Aeq}),  we can determine $(b_i,f_i,\chi_i,a_i)$ in terms of the horizon value  $(b_0,\chi_0,u_H)$. Together with the boundary conditions (\ref{bdycons}), the solutions of the full back-reaction can be obtained. 

Figure \ref{Bxyback} shows a calculation result with the back-reaction in alternative quantization of $B_{xy}$ order parameter field. By the given parameters, we numerically get $B^{(-1)}_{xy} = 1.3\times10^{-2},B^{(1)}_{xy} = 8.6 \times 10^{-6}$ with $T = 0.10$ . We also emphasize that in the alternative quantization, the source term of the order parameter is small. So that the effect of the symmetry breaking is located near $(\omega,k) = {\bf 0}$.  To observe the symmetry breaking effect, we amplify the coupling of the fermions by introducing a magnification constant, denoted as $\lambda$. This results in the transformation: $B_{xy}\psi^{(1)}\Gamma^{xy}\psi^{(2)} \rightarrow \lambda B_{xy}\psi^{(1)}\Gamma^{xy}\psi^{(2)}$.

The obtained results reveal the following: i) The degrees of freedom near $k_\perp = |\lambda b|$, spread out but the pole singularity structure remains stable. On the other hand, the singularity structure along the branch-cut shows deformation. ii) Regardless of fermion quantization, the main spectral functions maintain the same feature compared to the probe limit results. These results emphasize the reliability of probe-limit results, providing the advantage of avoiding complexities and excessive time calculations associated with full back-reaction calculations.


\section{Discussion}
In this paper, we found the analytic expressions of the Green's function of fermions under the various types of symmetry breaking:  vector and tensor as well as a few types of scalars.  We classified the propagator according to the types of singularities: Some of them have branch cut types but some of them have pole types. By having the analytic expressions, although it is in the probe limit, we now understand why various dimensional flat bands exist and why they have finite regions of support.

Our setup refers to the order parameter field configuration with zero condensation, $\langle \mathcal{O}_{\Phi} \rangle = 0$.  For the scalar condensation in $\text{AdS}_4$, an analytic study has already been made and reported \cite{Strangemetal}, but in the context of $\text{AdS}_5$,  the presence of the condensation term gives the Dirac equations 
nontrivial dependence of $u$, making the solvability unavailable for all types of order parameter fields. This is the reason why analytic calculation for condensation was not considered in the present work.

To support the analytic results which are obtained from the probe background and also with only source type order parameters, we performed the numerical analysis to find solutions of coupled system of gravity with space-like antisymmetric 2-tensor and use it to calculate the spectral function of the fermions. Comparing the spectral functions of fermions with and without the back-reaction,  we observed a qualitative agreement in the structural features. In other types of the order parameter, we do not show the results in this work, but we also found such qualitative agreement in the most cases, which will be extended to future projects. 

One should also notice that our analytic results are associated with the zero temperature results, which is an extremal limit, so they should not compare with the result at high temperatures. The singularity structure significantly changes as we change the temperature. The most important remark is that our numerical spectral functions provide slight deformation and a reduction in sharpness compared to the zero-temperature limit, which is nothing but the effect of the back reaction. However, we observed the  different effect of the back reaction on pole and branch-cut type singularity.

In the cases of the pole-type singularity or flat bands, we observe the negligible back reaction and high stability of the flat bands, which the structure of the singularities quanlitively remain and closely match with our analytic results. Evidently, in the cases where the rotational symmetric flat band is present: $B_{xy}^{(SS)}$, the simple pole singularity structure is remarkably stable. This observation agrees with our previous work which interpreted the pole spectrum as a topological mode. However, we observed spreading out of density of state near $k_\perp = |\lambda b|$ which is the back-reaction effect.

On the other hand, in the cases of the branch-cut type singularity spectrum, the singularity structure is slightly different from our analytic results. For $B_{xy}^{(SA)}$, we observe that the singularity near $k_\perp = |\lambda b|$  is more fuzzy and reshapes compared to the analytic result. However, it still qualitatively remains the main feature.

We now list a few further future projects apart from removing the above limitations. 
First, it would be interesting to discuss the presence of branch-cut type of singularity in the propagator in view of the non-fermi liquid. 
Second, it should be possible to discuss the topology of the various spectral functions. 
This would give a precise answer to the question of what happens to the topology in the limit where the quasi-particle disappears.  We hope we can come back to this issue in the near future. Finally, notice that we are in lack of chiral $\Gamma^5$ matrix to represent the chirality of the boundary. 
For this reason, we did not discuss the chiral dynamics in this paper. We think that it must be done by introducing another flavor of fermion to double the degrees of freedom. 
\clearpage
\appendix
\begin{center}
	{\Large \bf Supplementary Materials}
\end{center}
\begin{table}[t!]
	\renewcommand{\arraystretch}{2.0}
	\centering
	\resizebox{.6\textwidth}{!}{%
		\begin{tabular}{|c|c|}
			\hline
			\multicolumn{1}{|c|}{Interaction types} &  Duality  \\ \hline
			\multicolumn{1}{|c|}{$M_0$}               &  $\mathbb{G}_{M_0, AdS4}^{(SS)} =
			\mathbb{G}_{M_0, AdS5}^{(SS)}\big|_{k_z = 0} \eqtr \mathbb{G}_{M_{50}, AdS4}^{(SA)} $                           \\
			\multicolumn{1}{|c|}{$B_x$}               &   $\mathbb{G}_{B_x, AdS4}^{(SS)} =
			\mathbb{G}_{B_x, AdS5}^{(SS)}\big|_{k_z = 0} \eqtr \mathbb{G}_{B_{5x}, AdS4}^{(SA)} $                          \\
			\multicolumn{1}{|c|}{$B_{t}$}             &  $\mathbb{G}_{B_t, AdS4}^{(SS)} =
			\mathbb{G}_{B_t, AdS5}^{(SS)}\big|_{k_z = 0} \eqtr \mathbb{G}_{B_{5t}, AdS4}^{(SA)} $                           \\
			\multicolumn{1}{|c|}{$B_{xy}$}            &  $\mathbb{G}_{B_{xy}, AdS4}^{(SS)} =
			\mathbb{G}_{B_{xy}, AdS5}^{(SS)}\big|_{k_z = 0} \eqtr \mathbb{G}_{B_{tu}, AdS4}^{(SA)} $                           \\
			\multicolumn{1}{|c|}{$B_{tx}$}            &   $\mathbb{G}_{B_{tx}, AdS4}^{(SS)} =
			\mathbb{G}_{B_{tx}, AdS5}^{(SS)}\big|_{k_z = 0} \eqtr \mathbb{G}_{B_{ux}, AdS4}^{(SA)} $                          \\
			\multicolumn{1}{|c|}{$B_{ux}$}            &  $\mathbb{G}_{B_{ux}, AdS4}^{(SS)} =
			\mathbb{G}_{B_{ux}, AdS5}^{(SS)}\big|_{k_z = 0} \eqtr \mathbb{G}_{B_{tx}, AdS4}^{(SA)} $                             \\
			\multicolumn{1}{|c|}{$B_{tu}$}            &   $\mathbb{G}_{B_{tu}, AdS4}^{(SS)} =
			\mathbb{G}_{B_{tu}, AdS5}^{(SS)}\big|_{k_z = 0} \eqtr \mathbb{G}_{B_{xy}, AdS4}^{(SA)} $                            \\
			\multicolumn{1}{|c|}{$B_u$}               &   $\mathbb{G}_{B_{u}, AdS4}^{(SS)} =
			\mathbb{G}_{B_{u}, AdS5}^{(SS)}\big|_{k_z = 0} \eqtr \mathbb{G}_{B_{u}, AdS4}^{(SA)} $                            \\
			\multicolumn{1}{|c|}{$M_{50}$}            & $\mathbb{G}_{M_{50}, AdS4}^{(SS)} =
			\mathbb{G}_{B_{z}, AdS5}^{(SS)}\big|_{k_z = 0} \eqtr \mathbb{G}_{M_{0}, AdS4}^{(SA)} $                             \\
			\multicolumn{1}{|c|}{$B_{5x}$}            &   $\mathbb{G}_{B_{5x}, AdS4}^{(SS)} =
			\mathbb{G}_{B_{zx}, AdS5}^{(SS)}\big|_{k_z = 0} \eqtr \mathbb{G}_{B_{x}, AdS4}^{(SA)} $                           \\
			\multicolumn{1}{|c|}{$B_{5t}$}            &    $\mathbb{G}_{B_{5t}, AdS4}^{(SS)} =
			\mathbb{G}_{B_{zt}, AdS5}^{(SS)}\big|_{k_z = 0} \eqtr \mathbb{G}_{B_{t}, AdS4}^{(SA)} $                         \\
			\multicolumn{1}{|c|}{$B_{5u}$}            &   $\mathbb{G}_{B_{5u}, AdS4}^{(SS)} =
			\mathbb{G}_{B_{zu}, AdS5}^{(SS)}\big|_{k_z = 0} \eqtr \mathbb{G}_{B_{u}, AdS4}^{(SA)} $                            \\ \hline
		\end{tabular}%
	}
	\caption{\small The daulity of 2-flavors fermions Green's function between $\text{AdS}_4$ and $\text{AdS}_5$. Remember that the dualities between the different quantization choice in $\text{AdS}_4$ $(\eqtr)$ are just only the trace eqivalence of the Green's functions not for the component level. Therefore, the full expressions of 4 by 4 Green's function are different.}
	\label{table:1}
\end{table}
\section{$\text{AdS}_4$ Green's functions and the spectral features dualities}\label{app4}
Even the spectral functions for $\text{AdS}_4$ were studied in our previous work but the analytic results have not been completely reported yet. However, we found the duality between $\text{AdS}_4$ and $\text{AdS}_5$ Green's functions which we will show in this section. We follow the gamma matrix convention for $\text{AdS}_4$ in \cite{Nonfermi,AdS4,Lieb,ABC,Yuk:2022lof}. 
\begin{align}
	\Gamma^{\underline{t}} &= \sigma_1 \otimes i\sigma_2, \quad \Gamma^{\underline{x}} = \sigma_1 \otimes \sigma_1,\quad \Gamma^{\underline{y}} = \sigma_1 \otimes \sigma_3,\quad
	\Gamma^{\underline{u}} = \sigma_3 \otimes \sigma_0,\quad
	\Gamma^{5} = i\Gamma^{\underline{t}}\Gamma^{\underline{x}}\Gamma^{\underline{y}}\Gamma^{\underline{u}}.
\end{align}
Under this convention, $\Gamma^5 \equiv \Gamma^{\underline{z}}$  in our main $\text{AdS}_5$ context, so that the bulk gamma matrices can be decomposed as follows,
\begin{align*}
	\Gamma^{\underline{\mu}} = \begin{pmatrix}
		0  & \gamma^{\underline{\mu}}\\
		\gamma^{\underline{\mu}} & 0
	\end{pmatrix},~~ \Gamma^{\underline{\mu\nu}} = \begin{pmatrix}
		\gamma^{\underline{\mu\nu}} & 0 \\
		0& \gamma^{\underline{\mu\nu}} 
	\end{pmatrix},
	\Gamma^{\underline{\mu u}} = \begin{pmatrix}
		0 & -\gamma^{\underline{\mu}} \\
		\gamma^{\underline{\mu}} & 0 
	\end{pmatrix},
\end{align*}
the structure of the gamma matrices shows us that the result of Green's functions will be the same as $\text{AdS}_5$ by removing complex conjugates in the expressions and eliminating the third momentum $k_z$. The reason is that the differential equations remain the same as before. For pseudo-interaction types, however, they might be confused due to lack of the fifth gamma matrix in $\text{AdS}_5$. According to $\Gamma^5, \Gamma^{5x},\Gamma^{5t}, \Gamma^{5r}$, they are equivalent to $\Gamma^z, \Gamma^{zx},\Gamma^{zt}, \Gamma^{zr}$, respectively, by setting $k_z = \bf{0}$. As a result, the duality  of the Green's functions between $\text{AdS}_5$ and $\text{AdS}_4$ are obtained. See table \ref{table:1}. 

Our results with quantization dualities of $\text{AdS}_4$ spectral features agree with our previous numerical study \cite{AdS4}. However, we emphasize the trace equivalence on the table means that the trace of the dual Green's functions is the same, but each element in the Green's function might vary. So, this is a crucial clue indicating that each interaction's topological properties are different even though they share the same spectral feature.

\acknowledgments
This  work is supported by Mid-career Researcher Program through the National Research Foundation of Korea grant No. NRF-2021R1A2B5B02002603    and the Basic research Laboratory support program RS-2023-00218998,  and the brain Link program NRF-2022H1D3A3A01077468. 
We  thank the APCTP for the hospitality during the focus program, where part of this work was discussed.

\bibliographystyle{JHEP}
\bibliography{reference.bib}
\end{document}